\documentclass[pre,twocolumn,aps,showpacs,floatfix]{revtex4}
\usepackage{amssymb}
\usepackage{graphicx}
\usepackage{amsmath}
\usepackage{times}
\usepackage{latexsym}
\usepackage{subfigure}
\usepackage{color}
\usepackage{CJK}
\usepackage{appendix}

\usepackage[normalem]{ulem}

\usepackage[colorlinks,bookmarks=false,citecolor=blue,linkcolor=red,urlcolor=blue]{hyperref}

\begin{document}


\title{Self-trapping dynamics in a 2D optical lattice}

\author{ Shuming Li, Salvatore R. Manmana$^{*}$, Ana Maria Rey}
\affiliation{JILA, NIST, Department of Physics, University of Colorado, 440 UCB, Boulder, Colorado  80309, USA}
\affiliation{$^{*}$Institute for Theoretical Physics, University of G\"ottingen, D-37077 G\"ottingen, Germany}
\author{Rafael Hipolito}
\affiliation{School of Physics, Georgia Institute of Technology, Atlanta, GA 30332, USA}
\author{Aaron Reinhard$^{\dagger}$, Jean-F\text{\' e}lix  Riou, Laura A. Zundel, and David S. Weiss}
\affiliation{Physics Department,  The Pennsylvania State University, 104 Davey Lab, University Park, Pennsylvania 16802, USA}
\affiliation{$^{\dagger}$Department of Physics,  Otterbein University, 1 South Grove Street, Westerville, OH 43081, USA}
\date{\today}

\begin{abstract}

We describe  theoretical models {\color{black}for the} recent experimental observation of  Macroscopic Quantum Self-Trapping (MQST) in the
transverse dynamics of  an ultracold bosonic  gas  {\color{black} in  a 2D lattice.}
 The pure mean-field model based on the solution of coupled nonlinear equations  fails to reproduce the experimental observations.
 It greatly overestimates the initial expansion rates at short times and predicts a slower expansion rate of the cloud at longer times.
It also predicts the formation of a hole surrounded by a steep  square fort-like barrier which was not observed in the experiment. An improved theoretical description based on a simplified  Truncated Wigner Approximation (TWA){\color{black},}  which adds phase and number  fluctuations in the initial conditions{\color{black},}  pushes the theoretical results closer to the experimental observations but fails to quantitatively reproduce them. An explanation of the delayed expansion {\color{black} as a consequence of a new type of self-trapping mechanism}, where quantum correlations suppress tunneling even when there are no density gradients, is discussed and supported by numerical time-dependent Density Matrix Renormalization Group (t-DMRG) calculations performed in a simplified {\color{black} two coupled} tubes set-up.

\end{abstract}
\maketitle	
\section{introduction}\label{sec:introduction}

The  {\color{black}term ``self-trapping''} was first introduced by Landau {\color{black}to describe} the motion of an electron in a crystal lattice \cite{Landau}.
According to Landau, the  electron  gets dressed by the lattice polarization and deformation caused by its own motion. The resulting {\it polaron}
can be  localized or self-trapped in the presence of strong electron-lattice interactions. Nowadays, the concept of {\color{black}self-trapped} polarons  is  {\color{black}applied to}  various condensed matter systems such as semiconductors, organic molecular crystals,  {\color{black}high temperature cuprate superconductors}  and  colossal magnetoresistance manganates \cite{Shluger1993,Salje1995,Millis1995}.  {\color{black}Still}, the theory of  the dynamical coupling of a conduction electron to lattice phonons is a complicated  highly nonlinear problem which is difficult to tackle even  in one-dimensional situations  \cite{Wellein1998}. Many questions remain open.

A parallel  situation in which {\color{black}self-trapping} has been theoretically investigated and experimentally observed is in dilute ultra-cold bosonic gases {\color{black}\cite{Milburn1997,Smerzi1997,Raghavan1999,Trombettoni2001,Morsch2002,Smerzi2003,Albiez2005,Anker2005,Alexander2006,Xue2008,wuster2012}}.  In those systems{\color{black},} self trapping is {\color{black} caused by} a competition between the discreteness introduced by the lattice {\color{black}and} non-linear  effects.  {\color{black}However, in these systems the lattice} potential is rigid, imposed by optical fields, and the non-linearity has its roots in intrinsic interatomic interactions. The self-trapping mechanism is then  inherently a many-particle {\color{black} phenomenon,}   commonly referred to as Macroscopic Quantum Self -Trapping (MQST).

{\color{black}MQST has been  experimentally observed with cold atoms in double well setups \cite{Albiez2005} and in 1D lattice geometries \cite{Anker2005}.  MQST can be understood as wave-packet} localization  caused by the interaction-induced suppression of tunneling  in the presence of chemical potential gradients. In those cases{\color{black},} a mean field description in terms of the so  called  time-dependent Gross-Pitaevskii  equation {\color{black} (GPE)  is} shown to be sufficient to  describe the experiments.

Recently, we reported \cite{Aaron} what seems to be a new type of  MQST effect  induced by quantum correlations  in  a gas of ultra-cold  ${}^{87}$Rb  atoms trapped in  a two dimensional optical lattice. The lattice potential {\color{black}created} coupled  arrays of one dimensional tubes.  A  striking  inhibition of the transverse expansion dynamics was observed as the depth of the 2D lattice was increased.
   In contrast to other self-trapping experiments, in this case  the cloud was allowed to freely expand along the axis of the tubes, so that the atom density decreased with time. The inhibited expansion persisted until the density became  too low to sustain MQST.

The observed transverse localization was not reproduced by a pure mean field analysis. When phase fluctuations among the tubes were added to the theory in an attempt to approximate the effect of correlations, the agreement with the experiment improved for shallow lattices. But for deep lattices, the calculated suppression of tunneling was less complete than what was observed in the experiment, even with maximal phase fluctuations. In Ref. \cite{Aaron} we proposed a possible mechanism for this effect that cannot be reproduced by modifying mean field theory. 1D gases develop quantum correlations to reduce their mean field energy. Atoms might be prevented from tunneling because of the mean field energy cost they would have to pay in the adjacent tube, with whose atoms they are not properly correlated. 

{\color{black} A simple understanding of this mechanism can be gained in the Tonks-Girardeau regime where due to quantum correlations  bosons fermionize \cite{Girardeau1960,Kinoshita2004,Paredes2004tg}. Instead of a macroscopically occupied single particle wave function, the many-body wave function is more like a set of spatially distinct  single-particle wave functions to avoid interactions. In this case it is clear that  tunneling between adjacent tubes can be suppressed due to the large interaction energy cost an atom would have to pay due to mismatched correlations between tubes, i.e., if  the only region that an atom can access via tunneling (which exponentially decreases with  distance) in the adjacent tube overlaps with  the wave function of an atom already present there. }

{\color{black}The succession of MSQT experiments from the double well  BEC (coupled 3D gases with no lattice), to coupled 2D gases in 1D lattices, to this work, coupled 1D gases in 2D lattices, is something of a microcosm of the way cold atom physics has developed. } As atoms become more confined and the coupling strength increases, the GPE ceases to capture all the important physics. Quantum correlations play an enhanced role, giving rise to new phenomena and posing a challenge to theory. It is natural to speculate that the new type of MSQT that appears in 2D lattice experiments could also be affecting transport in 3D lattice experiments.

In this paper we describe in detail various theoretical models developed {\color{black}to try to understand} the  observed behavior. We first use  a mean field formulation and discuss why the model fails to capture the experimentally observed dynamics. Next, we present an attempt to add quantum fluctuations to the mean field theory by means of the so called truncated Wigner approximation (TWA) {\color{black}(}See Ref. \cite{Polkovnikov2010} and references therein{\color{black})}. {\color{black}As we will explain below, a rigorous implementation of the TWA is challenging for the experimental conditions we want to model. Nevertheless, we show  it is possible to qualitatively capture some of the experimental features by using an ad-hoc approximation of the TWA (aTWA) which includes correlations between adjacent tubes, but neglects quantum correlations within each tube. The aTWA pushes the theoretical predictions closer to the experimental observations, especially for moderate lattice depths. The description breaks down and fails to quantitatively reproduce the deeper lattice dynamics.}  To further investigate the role of genuine quantum correlations as {\color{black}a} localization mechanism and to benchmark the validity of the  aTWA we study the expansion dynamics  in a simpler   two-tube  {\color{black} configuration using exact numerical time-dependent Density Matrix Renormalization Group (t-DMRG) methods} {\color{black}(}See Ref. \cite{Schollwock2005,Daley2004} and references therein{\color{black})}.  

The outline of this paper is as follows.  In section \ref{sec:MQSToverview}, we review  the concept of MQST, its treatment  in terms of a mean field model and the application of this method  to   prior experiments. In Sec. \ref{sec:MQST2D}, we introduce  our experimental set-up and develop  a mean field treatment of the dynamics  capable of dealing with the axial expansion along the tubes. We compare our  solutions, obtained by numerical evolution of the non-linear equations, with the experimental data.  To overcome the limitations of the mean field model, in section \ref{sec:beyondMFM}  we discuss ways to  incorporate quantum fluctuations using {\color{black}the aTWA} methods. In section \ref{sec:twotubes}, we use t-DMRG to study the problem of atoms tunneling between two tubes and {\color{black}use it to test} the parameter regime in which the GPE and the aTWA are valid. Those calculations {\color{black}show} the relevance of quantum correlations during the expansion dynamics. Finally{\color{black},} we conclude in Sec. \ref{sec:conclusion}.

\section{MQST: an overview}\label{sec:MQSToverview}

A dilute  bosonic gas forms a Bose Einstein condensate  (BEC) below a critical temperature. In atomic  BECs{\color{black},} most atoms occupy the same ground state{\color{black}, so} quantum fluctuations can be neglected to a good approximation. Therefore  the field-operator can be replaced by a c-number, $\hat{\Psi}(\vec{x})\rightarrow\Psi(\vec{x})$. The function $\Psi(\vec{x})$ is often called the ``condensate wave function''  or  ``order parameter''{\color{black}, and it evolves} {\color{black}according to  the time-dependent GPE} \cite{pethick2002},
\begin{equation}
i\hbar \frac{\partial \Psi}{\partial t}=-\frac{\hbar^2}{2M}\nabla ^2\Psi+\left(V_{ext}+g|\Psi|^2\right)\Psi,
\end{equation}
where  $V_{ext}$ is the external potential and $g=\frac{4\pi\hbar^2 a_s}{M}$  with $M$ the atomic mass and $a_s$ the scattering length.  The GPE has {\color{black}proven} to be a very powerful and successful {\color{black}way to describe} the dynamics  of BECs.

A BEC confined in a double well potential can experience MQST. This was  predicted  theoretically \cite{Milburn1997,Smerzi1997,Raghavan1999} and has been directly observed  experimentally  \cite{Albiez2005}. MQST was induced in the experiment by creating a population difference between the left and right well. For {\color{black}small} population imbalance, the atoms oscillate back and forth between the wells{\color{black},} as expected for non-interacting atoms{\color{black},} but beyond a critical population imbalance the  mean field interactions inhibit tunneling and   atoms are self-trapped.

At the mean-field level, the double-well system can be described by a two-state model,
\begin{eqnarray}
i\hbar\frac{d}{dt}\left(\begin{array}{c}
\psi_L\\
\psi_R
\end{array}\right)
=\left(\begin{array}{cc}
 U_L N_L +E_L^0 & -J \\
 -J&U_RN_R +E_R^0
\end{array}\right)\left(\begin{array}{c}
\psi_L\\
\psi_R
\end{array}\right),\label{gpe2wells}
\end{eqnarray}
where $\psi_L$ and $\psi_R$ are ``macroscopic" wave-functions of the particles in the left and right wells, $J$ is the tunneling  matrix element and $U_{L,R}$ describe the  interaction energy cost of having  two particles in the left or in the right well respectively. $ N_{L,R}=|\psi_{L,R}|^2$ are {\color{black}the numbers of particles in the two wells} and  $E_{L,R}$ are the respective zero point energies.  These parameters can be determined by the overlap integrals of the eigenfunctions of  isolated wells.

We set $E_L+E_R=0$ and  denote the zero point energy difference between the two levels as $\Delta E=E_R-E_L$.  Assuming {\color{black}that all particles initially} occupy one well, when $U_{L,R} N_{L,R}\rightarrow 0$,  Eq. (\ref{gpe2wells}) yields  {\color{black}Josephson}  oscillations of   the population with a frequency $\sqrt{4 J^2+\Delta E^2}$ and  an amplitude  $\frac{4J^2}{4J^2+\Delta E^2}$.  {\color{black}When $\Delta E\gg J$, the system is off-resonant and the atoms oscillate rapidly with a small amplitude around the initial configuration.}

When  $\Delta E=0$ and $\left| U_L N_L-U_R N_R\right|\gg J$, the atoms also  show reduced amplitude oscillations. In this case, however, the population imbalance  is self-locked to the initial value due to  MQST.  

MQST in 1D optical lattices has also been studied theoretically and observed experimentally {\color{black}\cite{Morsch2002,Anker2005,Smerzi2003,Trombettoni2001}}.
Atoms were first loaded in a 1D optical lattice {\color{black}with an additional dipole} trap, forming  arrays of 2D pancake-like BECs.  Then the dipole trapping beam along the lattice direction {\color{black}was} suddenly removed. Even though the 1D lattice system is more complicated than the double well, the physics {\color{black}responsible for the} MQST is similar {\color{black}in the two cases.} In the {\color{black}1D} lattice MQST manifests as a dynamical localization of an initially prepared wave packet. In contrast to the non-interacting case, in which a continuous increase of the width of the {\color{black}wave} packet with time is expected, interactions can stop the expansion. The MQST starts  at the edges of the cloud  where the density gradient is the largest. As the system evolves, atoms form  a hole at the center surrounded by  immobile steep edges. {\color{black}In Ref. \cite{Anker2005},  the MQST behavior was probed by directly imaging the atom spatial distribution.  At a critical value of interaction energy  a transition from {\color{black}diffusive} dynamics {\color{black}(monotonic expansion of the wave packet with time)} to MQST was observed.}

The local dynamics of MQST in lattices is  complex, but  the global dynamics, i.e., evolution of the root-mean-square (RMS) width of the wave packet {\color{black}can be} predicted analytically using a very simple model. {\color{black} A variational ansatz {\color{black}has been} used to predict  when MQST will occur and the result is solely determined by global properties of the gas \cite{Trombettoni2001}.  It  has been applied to 1D lattice systems  in which the lattice splits the BEC in an array of two dimensional ``disks" or pancakes \cite{Anker2005}.
The predictions of the model were found to be in qualitative agreement with the experimental observations.  Details of the variational method are presented in Appendix A.}

\section{MQST in a 2D lattice} \label{sec:MQST2D}

\begin{figure}
\includegraphics[width=3in]{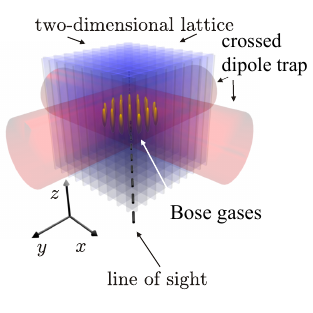}
\caption{(Color online) Schematic of the experimental set-up.  A BEC  of  $^{87}$Rb atoms  was initially prepared in the crossed-dipole-harmonic trap.  A 2D  optical lattice along the $x$-$y$ direction was then adiabatically  ramped on {\color{black}to create} an array of  {\color{black} quasi-1D} tubes.  The crossed dipole-trap was then turned off and the expansion dynamics in the presence of the 2D periodic potential were investigated by direct imaging.  The line of sight was {\color{black}at} 45 degrees {\color{black}from} the lattice directions (45 degrees from $x$ and $y$).  The 2D atom density distribution was {\color{black}recorded  after various  expansion times.} }\label{fig1}
\end{figure}

{\color{black}In previous work \cite{Aaron},}  we reported a similar yet {\color{black}richer} situation in which the expansion dynamics took place in an array of quasi-1D tubes created by a 2D lattice.    Atoms expanded along the tubes as they underwent their transverse dynamics. {\color{black}This made it possible to observe a self-trapping transition ``on the fly", as the overall density steadily dropped due to the axial expansion}. 

A BEC of $N \sim 3.5\times 10^5$ $^{87}\text{Rb}$ atoms was initially prepared in a crossed-dipole-harmonic trap with transverse frequency $\omega_{\perp}=2\pi\times 38$~Hz  and vertical confinement $\omega_{||}=2\pi \times 94 $~Hz. It was then loaded into a 2D square-lattice potential with lattice spacing $d=385$ nm.   The experimental set-up is shown in Fig. \ref{fig1}.    {\color{black} The {\color{black}blue-detuned} 2D  lattice was ramped up in time according  to {\color{black}$I(t)\propto [1-(t/\tau)]^{-2}$}, reaching a final lattice depth of $V_o=7.25E_R, 9.25E_R, 11E_R\ \text{or}\ 13E_R$ and creating an array of 1D tubes{\color{black}.  Here} $I$ is the lattice intensity, $\tau=4.15$ms is the time constant and $E_R=h^2/(8Md^2)$ is the recoil energy where $d$ is the lattice spacing.} After the initial preparation all the harmonic {\color{black}confining} potentials were suddenly  turned off  {\color{black}so that the {\color{black}atoms could expand} in} 
the 2D lattice. The expansion dynamics  {\color{black}were}  investigated by direct imaging. The line of sight was along a direction 45 degrees between the lattice directions (45 degrees from $x$ and $y$), and  the image of {\color{black}the}  2D  {\color{black} density-distribution squared was recorded as a function of time.} The  main finding was a suppressed  expansion rate of the {\color{black}RMS width of the density profile} along the lattice, consistent with  MQST. However, the expansion rate did not agree with the mean field predictions. Moreover, no  signature of MQST behavior  predicted by the mean-field theory was observed, such as the formation of  steep edges or  a hole at the center {\color{black}of the density profile.}

\subsection{Mean Field: Coupled GPEs}

For the 2D lattice configuration, the  total  external potential was given by  $V_{ext}(x,y)=V_{lat}(x,y)+V_{dip}(\vec{x})$ with  $V_{lat}(x,y)=V_{o}[\sin^2 ( \pi x /d)+\sin^2 ( \pi y /d)]$  the lattice potential and $V_{dip}(\vec{x})=\frac{1}{2}M\omega_{\perp}^{2}(x^{2}+y^{2})+\frac{1}{2}M\omega_{||}^{2}z^{2}$  the confinement introduced by the crossed-dipole-trap. The dipole confinement was turned off during  the expansion and therefore it was only {\color{black}relevant for determining the} initial conditions. A small  anti-trapping potential remained due to the Gaussian profile of the  laser {\color{black}beams that generated} the lattice. {\color{black}The parameters of the crossed dipole trap {\color{black}used} to set up the initial state in the theoretical model {\color{black}were} chosen so that the atom distribution matches the initial spatial extent and energy of the atoms in the experiment. {\color{black}The experimental parameters cannot simply be used because atoms in the experiment do not expand by as much as mean field theory predicts when the blue-detuned lattice is turned on.} The explanation for this may be related to the anomalous self-trapping that is the subject of this paper, but a full understanding of that effect will require future work. The suppressed atom expansion during the turn-on of the lattice might also mean that the initial atom correlations are not the same as in the mean field model.}

Assuming {\color{black}that} only the lowest band of the 2D optical lattices was populated, {\color{black}a} condition {\color{black}that was} satisfied in the experiment, we write $\Psi$ in terms of the lowest band Wannier orbitals,
\begin{equation}
\Psi=\sum_{m,n}W(x-d n)W(y-d m) \Phi_{nm}(z,t), \label{psi2d}
\end{equation}
where $\Phi_{nm}$  is the order parameter describing  the tube centered at lattice site $(n,m)$, and $N_{nm}(t)=\int dz |\Phi_{nm}(z, t)|^2$ is its corresponding atom number. The normalization condition
$\sum_{n,m}N_{nm}=N$ with $N$ the total number of atoms is satisfied.

{\color{black} 
By assuming $\Phi_{nm}(z,t)=\psi_{nm}(t)  \phi_{nm}[z,N_{nm}(t)]$ ($\int dz\left|\phi_{nm}[z,N_{nm}(t)]\right|^2=1$ and $N_{nm}=|\psi_{nm}(t)|^2$) and using a similar variational-method \color{black}to} the one described in Appendix A,  we can obtain the following  effective Hamiltonian,  
\begin{eqnarray}
H_{eff}
&=&N\left[-4J +\frac{9}{25}U_{eff} \left(\frac{2}{\pi r^2}\right )^{2/3}N^{2/3}\right]\nonumber\\
&=&N\left[-4J+\frac{9}{25}U\rho d^3\right]
\end{eqnarray}
{\color{black}where $r$ is the width of the wave packet in the lattice directions (in the lattice units).} Here we have defined the  2D interaction strength  as $U=\frac{g}{d} \int dx dy\ W^4(x)W^4(y)$. $U_{eff}=\left(\frac{9 M\omega_z^2 d^2U^2}{32}\right)^{1/3}$ plays a similar role as $U_{\alpha}$  in Appendix A, and $J=\int dx\ W(x-d) \left[\frac{\hbar^2}{2M}\frac{\partial^{2}}{\partial x^{2}}-V_{lat}\right]W(x)$.    $\rho=\frac{3}{4d^3}\left(\frac{2 N}{r^2\pi}\right)^{2/3}\left(\frac{2M\omega_z^2d^2}{3U}\right)^{1/3}$ is the 3D density  at the center of the cloud.
Based on the variational method, we predict that  the MQST  in the 2D optical lattice should take place  when  $U \rho d^3 /J \geq \frac{100}{9}$.

Although the variational method roughly describes the global dynamics and gives the threshold value $\rho_c$ of the MQST,  it {\color{black}only explicitly  accounts for the expansion  dynamics   across the lattice but not along the tube's direction}.  We first tried to adapt the variational methods to capture the expansion in both directions.  For that purpose, we  assumed  the density profile  along each tube $\phi_{nm}(z)$ to have either a Thomas-Fermi (TF) or a Gaussian shape with a width $R_{nm}(t)$ and a phase $\delta_{nm}(t)$. In the lattice direction on the other hand, the evolution of the amplitudes  $\psi_{nm}(t)$ at each lattice site was numerically computed. Unfortunately, we found that it  was  quantitatively accurate  {\color{black}only}  at very short  times. The reason is that (in the  experiment) atoms in the tubes  rapidly expanded  vertically  with a number-dependent expansion rate. {\color{black}Since the atoms tunneled  between neighboring tubes at the same time, these two processes destroyed  the assumed Gaussian or  TF shape in each tube  and  the  ansatz broke down.}

Since the variational method failed, we instead numerically solve for the axial dynamics. We neglect any temporal {\color{black}dependence} of the Wannier functions{\color{black},  which is} justified because the inter-well number/phase dynamics is much faster than the time associated with the change in shape of such functions.  Generically, the wave-field $\Psi(\vec{x},t)$ can still  be written in terms of Eqn. (\ref{psi2d}), {\color{black}so} we keep $\Phi_{nm}(z,t)$ as an arbitrary function. After integrating out the lattice directions,  we obtain the following Nonlinear Schr$\text{\" o}$dinger Equations and {\color{black} solve them numerically}.

\begin{widetext}
\begin{eqnarray}
i\hbar \dot{\Phi}_{nm}(z,t)&&=\left(-\frac{\hbar^{2}}{2M}\frac{\partial^{2}}{\partial z^{2}}+ V_{dip}(dn,dm,z)+U d |\Phi_{nm}(z,t)|^{2}\right)\Phi_{nm}(z,t)-J\left[\Phi_{n\pm1,m}(z,t)+\Phi_{n,m\pm1}(z,t)\right].\label{gpes}
\end{eqnarray}
\end{widetext}

{\color{black}To characterize the  MQST, we numerically compute  the  root mean square (RMS) transverse width, $x_{rms}$,  of a vertical slice of the cloud centered at z=0 and with a width corresponding to  {\color{black}$\pm 10\%$} of the total axial width (characterized by the Thomas-Fermi radius, $R_z$).    The reason for selecting a slice instead of integrating over the whole cloud is to consider  a sample of approximately constant density. }  Theoretically $x_{rms}$ is directly calculated as:  $x_{rms}=\sum_m \rho_m m^2/\sum_m \rho_m$ where $\rho_m= \int_{-Rz/10}^{Rz/10} dz  \sum_n|\Phi_{nm}(z,t)|^2$.
 
\begin{figure}[htb]
\subfigure[]{\includegraphics[width=1.6in]{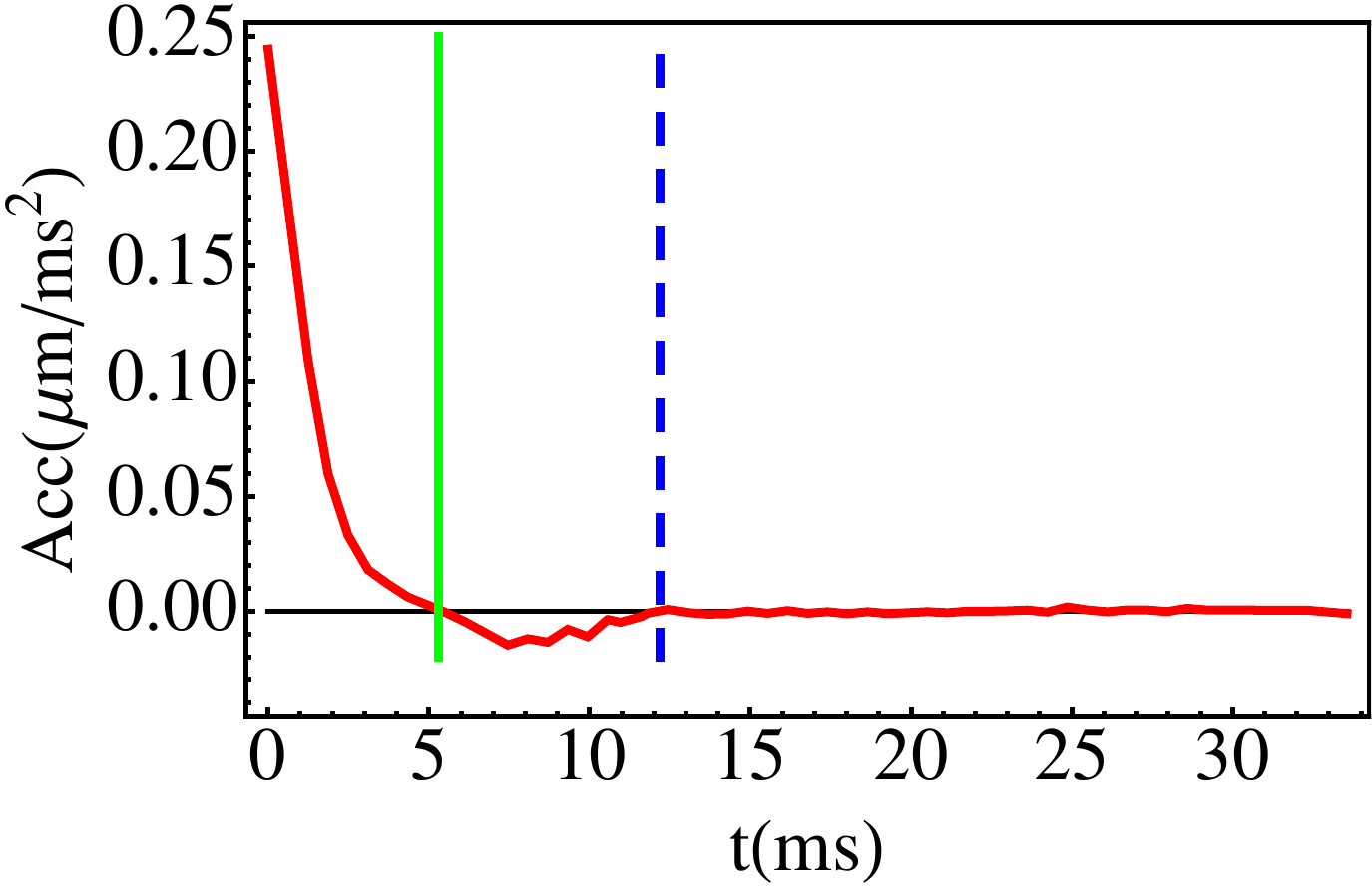}}\quad
\subfigure[]{\includegraphics[width=1.63in]{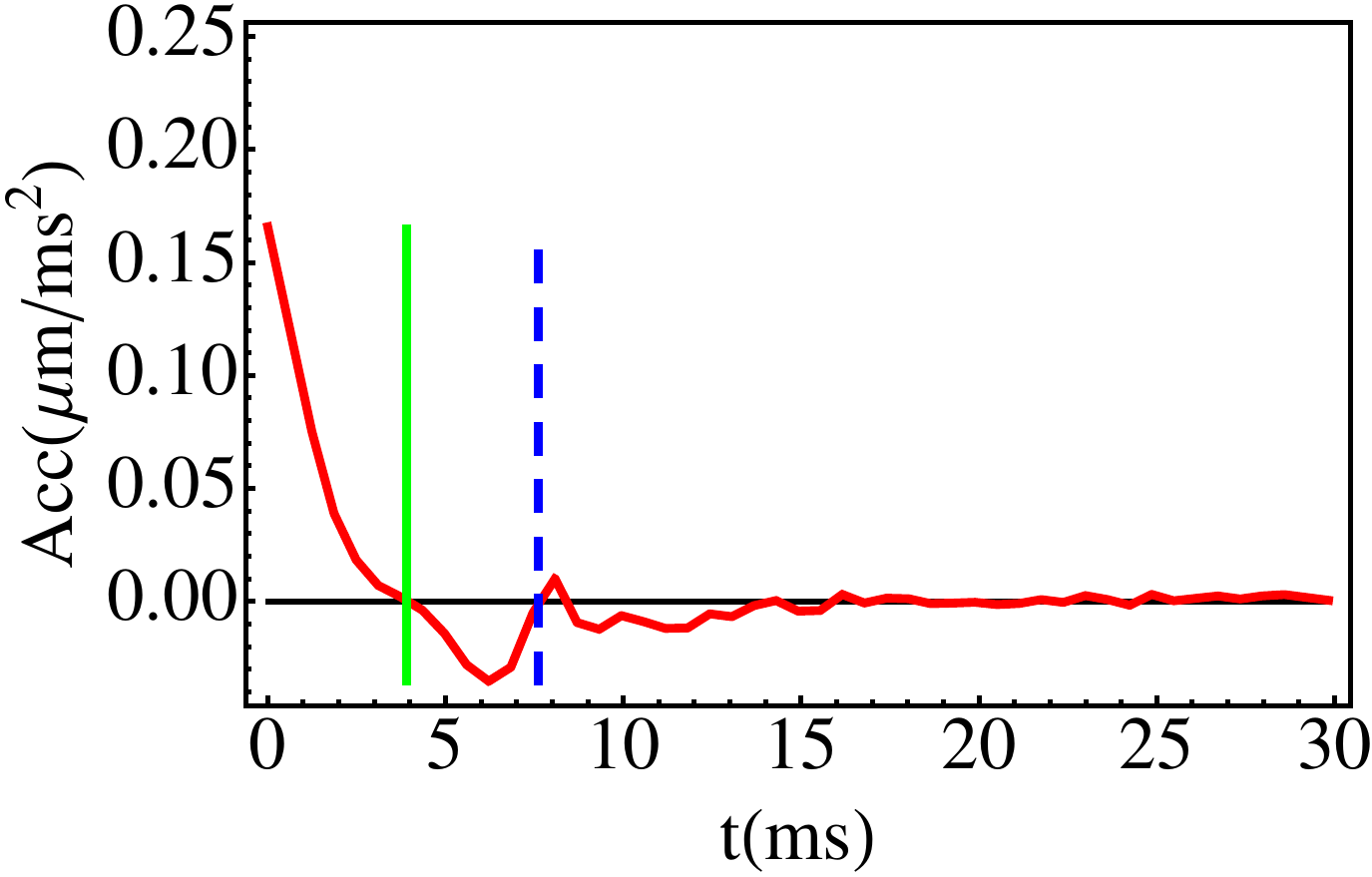}}\\
\subfigure[]{\includegraphics[width=1.6in]{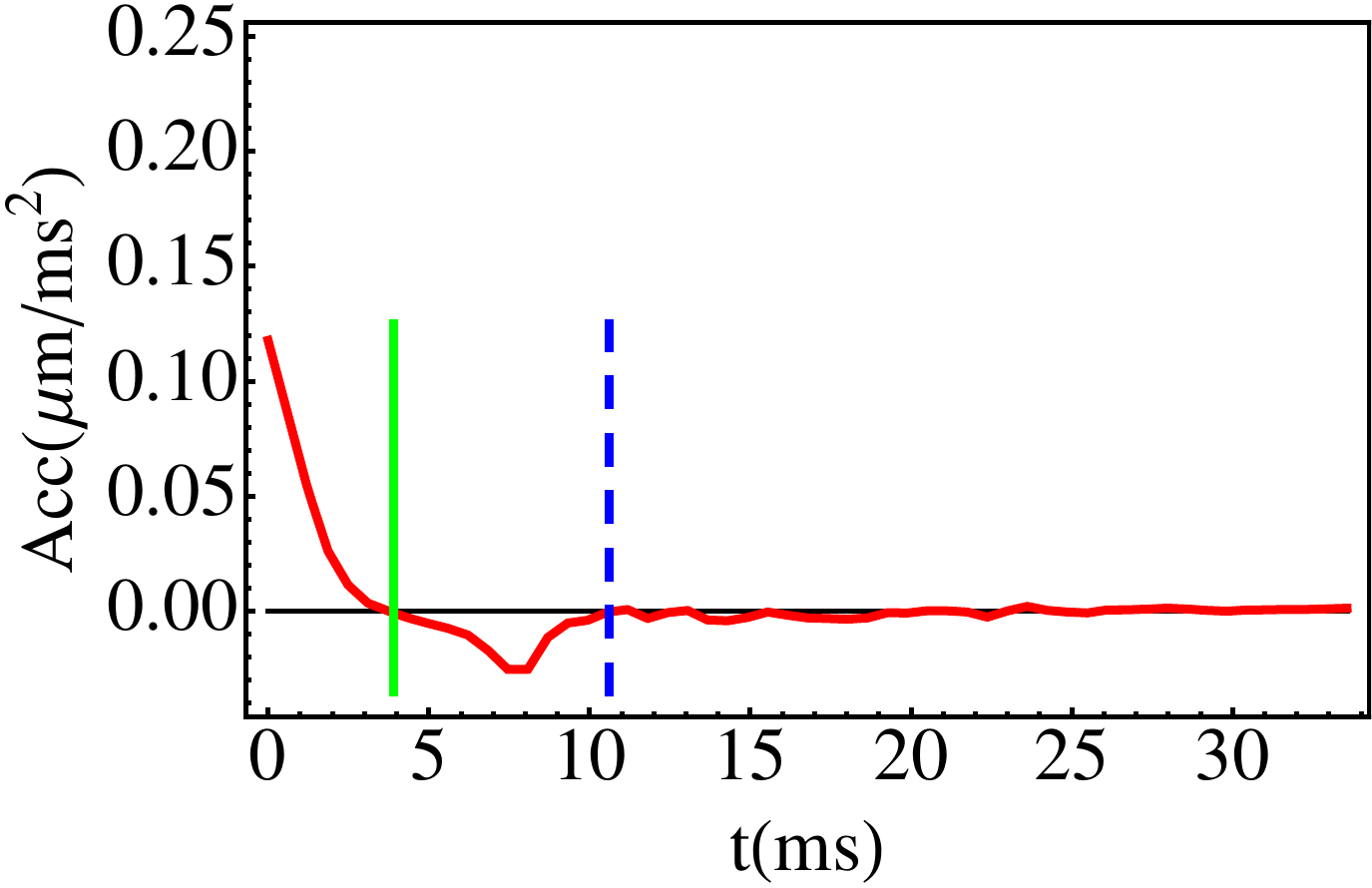}}\quad
\subfigure[]{\includegraphics[width=1.65in]{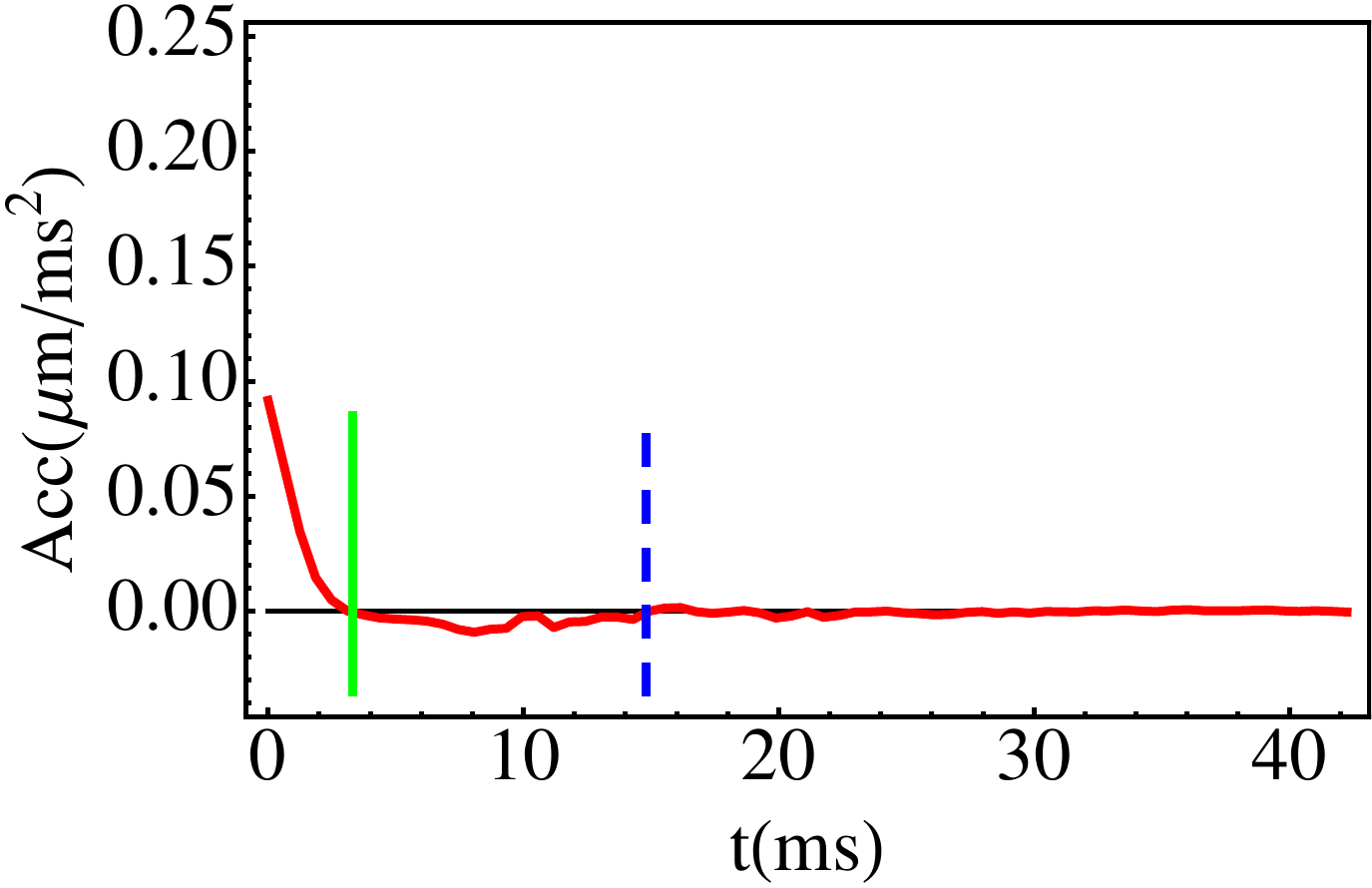}}
\caption{(Color online) The  acceleration  of $x_{rms}$ ($\ddot{x}_{rms}$) {\color{black}obtained from the mean field calculations} for all lattice depths used in the experiment{\color{black}: (a) 7.25$E_R$, (b) 9.25$E_R$, (c) 11$E_R$ and (d) 13$E_R$.}  From the sign of $\ddot{x}_{rms}$, the evolution can be separated  into three different regimes indicated by green solid lines and blue dashed lines. The  three regimes are  discussed in detail in the text.   {\color{black}MQST is signaled in the $x_{rms}$  as a negative acceleration. It starts at the green solid line and  stops  at $\sim t_c$ (indicated by the blue dashed line)  when the acceleration-curves cross 0 from below.}  The values of $U \rho(t_c)/J a^3$ for the  four depths in consideration at $t_c$ are $\{0.4, 3.0, 3.0, 3.0\}$. Those correspond to $\rho(t_c)/J=\{ 60, 350, 320, 300\}$ $\mu \rm{m}^{-3}$$E_R^{-1}$ respectively. } \label{acc_eta0}
\end{figure}

The  expansion  along the tubes {\color{black}makes} MQST a {\color{black}time-dependent} phenomenon{\color{black},}  since as the atoms freely {\color{black}expand}, the density {\color{black}decreases} with time until the system {\color{black}becomes} too dilute to sustain self-trapping. To determine  the critical value of $ \rho/J $ at which MQST {\color{black}stops,}  we numerically {\color{black}calculate}  $\ddot{x}_{rms}$. {\color{black} We construct an approximation function that interpolates the numerical values of $x_{rms}(t_i)$ at recorded time $t_i$. As shown in Fig. \ref{acc_eta0}, the curves of $\ddot{x}_{rms}$ are not completely smooth, especially at long times,  due to accumulated numerical errors. }  {\color{black}We define the critical time $t_c$ to be the time}  when  the acceleration first {\color{black}crosses} zero  from below, i.e.,  it is negative for times shorter than the critical time due to the MQST (See Fig.~\ref{acc_eta0}). {\color{black} Note that $t_c$ as well as $ \rho(t_c)/J $ are only roughly estimated. } 

The variational method {\color{black}gives} the criterion, $U \rho(t_c)/J d^3\approx 11$. Numerically, we {\color{black}found} $U \rho(t_c)/J d^3= \{ 0.4, 3.0, 3.0, 3.0\}$ for lattice depths $7.25$, $9.25$, $11$ and $13 E_R${\color{black},}  which correspond to $\rho(t_c)/J= \{60, 360,320, 300\}\ \mu \rm{m}^{-3}$ $E_R^{-1}$  respectively.  Although the numerical values of $ \rho(t_c)/J $  are smaller than the value predicted by the variational method,  {\color{black}they} are almost constant{\color{black},} except for the  {\color{black}one associated to the} {\color{black}$7.25 E_R$ curve}, consistent with the general concept of a self-trapping threshold.

There are three distinct regimes clearly observed in $\ddot{x}_{rms}$ which we {\color{black}use} to characterize the dynamics.   We will discuss those different regimes in detail  below.
\begin{figure}[htb]
\includegraphics[width=3.2in]{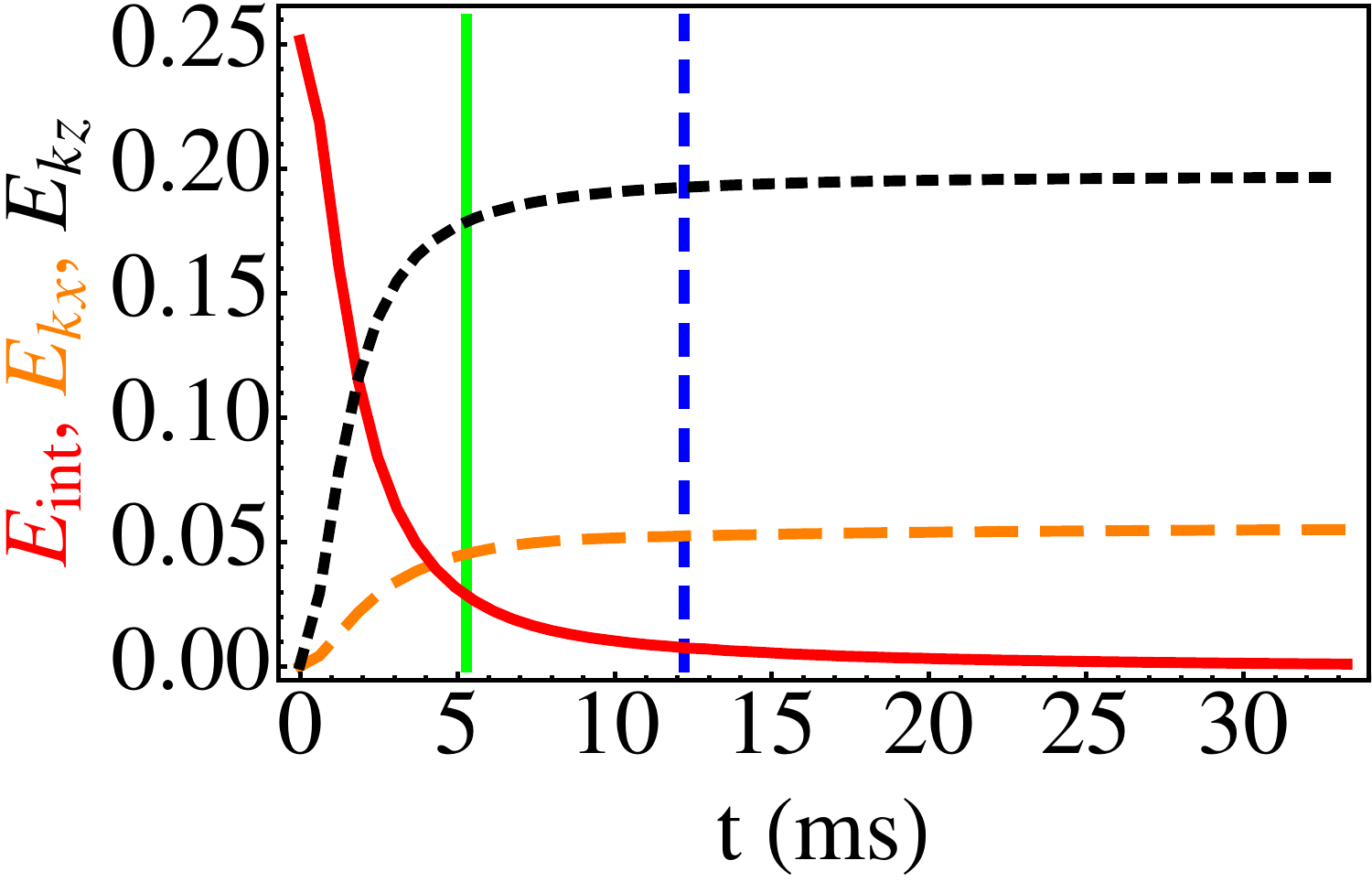}
\caption{(Color online) Time evolutions of:  the kinetic energy along the lattice direction {\color{black} ($E_{kx}$, 
orange dashed line)}, the  kinetic energy along $z$ {\color{black}($E_{kz}$,  black dotted line)},  and the  interaction energy {\color{black}($E_{int}$,  red solid line)}. This plot {\color{black} is computed using the mean field model} for  {\color{black}a} $7.25E_R$ lattice{\color{black}. The vertical lines are at the same positions as in Fig. \ref{acc_eta0} (a)}.}\label{energy}
\end{figure}

\begin{figure*}[htb]
\subfigure[]{\includegraphics[width=2.2in]{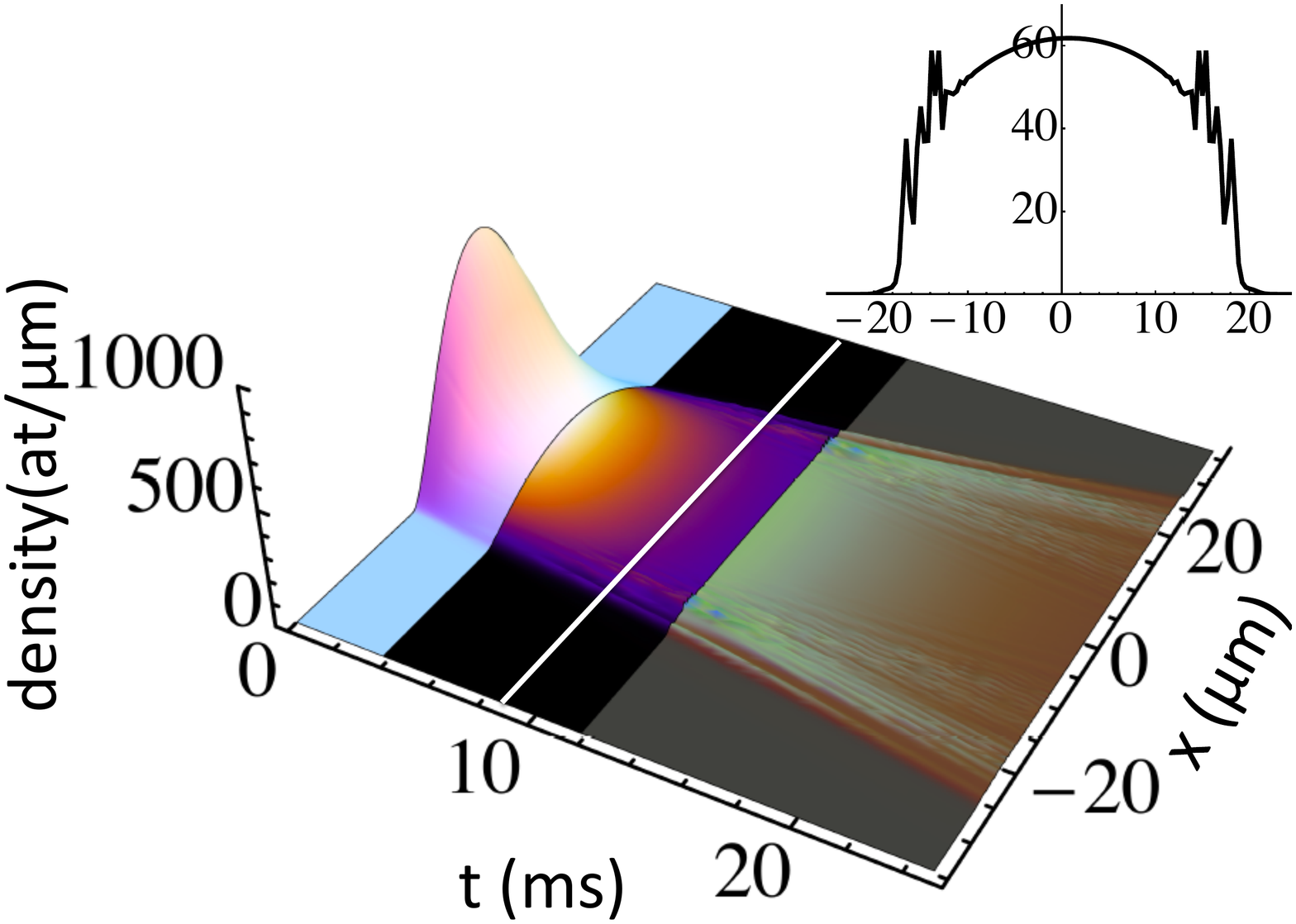}}
\subfigure[]{\includegraphics[width=2.2in]{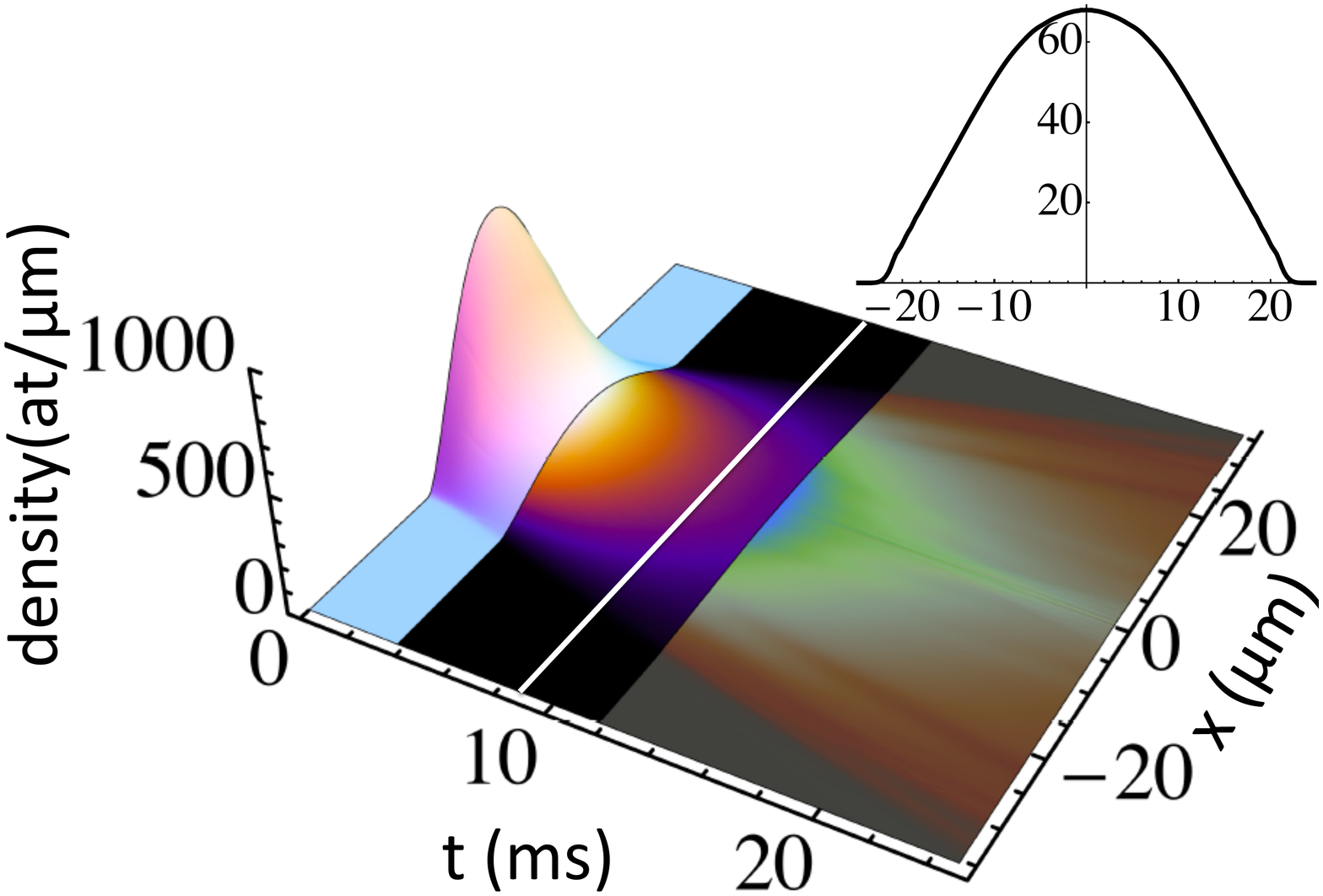}}
\subfigure[]{\includegraphics[width=2.2in]{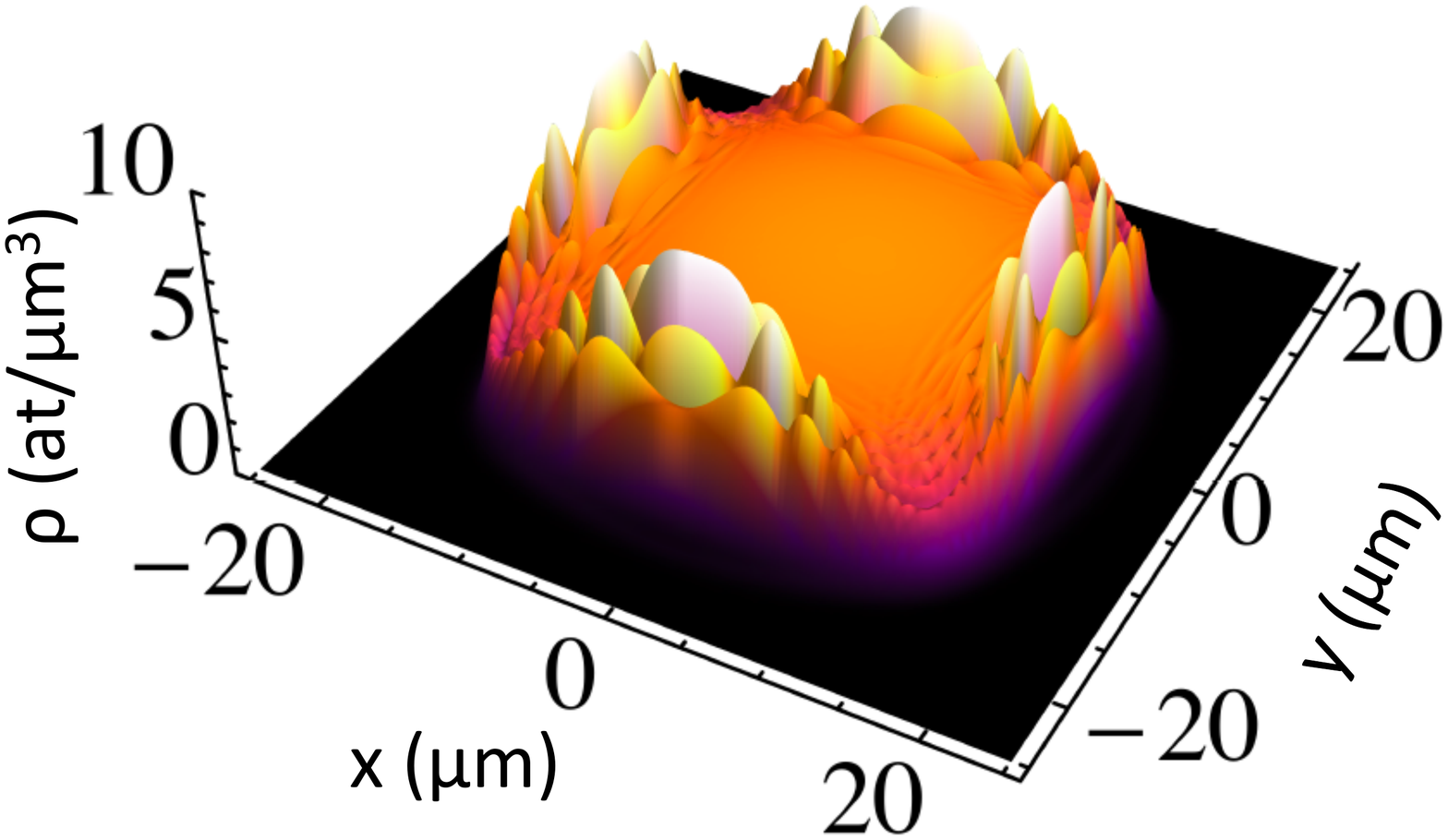}}
\caption{(Color online) Expansion dynamics {\color{black}predicted by the mean field calculations} for the lowest lattice depth (7.25 $E_R$): Panel (a) displays the  evolution of the density profiles at $z=0$,  after integrating along one transverse direction, $\tilde{N}(x,t)$. Panel (b) displays the same density profiles, but after integrating along a direction that is 45 degrees between the $x$ and $y$ axes (as done in the experiment). Panel (c) shows transverse density-profiles at $z=0$, $\tilde{N}(x,y)$,  during the MQST regime at $t=8.7\text{ms}$. {\color{black}The insets} in Panel (a) and (b) are intersecting profiles at $t=8.7 \text{ms}$ marked by white lines in the 3D density plots.}\label{density3d}
\end{figure*}

\subsection{General Behavior: Three Different Regimes }

Based on the sign of $\ddot{x}_{rms}$, we classify the dynamical evolution of the system into three different regimes: Initial expansion (Exp), MQST, and Ballistic expansion (BE).
Each  regime  has a distinct behavior which can be  further characterized by  other observables such as the ratio between interaction {\color{black}and} kinetic-energies  (Fig.~\ref{energy}) and the shape of the density profile (See Fig.~\ref{density3d}).

$\bullet$ {\it  Initial expansion (Exp)}:

The initial expansion regime describes {\color{black}early-times} during which the atoms at the center of the cloud expand {\color{black}transversely}. We determine this regime by looking at the time period {\color{black}over} which  the  transverse RMS-width of the cloud expands with a non-linear rate (positive acceleration, $\ddot{x}_{rms}>0$).  During this time the density-profile remains  smooth. The interaction energy decreases very rapidly  and is converted into kinetic energy along the axial and transverse directions.  In contrast to  a  non-interacting system{\color{black},} in which the initial potential energy in the trap is converted to kinetic energy while  the width of the quasi-momentum  distribution {\color{black}stays} constant, the mean field  interactions cause a broadening of the quasi-momentum distribution whose width grows until it reaches values of quasi-momenta at which the effective mass becomes negative. At this point{\color{black},}  sharp peaks in the density profile develop.

For {\color{black}the  $7.25  E_R$ lattice}, the initial-expansion takes place within the first 5ms [See Fig.~\ref{acc_eta0}~(a)] after having turned off the parabolic-confinement.
In Fig.~\ref{density3d} (a)  and ~\ref{density3d} (b), we {\color{black}also} show the time evolution of a horizontal slice of the cloud $z = 0$, after having integrated  along one  {\color{black}transverse} direction (along $x$), and a direction 45 degrees from the {\color{black}transverse} direction (45 degrees from $x$ and $y$), {\color{black}which is the line of sight direction in the experiment. }

$\bullet$ {\it Macroscopic Quantum Self-Trapping (MQST):} At intermediate times ($t\sim 5-12$ms for the lowest lattice depth) MQST is signaled {\color{black}by} the RMS-width of the cloud.
During the MQST regime the interaction energy {\color{black} decreases slowly and remains comparable to the kinetic energy} of the atoms in the lattice. The interplay between interatomic interactions and the atomic-density-gradient at the cloud  edges {\color{black} prevents atoms from tunneling outward.} In this regime  {\color{black}the atoms start to pile up at  the edges, and a  hole forms at the center of the cloud} [see Fig.~\ref{density3d}~(c)].  The radial expansion slows-down or even stops as {\color{black}illustrated} by  a negative $x_{rms}$ acceleration. The slowing-down can  also be linked to the population of a large number of  quasi-momentum states with negative effective-mass.

\begin{figure}[htb]
\vspace{0.5cm}
\subfigure[]{\includegraphics[width=2.8in]{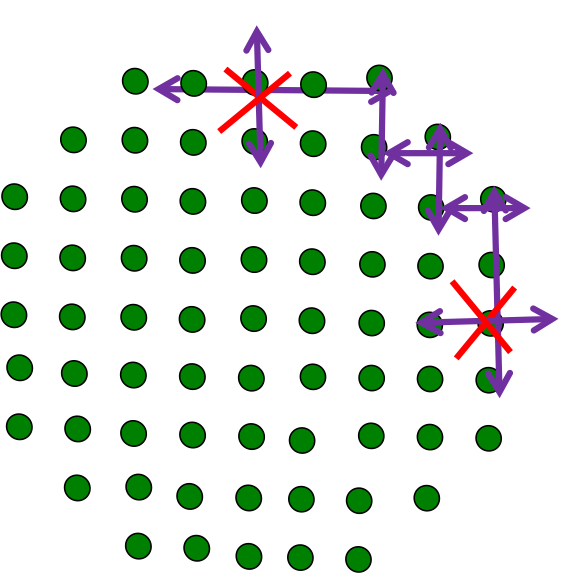}}\\
\subfigure[]{\includegraphics[width=3in]{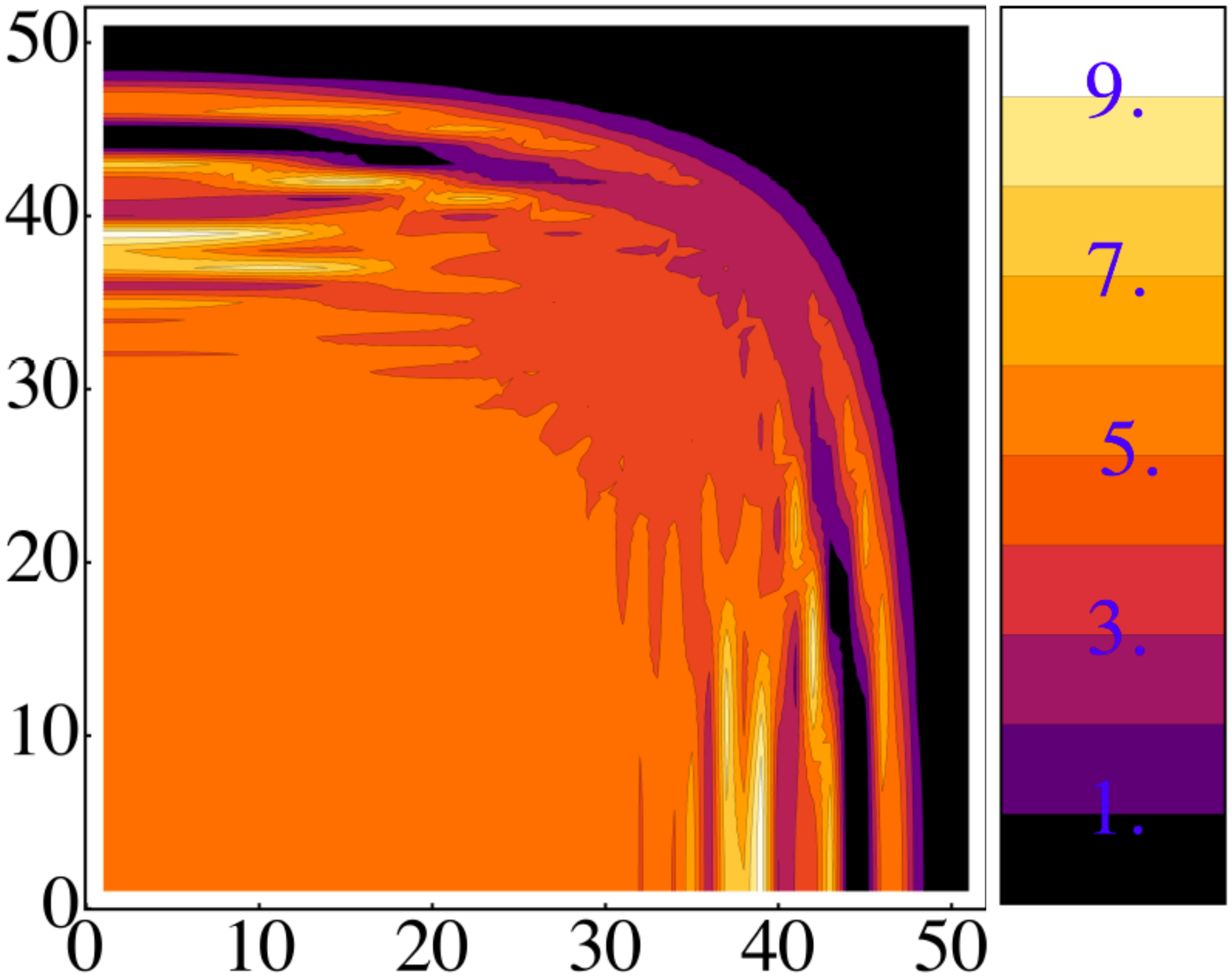}}
\caption{(Color online) Panel (a) is a schematic picture of the tunneling in 2D optical lattices. In 2D, no site is self-trapped in all directions so atoms are not fully frozen. {\color{black} The arrows show the  directions that atoms may tunnel while the crosses mean that tunneling along those directions is forbidden by MQST.}  Panel (b) is a density contour plot of a square fort-like barrier developed during the MQST regime.     }\label{2dflow}
\end{figure}

{\color{black} Global observables, such as the RMS-width are  easily  measured and characterized. They encode some important signatures of MQST.  However, MQST is an inherently local effect,  and important information can be hidden in the global probes.  For example, although the formation of steep edges and a hole is shared by the MQST phenomena in both 1D and 2D lattices and is observable in the RMS-width,  there are more features in the 2D system. 

In  the {\color{black}MQST} regime, the atom distribution in the lattice is no longer smooth, and its profile evolves  from having a circular shape  to a square one.   Atoms at $45^o$ to the lattice axes are only self-trapped at larger radii, since the corresponding  density gradients are smaller than in the lattice directions. This gives rise to the  square fort-like barrier around the edges, {\color{black} seen in} Fig.~\ref{2dflow}.  The formation of steep edges also happens in the 1D system, but in 2D there is always a direction along which atoms can tunnel. Since they are not fully frozen,  the self-trapped edges are not stationary as in the 1D case, but instead they  evolve with time.

 The  shape of the integrated density-distribution strongly depends on the direction along which the integration is performed (See Fig. \ref{density3d}(a) and Fig. \ref{density3d} (b)). For example,  while the peaks at the edges, signaling   the hole at the center of the cloud, are clearly visible  in Fig. \ref{density3d}~(a), when the imaging direction is along the lattices, they are  barely visible  when the imaging direction is  along  the diagonals as in Fig. \ref{density3d}~(b).
 }

$\bullet$ {\it Ballistic  Expansion Regime (BE):}
Because atoms expand axially along the tubes, the interatomic interactions decrease as time evolves and  at some point they are no longer strong enough to enforce the self-trapping mechanism. This determines the interruption of the MQST and the onset of the ballistic regime.
In this regime mean-field interactions have decreased so significantly  that an  analysis in terms of single-particle eigenstates and eigenmodes can be carried out. 

\subsection{Comparison with the experiment}

The experimentally  measured RMS-widths are slightly different from the  $x_{rms}$ defined above.  {\color{black} In the experiment the density-square distribution, instead of the density,  was directly measured. } To mimic the RMS-widths measured in the experiment, we define another RMS-width, denoted as $\sigma_{n^2}$.
To get $\sigma_{n^2}$, we square the 1D distribution $\rho_m$ {(\color{black}$\rho_m$ is obtained by integrating the 3D density distribution along {\color{black}a line 45 degrees from a lattice direction} over a slice centered at $z=0$)} and convolve it with a Gaussian of RMS-width $1.7 \mu \rm{m}$ {\color{black} (the imaging system resolution)} to get the convolved-density-square-distribution $n^2_{m}$ or $\tilde{\rho}_m$.  The convolution procedure mimics the finite-size-resolution of the {\color{black} imaging system}. Finally, $\sigma_{n^2}$ is directly calculated by the definition: $\sigma_{n^2}=\sum_m \tilde{\rho}_{m} m^2/\sum_m \tilde{\rho}_m$. This process removes fine features from the theoretical curves that are not resolvable in the experiment.

\begin{figure}[htb]
\vspace{0.5cm}
\subfigure[]{\includegraphics[width=1.6in]{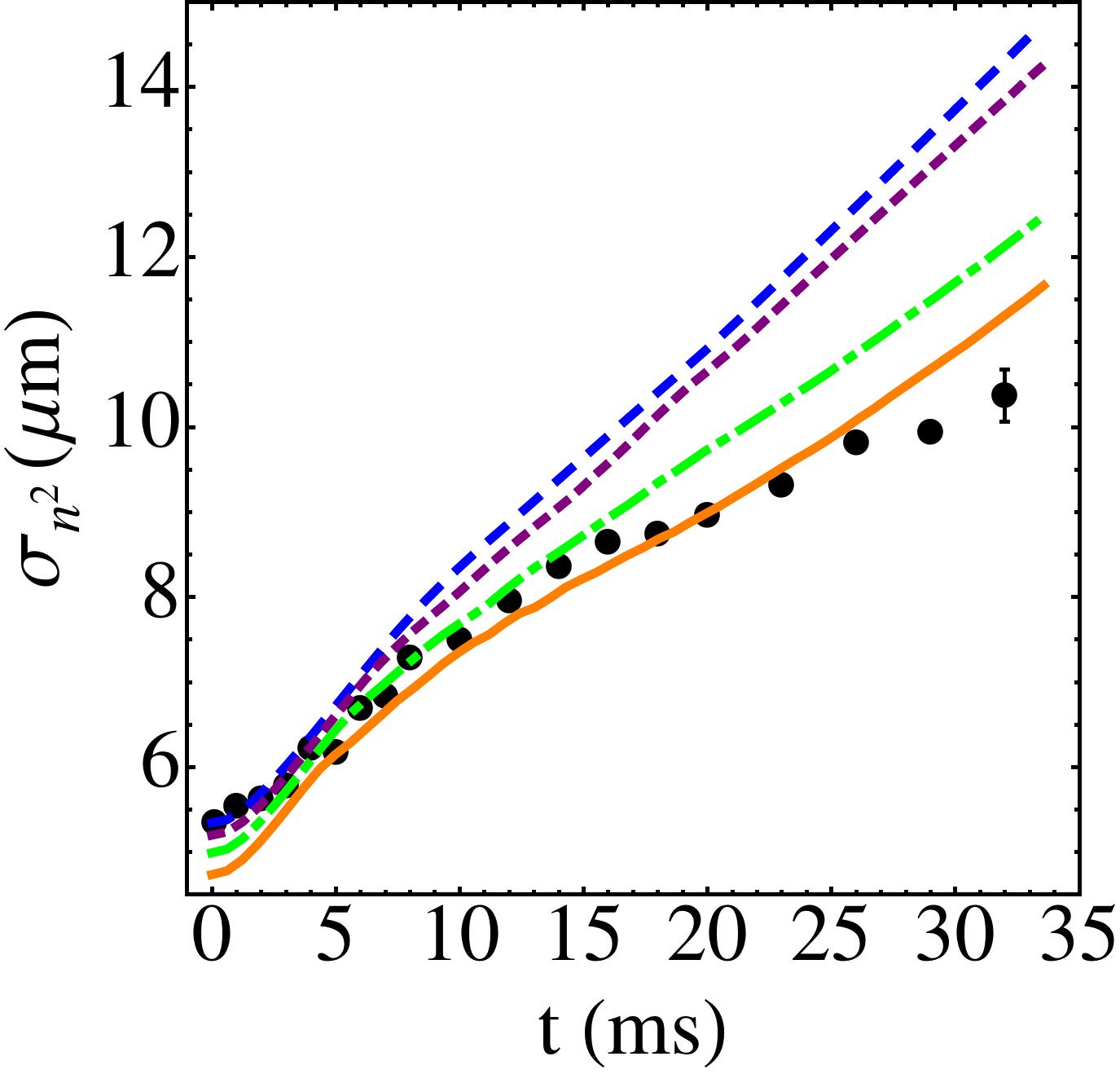}}
\subfigure[]{\includegraphics[width=1.6in]{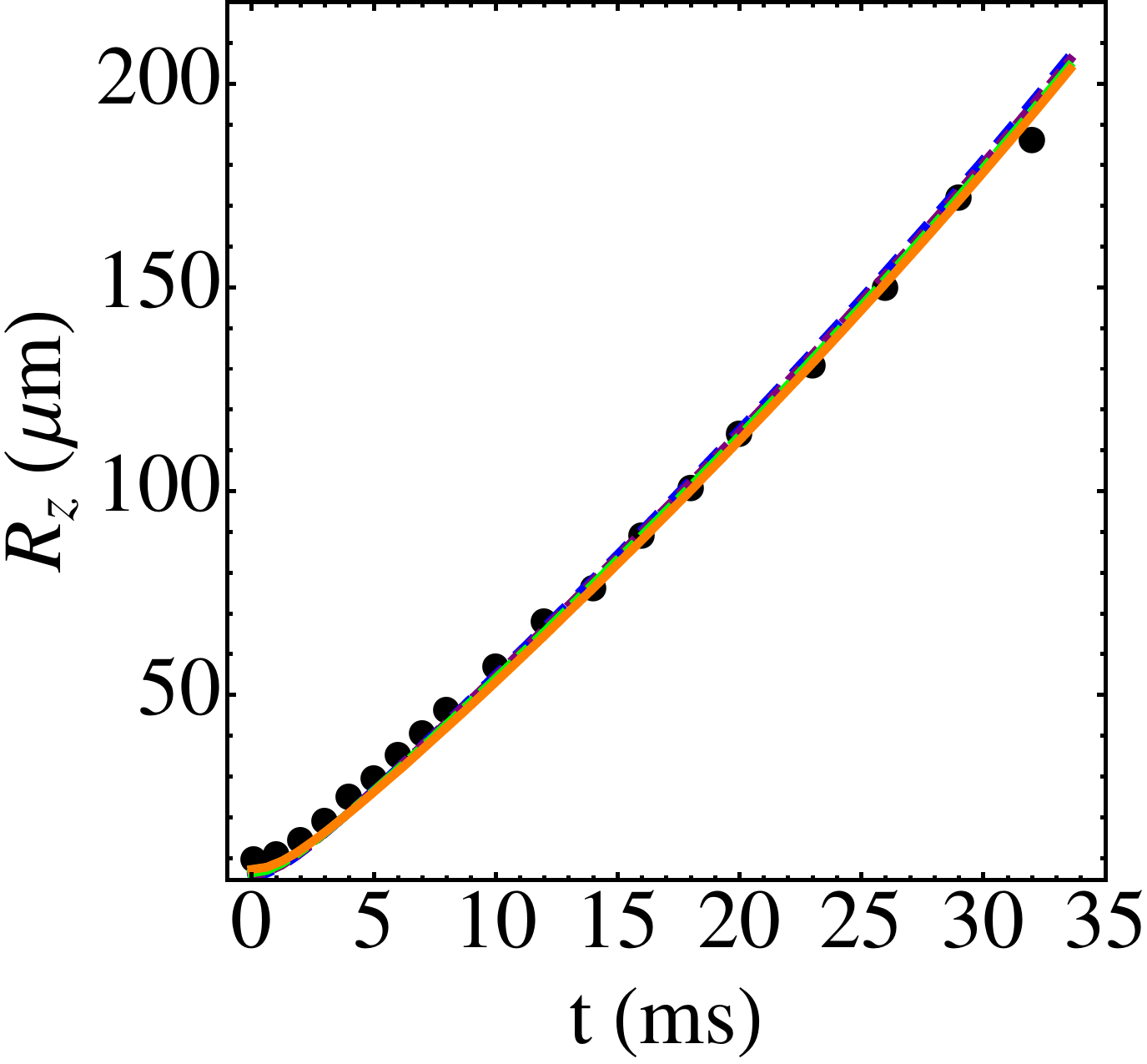}}
\caption{(Color online) {\color{black}The  evolution of  (a) $\sigma_{n^2}$ and (b) $R_z$  for $7.25E_R$ from mean field calculations {\color{black}and the experiment}.   Each black point is the average of 10 experimental measurements.   In Panel (a), the error bars represent random uncertainty, but an overall systematic uncertainty of $0.5\mu$m  associated with the imaging resolution is not included.  {\color{black}To illustrate the dependence of the mean-field  dynamics on the initial width, we show the  evolution of $\sigma_n^2$ for different initial conditions for the $7.25 E_R$ lattice.}   Each line  corresponds to a particular initial condition, {\color{black}subject} to the constraint of matching axial expansion rates  at long times. In Panel (b),  error bars are not shown, since we are mainly interested in the long time dynamics when $R_z$ is very large and the error bars are negligible. Even with different initial conditions, the expansion rates along $z$ are almost the same, therefore in Panel (b) all lines are almost on top of each other.}} \label{measurable_7p25}
\end{figure}

\begin{figure}[htb]
\vspace{0.5cm}
\subfigure[]{\includegraphics[width=1.6in]{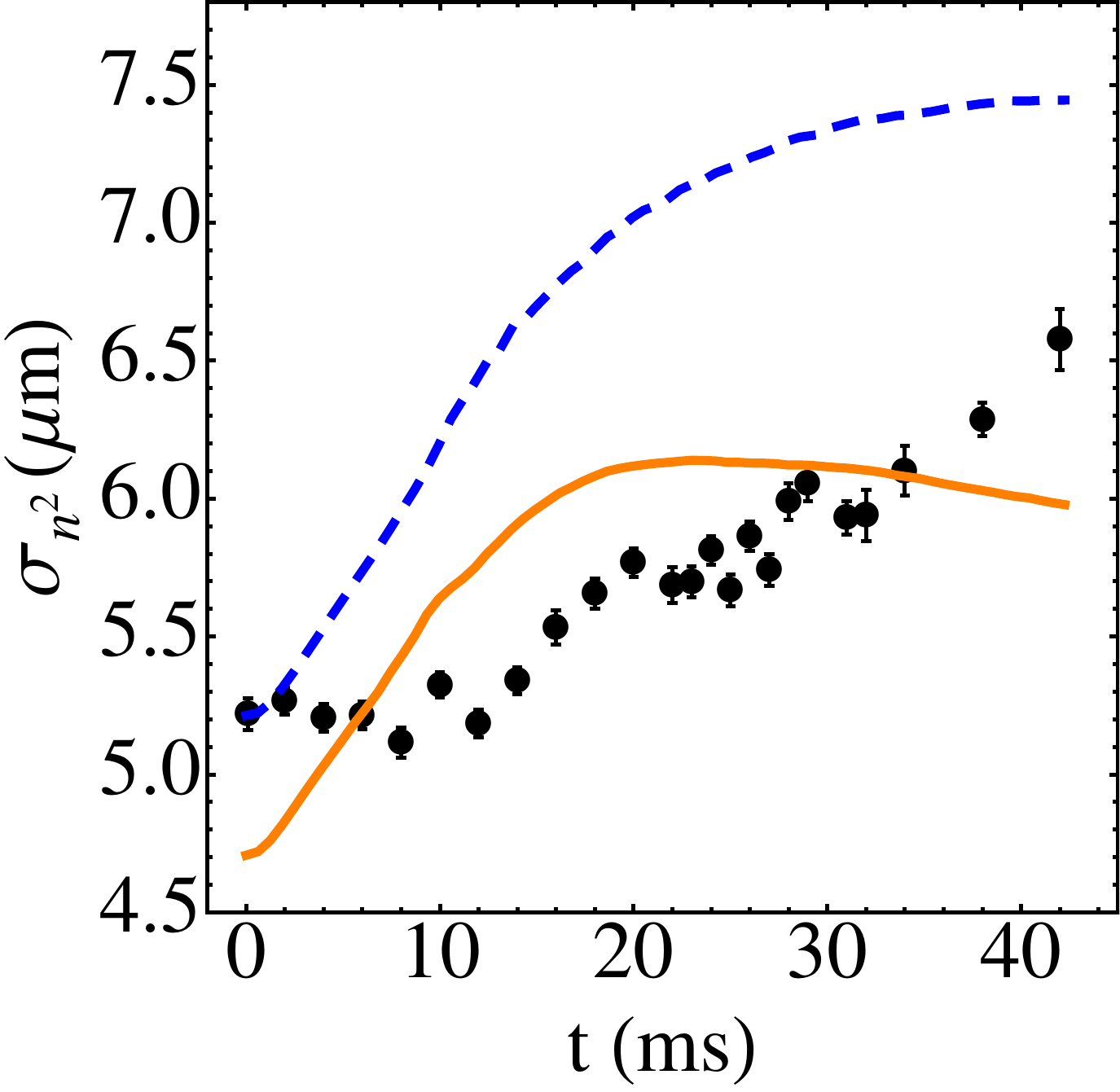}}
\subfigure[]{\includegraphics[width=1.6in]{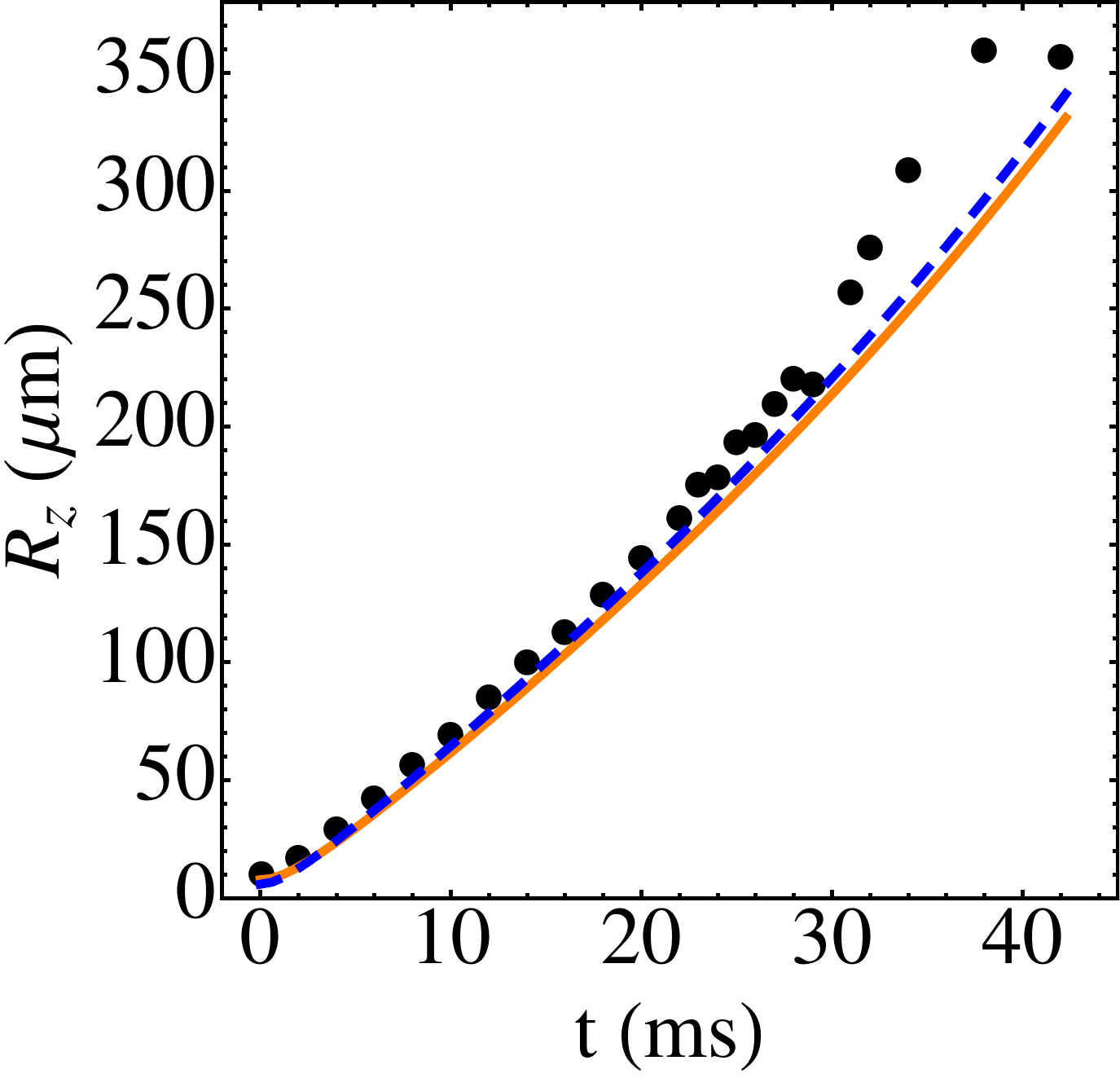}}
\caption{(Color online) {\color{black}  Time evolution of  (a) $\sigma_{n^2}$ and (b) $R_z$  for $13 E_R$ from mean field calculations {\color{black}and the experiment}.  As in Fig. \ref{measurable_7p25},  the solid and dashed lines are simulation results and each corresponds to a particular initial condition.  The  long-time dynamics of  $\sigma_{n^2}$ is sensitive to the initial width of the cloud also at $13 E_R$.  The black points are experimental data.} }\label{measurable_13}
\end{figure}

Another observable measured in the experiment was  the Thomas-Fermi radius $R_z$ of the atomic density along $z$. It was  obtained by fitting the atomic density distribution, after integrating along the {\color{black}transverse} directions, to  $ \frac{3}{4 R_z}  \max[0,1 - (z/R_z)^2]$.
Because {\color{black}the initial width of the cloud is sensitive to the loading procedure} {\color{black}and because the mean field treatment tends to overestimate the initial width of the cloud compared to the one measured in the experiment}, we use  different initial sizes to compare with the experiment{\color{black}, subject} to the {\color{black}constraint} of matching  axial expansion rates at long times. The latter just ensures that we {\color{black}are} using the correct total energy, which is  conserved during the evolution.
Comparisons between the numerical simulations and the experimental data are shown in Figs.~\ref{measurable_7p25} and \ref{measurable_13}.  In general, the  theory overestimates the expansion rate of the RMS-widths. {\color{black} Only} for $7.25 E_R$ {\color{black}does} the mean-field theory roughly {\color{black}capture} the experimental observations at short times. {\color{black} For} deeper lattice depths, such as $13 E_R$, the mean-field theory fails to reproduce the observed dynamics.

The  long-time dynamics of  $\sigma_{n^2}$ {\color{black}predicted by the mean field method}  is sensitive to the {\color{black}initial width} of the cloud.  When the initial RMS-width is larger,  the MQST boundary is further away  from the center and  thus it takes more time for the  RMS-width to reflect the presence of MQST at the edges. {\color{black}In contrast, the} MQST is more visible {\color{black}for}  smaller initial RMS-widths, as shown in Figs.~\ref{measurable_7p25} and \ref{measurable_13}.

The plots in Fig.~\ref{density3d} share the same {\color{black}initial condition} as the dashed  blue line in Fig.~\ref{measurable_7p25}.  With this {\color{black}initial condition},  the MQST effect is weak  and the self-trapping {\color{black}signatures,} such as peaks at the edges [Fig.~\ref{density3d} ~(a)]  {\color{black}and} a hole in the middle [Fig.~\ref{density3d}~(c)], are not very pronounced.  {\color{black} These self-trapping signatures were not seen in the experiment even when imaged from above.}  {\color{black} These characteristic signatures might be visible in  future experiments with {\color{black}smaller} lattice depths,  smaller initial RMS-widths  and  {\color{black}larger values of $R_z$}}.

\section{ Beyond Mean-Field Model}\label{sec:beyondMFM}

As shown in Sec. \ref{sec:MQST2D}, the mean-field treatment {\color{black}does} not correctly describe the experiment.  A fundamental limitation of this  treatment  is that it assumes all the atoms are in the condensate and  neglects condensate depletion due to  quantum correlations or thermal fluctuations (coming from an initial thermal component or non-adiabatic effects as the lattice is turned on). Condensate depletion  modifies the dynamics predicted by the single mode approximation. In an attempt to include quantum fluctuations, we use an approximate TWA \cite{Polkovnikov2010,Blakie2008,Sinatra2002}{\color{black}. The TWA}  incorporates leading order corrections to the  dynamics expanded around the classical  (GPE) limit. The whole idea of the TWA is that the expectation value of a quantity at time $t$ can be determined by  solving the classical equations of motion from 0 to $t$ and   sampling over the Wigner distribution at time $t = 0$. The TWA  is guaranteed to be accurate at short times. For longer times, there {\color{black} are also} higher order  corrections to the classical trajectories.  In a 3D lattice (0D condensate in each well) there are two limiting regimes where the  Wigner function can be easily found and {\color{black} has been shown to capture the exact dynamics well} \cite{Polkovnikov2010}. One is the very  weakly interacting regime in which the system is almost an eigenstate of the noninteracting Hamiltonian{\color{black},}  a product of coherent states.  The other case is the strongly interacting regime with commensurate filling, in which the system is a Mott insulator. In this case the many body wave function is mostly  a product Fock state $\prod_{j=1}^L|n\rangle_i$ with $n$ the filling factor. The semiclassical wave function is  $\prod_{j=1}^L \sqrt{n} \exp{ ({\rm i}\phi_j)}$  and the  corresponding  Wigner function {\color{black} is} characterized by independent random variables $\phi_j$ uniformly distributed between $(0,2 \pi)$.

The case that we are dealing  {\color{black}with} now has the complication that not only  {\color{black}is it} in the intermediate regime where {\color{black}neither}  of the two limiting cases holds, but also{\color{black},}  instead of a 0D condensate within a lattice site we have a {\color{black} quasi-1D} gas. Finding the Wigner function  therefore is not a trivial task.
Prior work done for two coupled 1D tubes \cite{Rafael2010} has already shown the importance of {\color{black}taking} into account quantum correlations for the proper characterization of the dynamics.  {\color{black} Consequently, we expect  that beyond-mean-field} corrections will play  an important role in our system.

With this goal in mind and {\color{black}subject} to the limitation  that the pure mean field dynamics  involve the propagation of thousands  of coupled GPEs, we implement the TWA in an ad-hoc way which we refer to as  aTWA.  We account for {\color{black} initial} phase fluctuations within each tube  by adding random phase factors: $\phi_{nm}(t=0)\rightarrow \eta \cdot \theta_{nm}$, where $\theta_{nm}$ is a uniformly distributed random variable between $0$ and $2\pi$, and $\eta$ parameterizes the strength of the phase fluctuations.  When $\eta=0$, the initial conditions correspond to a fully coherent array of 1D gases and   when $\eta=1$, we have an initially fully  incoherent array. $0<\eta<1$ corresponds to a partially coherent array.  {\color{black} Since $\eta$ is not determined in the theory, we compare the results for different $\eta$ to experiment, and choose the most similar one.}  We average the results over at least $20$ initial configurations.

Besides phase fluctuations, we also study the effect of  including number fluctuations {\color{black}in} the initial conditions. The results obtained  using  the  aTWA are described in detail in the following subsections.

\begin{figure}[htb]
\includegraphics[width=3.2in]{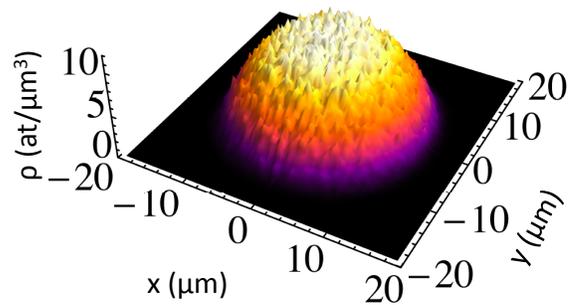}
{\color{black} \caption{(Color online)  Transverse density profiles at $z=0$ during the MQST regime at $t=8.7\text{ms}$ for the $13 E_R$ lattice obtained from the aTWA results.}\label{rhoxynoise}}
\end{figure}

The aTWA considerably {\color{black}improves} the agreement between theory and experiment. The value of $\eta$ {\color{black}that} optimizes the agreement with the experiment increases with increasing lattice depth. This is expected{\color{black},} since as  tunneling between tubes becomes weaker, phase coherence is suppressed.
The noise introduced by those random phases quickly  generates local self-trapping everywhere{\color{black},}  suppressing  the transverse dynamics until, due to the axial expansion,  interactions drop so significantly that the system goes to the ballistic expansion regime.  Noise also tends to  suppress  the hole formation. {\color{black} See Fig. \ref{rhoxynoise}.}

\subsection{Comparison between theory and experiment: phase fluctuations}

\begin{figure}[htb]
\subfigure[]{\includegraphics[width=1.7in]{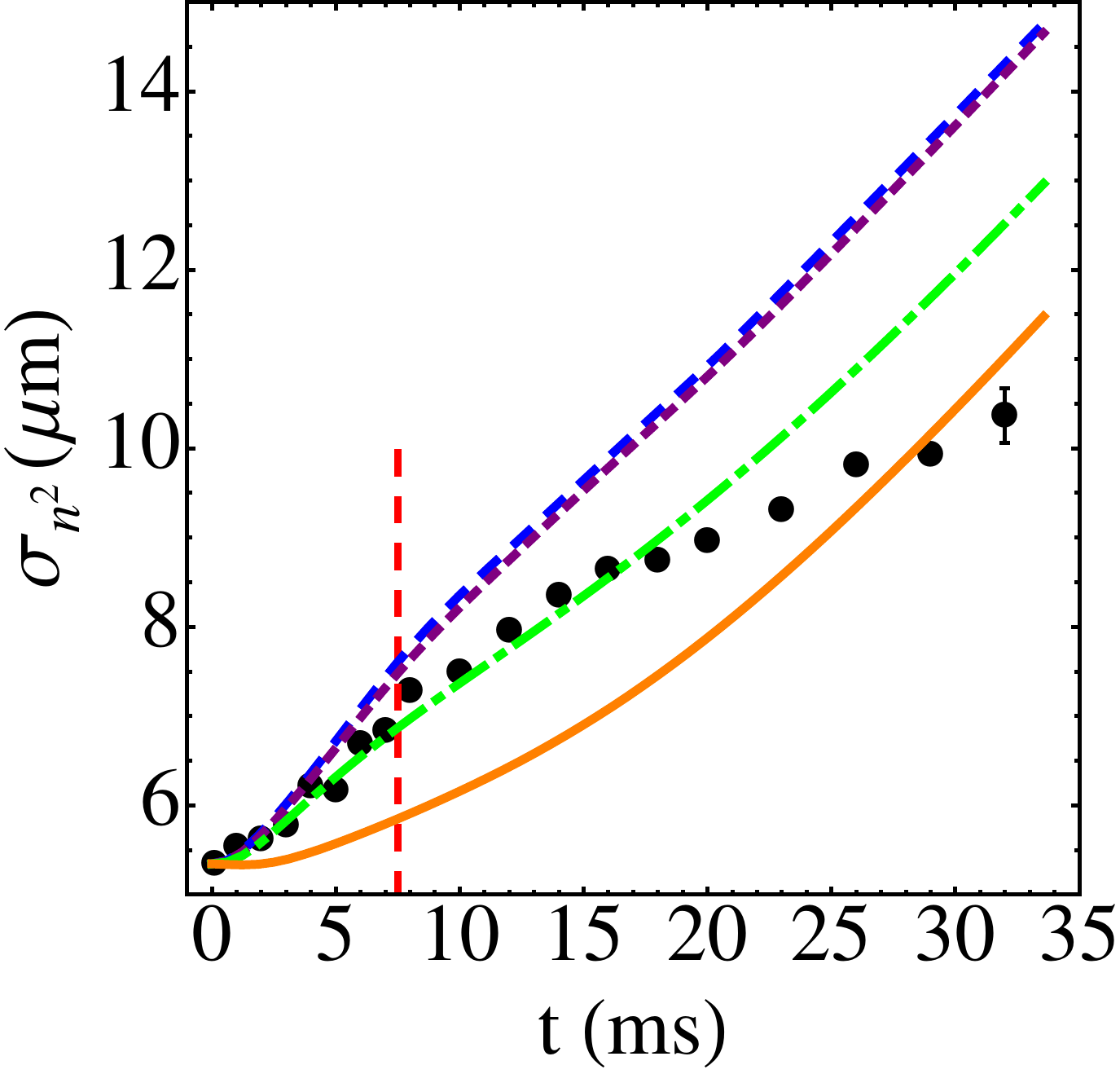}}
\subfigure[]{\includegraphics[width=1.6in]{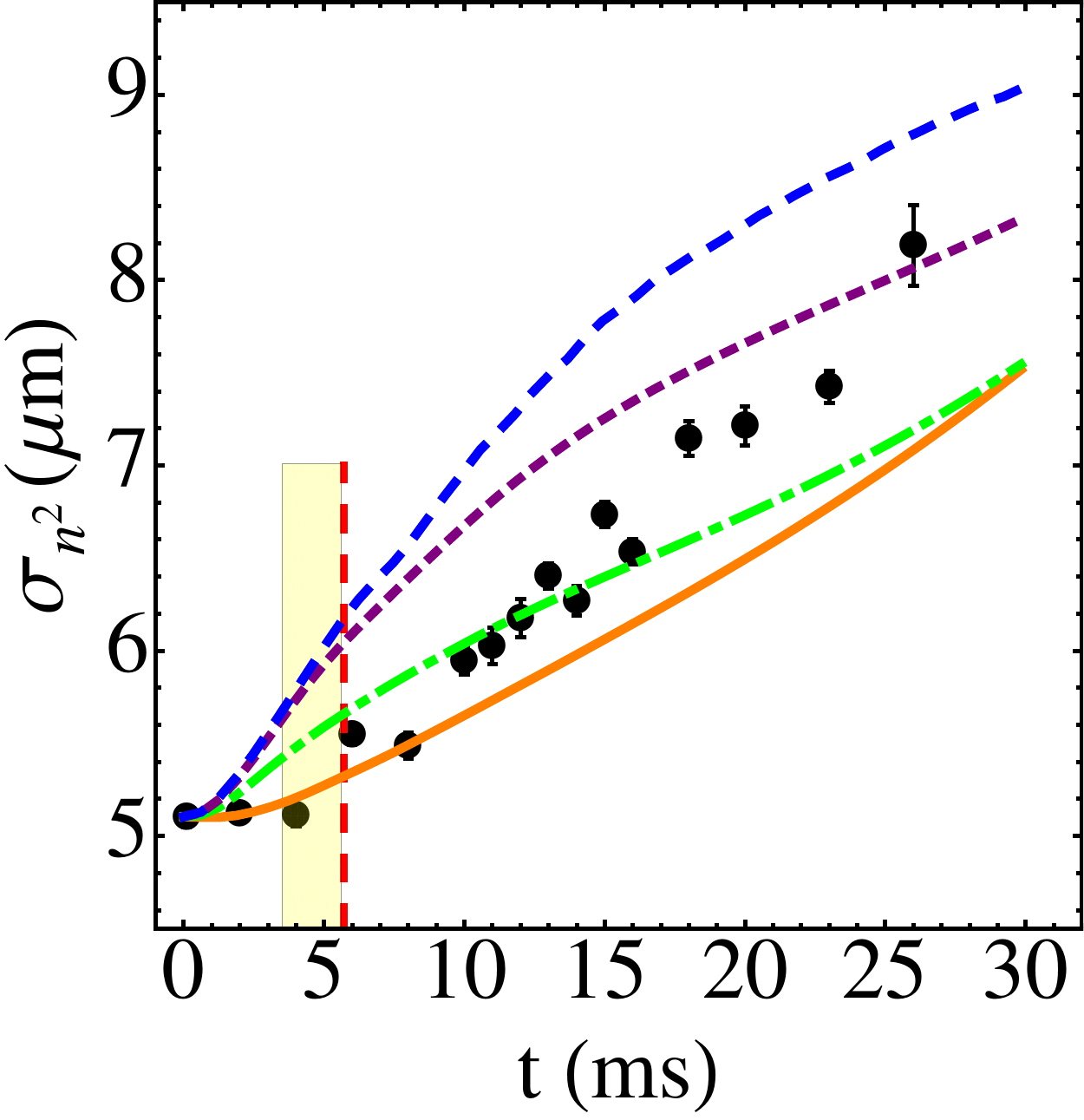}}\\
\subfigure[]{\includegraphics[width=1.6in]{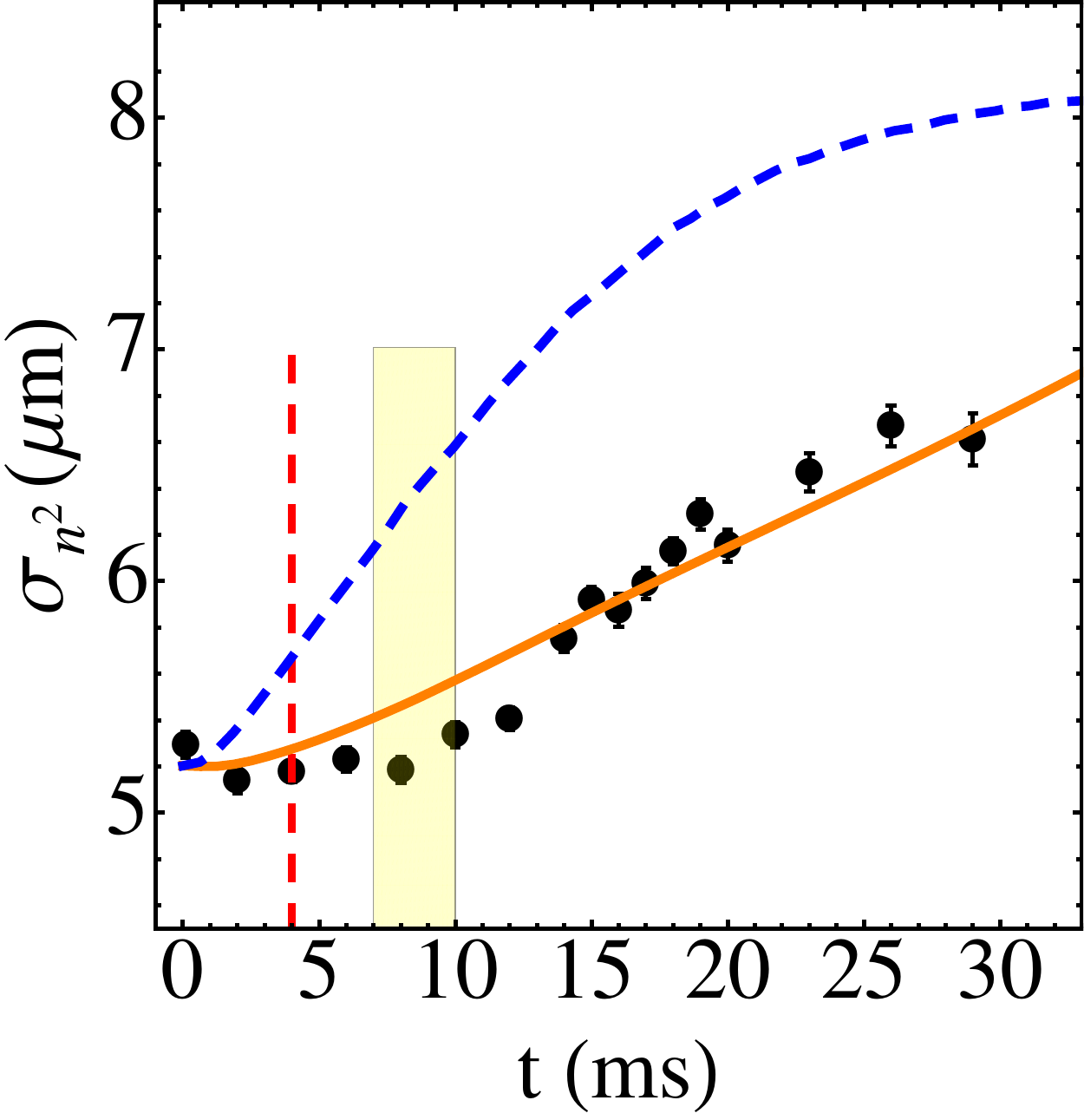}}
\subfigure[]{\includegraphics[width=1.7in]{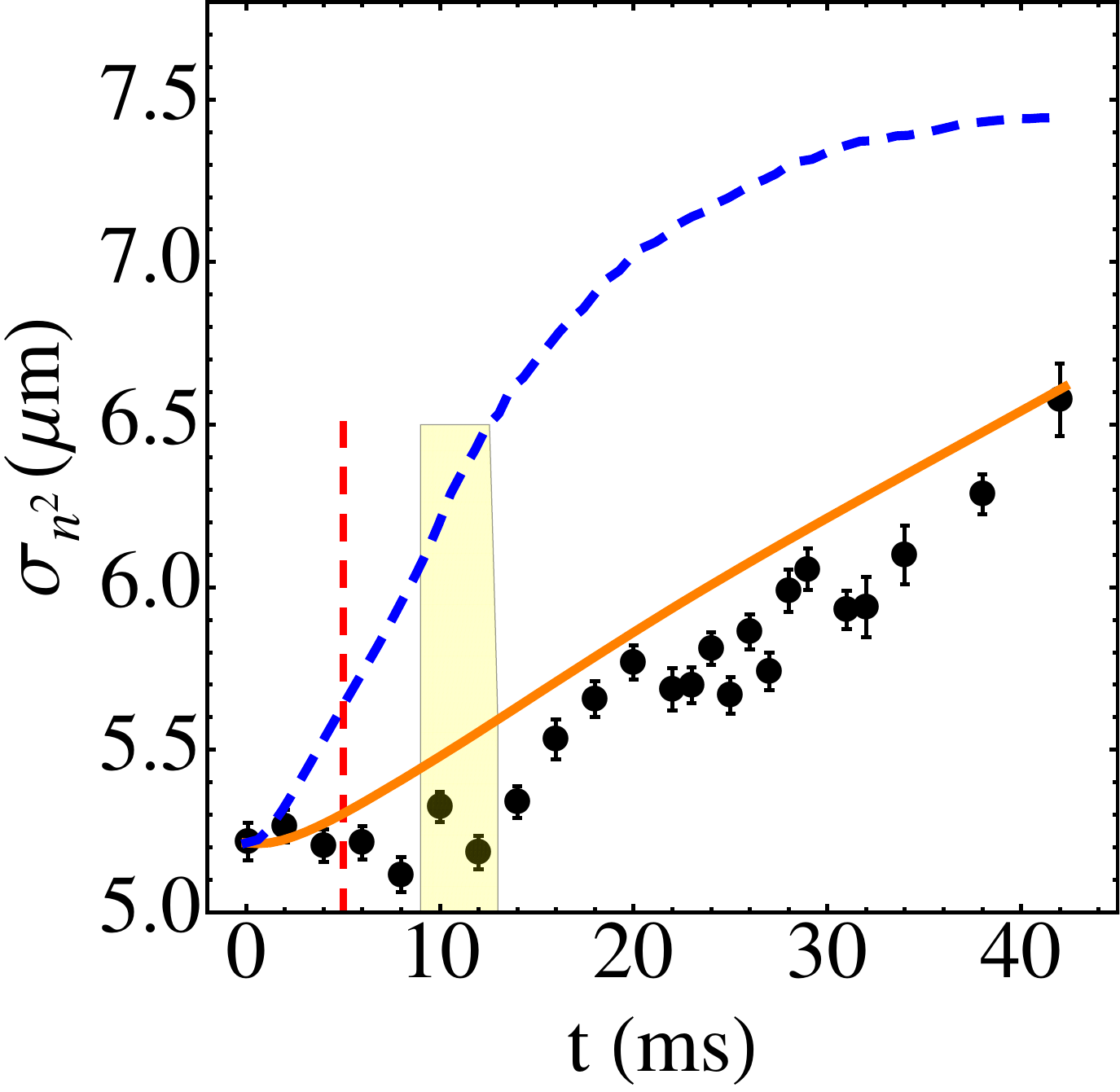}}
\caption{(Color online) {\color{black} Mean-field, aTWA and experimental results for the} evolution of $\sigma_{n}^2$ for all lattice depths used in the experiment.  {\color{black} The lattice depths are (a) 7.25$E_R$, (b) 9.25$E_R$, (c) 11$E_R$ and (d) 13$E_R$. }  At low lattice depth, a small  $\eta$ seems to account for the observed behavior at short times.     The dashed blue line is the result of a mean-field calculation of the dynamics with no random phase between the tubes ($\eta=0$). The dotted purple line is for $\eta=0.2$, the dashed-dotted green lines are for (a)  $\eta=0.4$ and (b) $\eta=0.5$,  and the solid {\color{black} orange} line is for $\eta=1$.  The dashed vertical red lines indicate the transition from  MQST to BE predicted  {\color{black} by the aTWA}. The yellow shadow regions  indicate the $t_c$ inferred from  the experimental data.} \label{xrmsall}
\end{figure}

Phase fluctuation among tubes substantially suppresses the expansion dynamics. To illustrate this effect, consider the toy model of a double-well with initial relative phase difference $\phi(0)$.  The  {\color{black} critical value of $\Lambda$ ($\Lambda$ is proportional to the ratio of the on-site interaction energy and the tunneling matrix element between two wells) which determines  the MQST-to-{\color{black}diffusive} transition} depends on $\phi$ as,

 \begin{equation}
\Lambda_c=2\left(\frac{\sqrt{1-z(0)^2}\cos[\phi(0)]+1}{z(0)^2}\right),
\end{equation}
{\color{black}where} $z$ is the fractional population difference between the two wells, $z=\frac{N_L-N_R}{N_L+N_R}$.  In this case one can see that even  when the initial population imbalance is small, $z(0)\ll 1$, the system can become self-trapped if $\phi(0) \to \pi$ ($\Lambda_c\rightarrow 1$).
In other words the critical value of $\Lambda_c$ {\color{black} decreases} with increasing  $\phi(0)$.

 \begin{figure}[htb]
\subfigure[]{\includegraphics[width=1.65in]{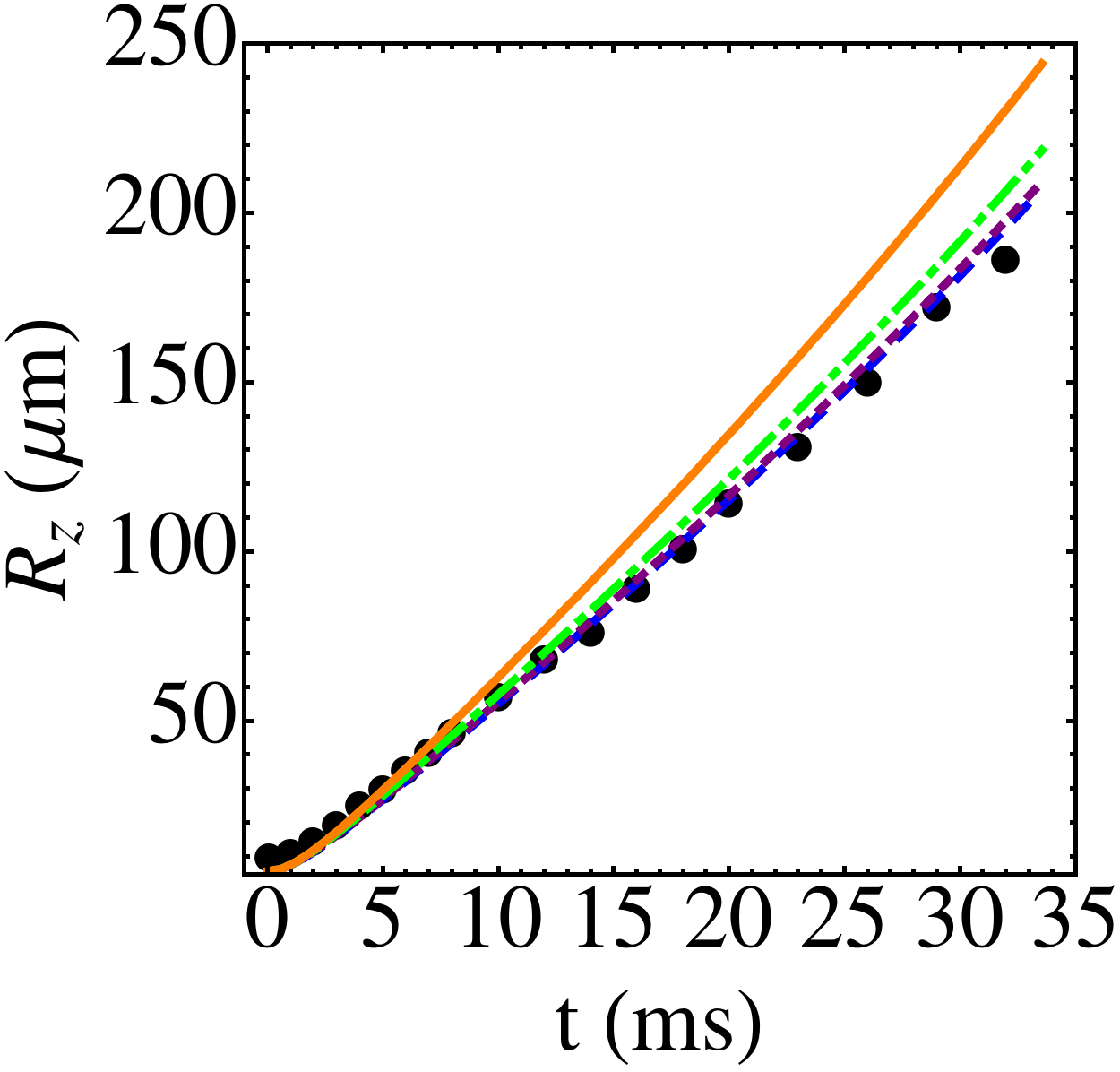}}
\subfigure[]{\includegraphics[width=1.6in]{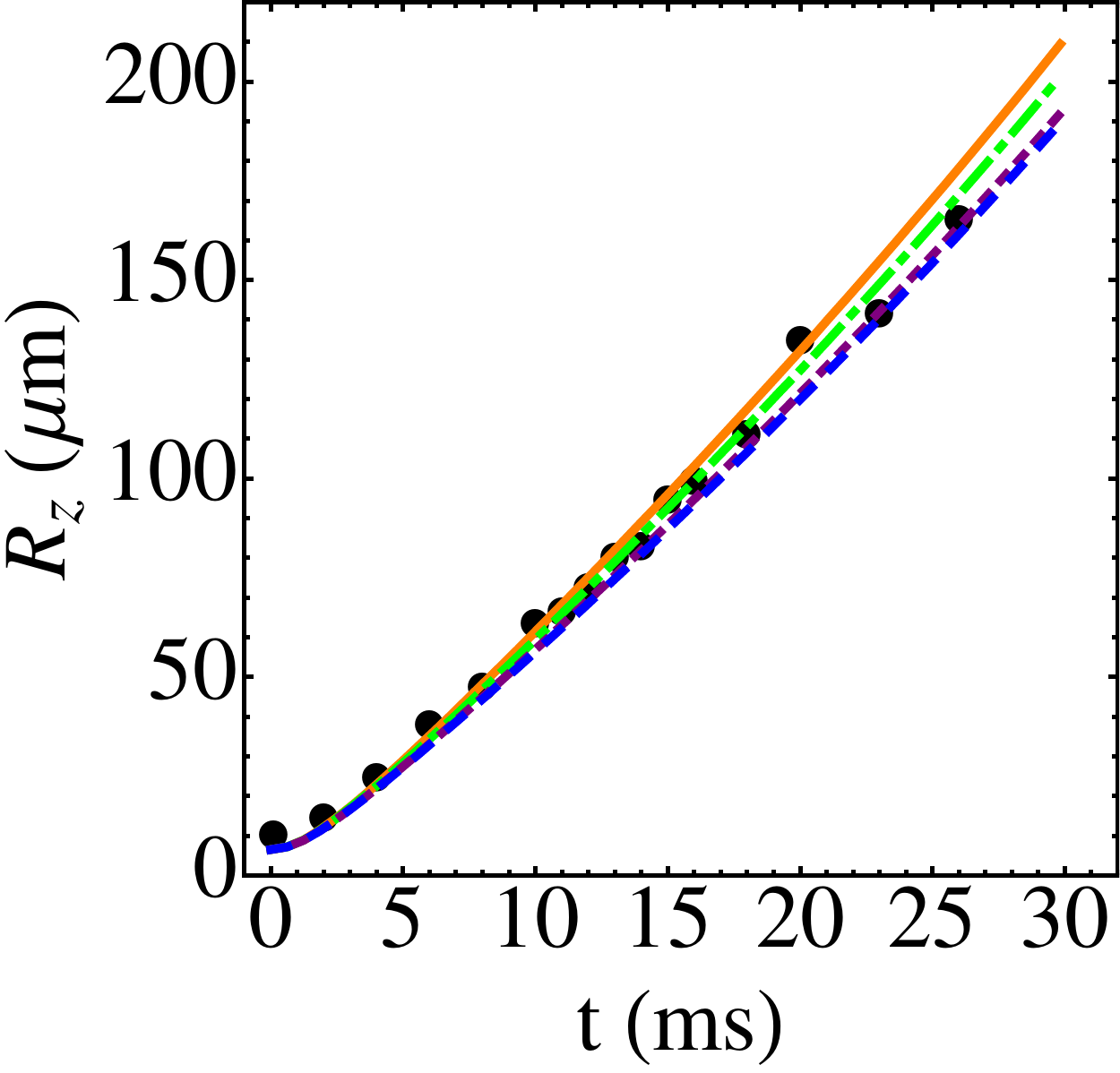}}\\
\subfigure[]{\includegraphics[width=1.6in]{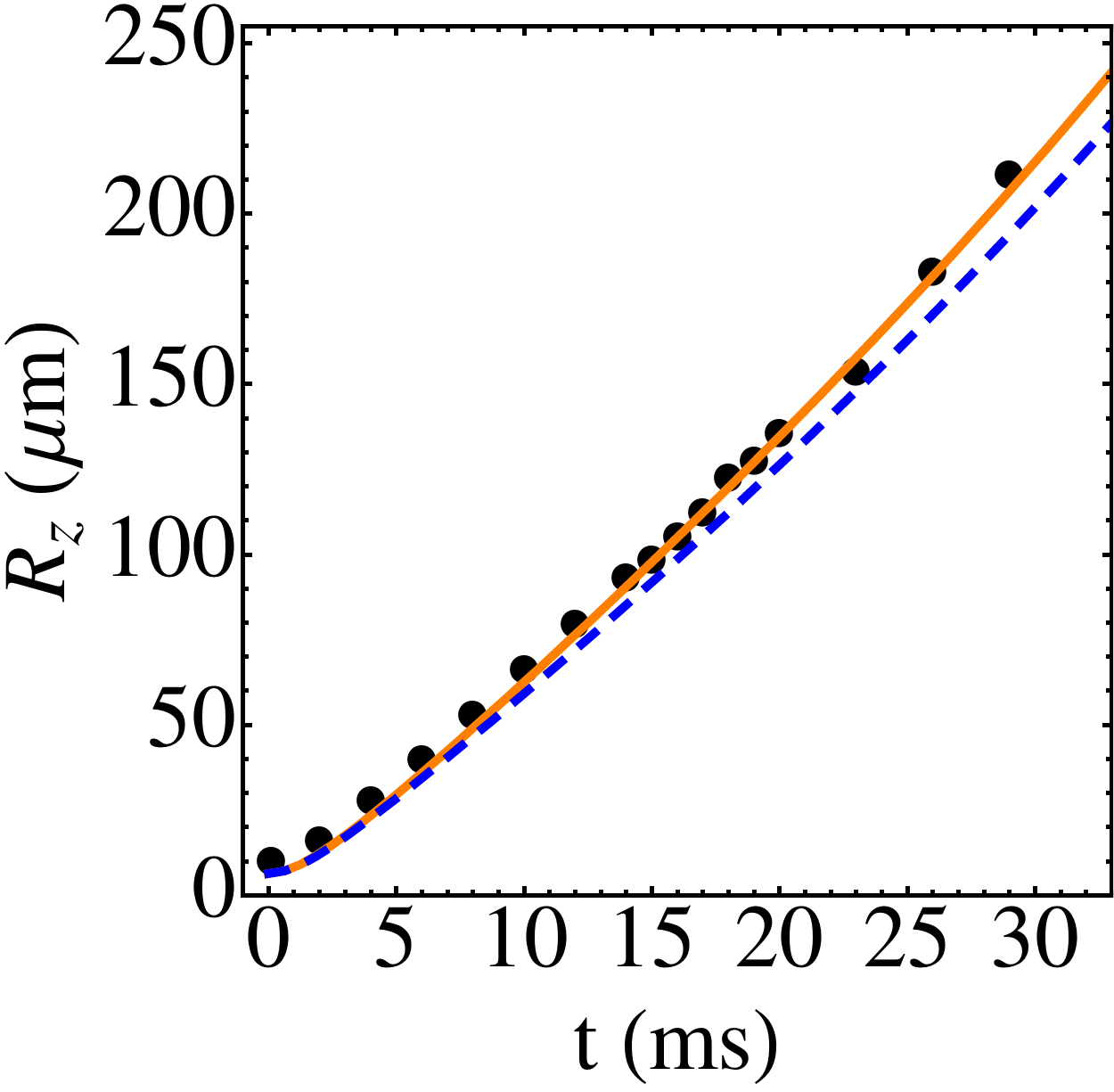}}
\subfigure[]{\includegraphics[width=1.6in]{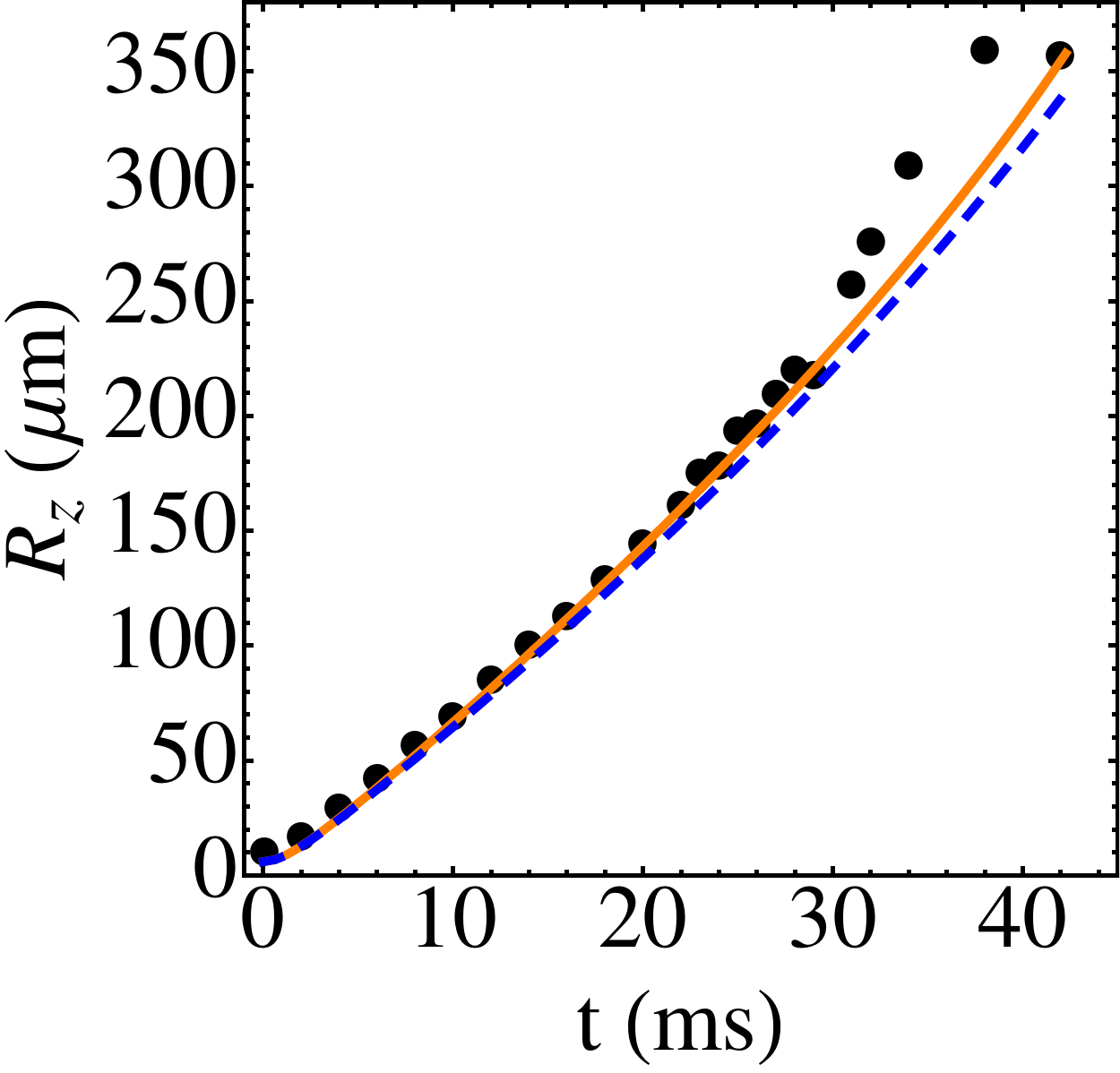}}
\caption{ (Color online) Evolution of  $R_z$   for all lattice depths used in the experiment. {\color{black}The dots  show experimental data and the solid and dashed lines show the predictions obtained from the aTWA for  different $\eta$.}  {\color{black} The lattice depths are (a) 7.25$E_R$, (b) 9.25$E_R$, (c) 11$E_R$ and (d) 13$E_R$. }  $R_z$ is obtained  by fitting the density  along $z$  to a Thomas-Fermi profile.  {\color{black}The theoretical curves use the same parameters {\color{black}as} the ones {\color{black}shown} in Fig. \ref{xrmsall}}.} \label{rzall}
\end{figure}

Our simulation agrees with  this expectation.  Fig.~\ref{xrmsall}-\ref{rzall}  show the {\color{black}transverse} RMS $\sigma_{n^2}$ and the vertical 
Thomas-Fermi radius $R_z$. {\color{black} The aTWA treatment agrees  much better with experimental observations,  especially for the deepest lattices.}  In Fig. \ref{xrmsall},  we can see that the greater  the $\eta$  the slower is the  expansion rate of $\sigma_{n^2}$.  When $\eta=1$, i.e., {\color{black}when} the phase
fluctuations are maximal{\color{black} ,}  the expansion rate of $\sigma_{n^2}$ reaches {\color{black}its} smallest value.

 \begin{figure}[htb]
\subfigure[]{\includegraphics[width=1.6in]{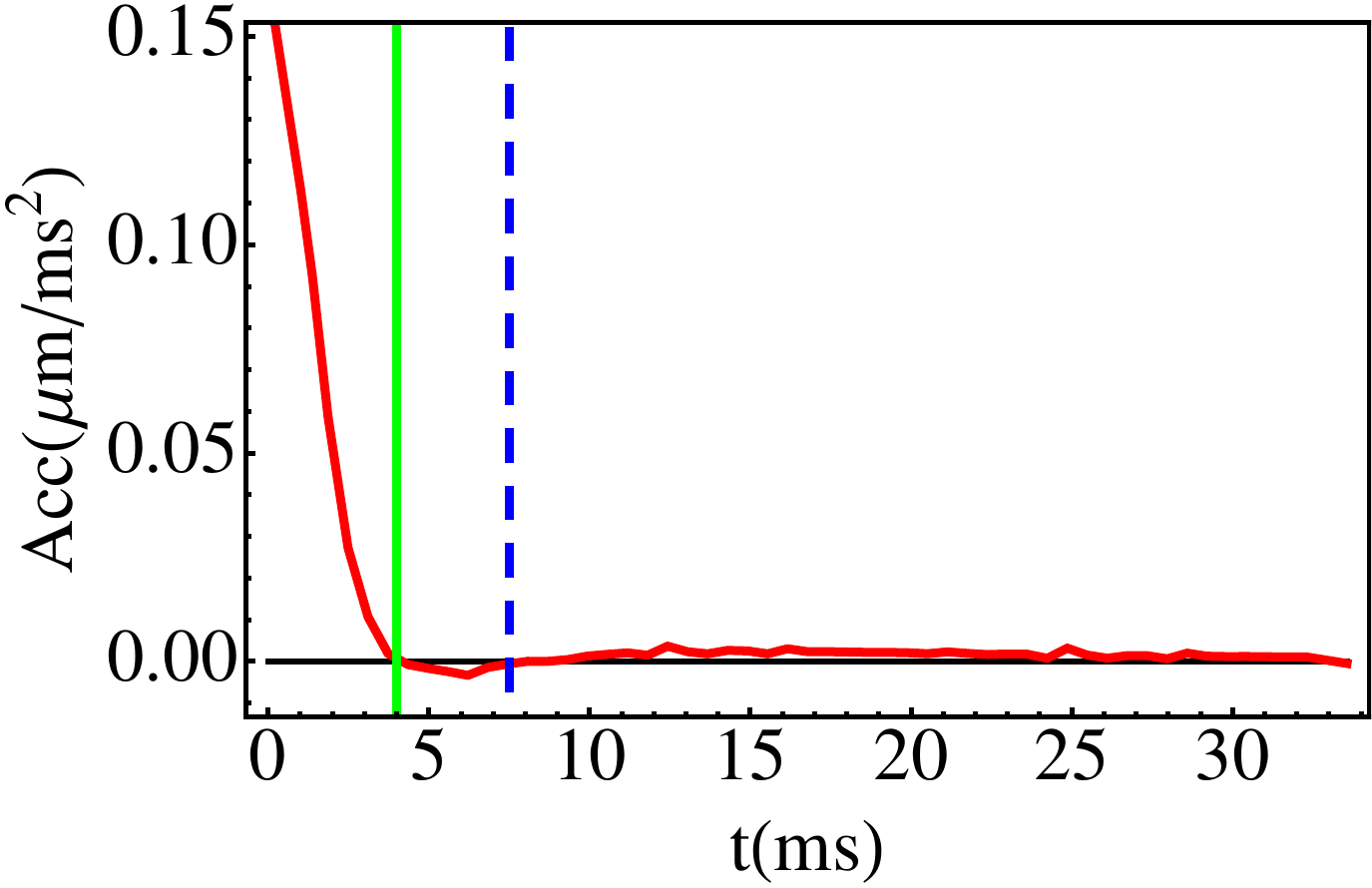}}\quad
\subfigure[]{\includegraphics[width=1.6in]{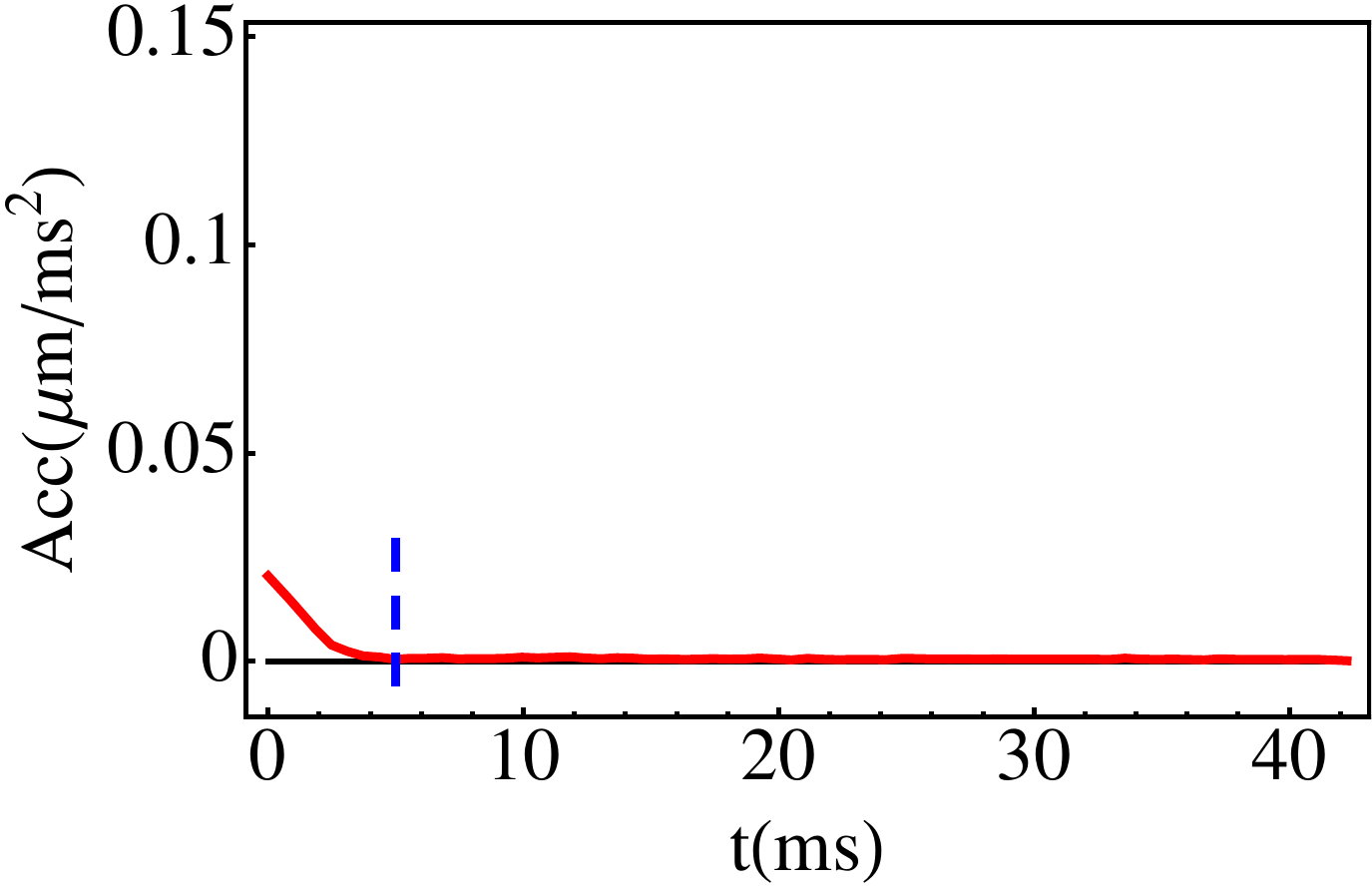}}
\caption{ (Color online) $\ddot{x}_{rms}$  {\color{black}calculated using the aTWA} for different  lattice depths {\color{black}and for the} corresponding optimal $\eta$,  {\color{black} (a) $7.25 E_R,\eta=0.4$  and (b) $13E_R,\eta=1$. }  The boundaries between MQST and BE are indicated by {\color{black}the} blue dashed lines. {\color{black} In panel (a), the boundary between EXP and MQST is {\color{black}indicated} by the green solid line.  {\color{black}In panel (b), because  $\ddot{x}_{rms}$  is almost 0 at all times, it can be interpreted as the self-trapping starts at $t=0$ ms.}}} \label{acceta1}
\end{figure}

\begin{figure}[htb]
\includegraphics[width=3.4in]{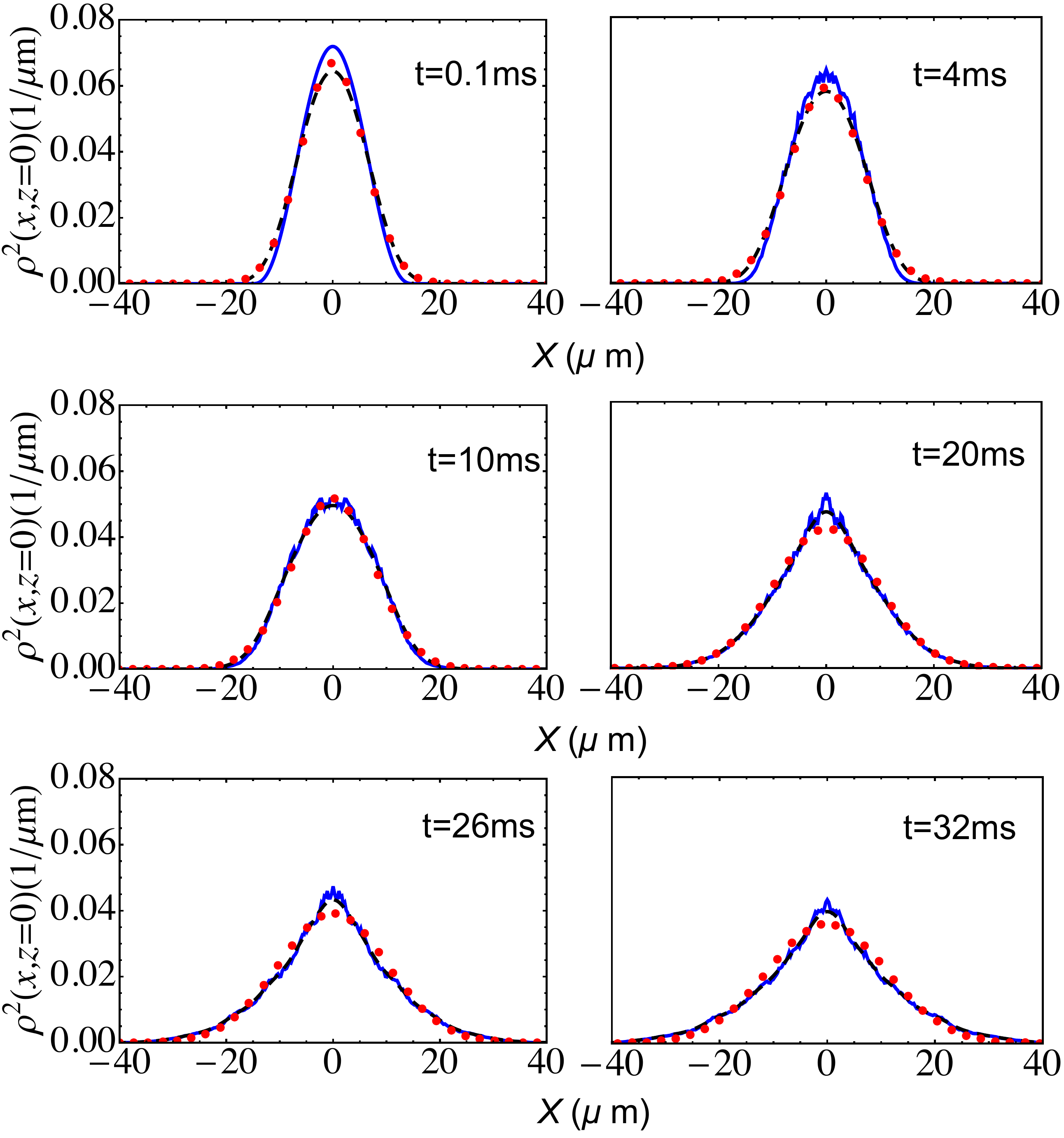}
\caption{(Color online) {\color{black}Transverse density profiles:  The solid/dashed lines are obtained  from the aTWA results for the $V=7.25 E_R$ lattice and for $\eta=0.4$.} The profiles are  taken at $z=0$ and viewed transversely at an angle $45^o$ from the lattice axis and normalized to $1$. The solid blue lines indicate the theoretical profiles before convolution and the dashed black lines after convolution. The red solid  {\color{black}points} are averaged experimental data  of 10 measurements. } \label{diagonal7}
\end{figure}

Although the modeled curves with the appropriate $\eta$ are never too far from the data,  for {\color{black}$V_0\geq11E_R$},  the experiment shows a {\color{black}more delayed and more sudden} onset of ballistic expansion than the model. For $V_0=7.25E_R$, $\eta\approx 0.4$ best fits the early evolution, when there is MQST. For larger $V_0$, $\eta=1$ clearly fits best {\color{black} for  the short  evolution time}.  For $V_0=7.25E_R$, $\eta\approx 0.4$, the three regimes predicted by the pure mean field model are still visible{\color{black},} although   $\ddot{x}_{rms}$ becomes barely negative in the MQST regime. On the other hand, for the deeper lattices, the random initial phases cause a fast development of site-to-site  density fluctuations and the formation of localized domains  randomly distributed throughout the tube array.  In contrast to the pure mean field model, localization occurs without the formation of sharp edges, there is no reflection from them and  $\ddot{x}_{rms}$ never becomes negative. Instead, it asymptotically approaches zero from above as shown in  Fig. \ref{acceta1}. MQST  always  disappears  when the density  drops below a critical value that depends on $V_0$.  The model predicts the following  values for the transition to a ballistic regime, $\rho(t_c)/J=\{ 840,1900,1970\}\mu {\rm  m}^{-3} E_R^{-1}$ for $V_0/E_R = \{9.25,11, 13\}$. {\color{black}We can compare these values with the ones measured in the experiment  which occur when the ratio $\rho(t_c)/J = \{ 1190\pm310, 860\pm 190, 910 \pm 210\} \mu {\rm m}^{-3} E_R^{-1}$ for $V_0/E_R = \{ 9.25,11, 13\}$ respectively.  We can see that  the ratio of $\rho(t_c)/J $ in the experiment  is roughly constant. However, while the values of $\rho(t_c)/J$ predicted by the aTWA  are almost the same for $11E_R$ and $13 E_R$,  $\rho(t_c)/J$  for $9.25 E_R$ is {\color{black}only  half of} that value.  {\color{black} On the other hand, the actual $\rho(t_c)/J$ value predicted by the aTWA at $9.25E_R$ is closer to the observed value. }}

\begin{figure}[htb]
\includegraphics[width=3.4in]{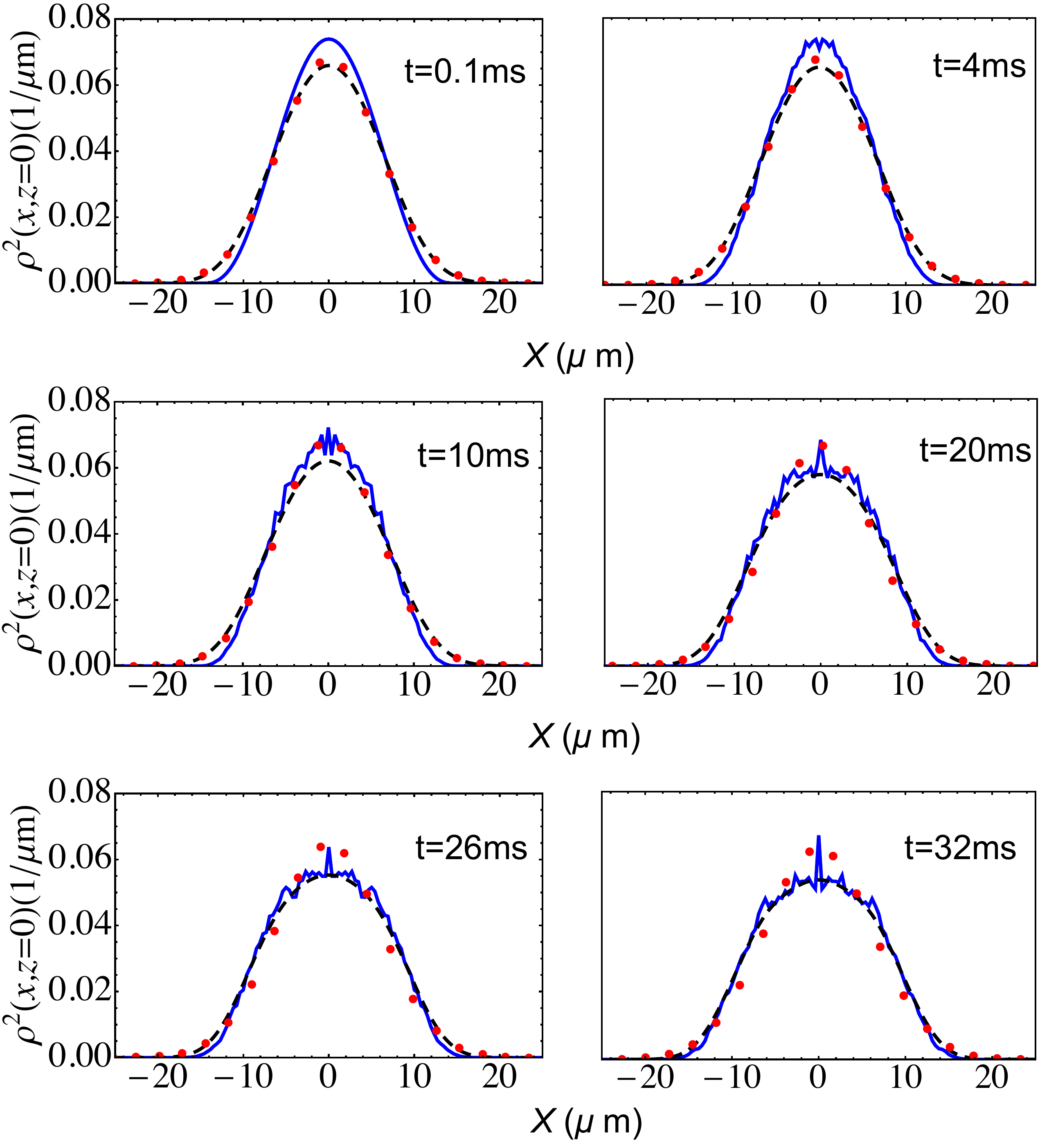}
\caption{(Color online) {\color{black}Transverse density profiles: The solid/dashed lines are obtained from the aTWA results} for the $V=13 E_R$ lattice and for $\eta=1$. The profiles are taken at $z=0$ and viewed transversely at an angle $45^o$ from the lattice axis and normalized to $1$. The solid blue lines indicate the theoretical profiles before convolution and the dashed black lines after convolution. The red points are averaged experimental data of 10 measurements.  } \label{diagonal13}
\end{figure}

Figures~\ref{diagonal7}-\ref{diagonal13} show the density profiles measured experimentally for the shallowest and deepest  lattices and the corresponding theoretically computed profiles. For the  $7.25E_R$  lattice,  one can see a good agreement  after the theory curves are convolved  to account for the imaging system resolution. However, the  theoretical distributions are not well fitted by a Gaussian at longer times. The experimental distributions on the other hand always retain a Gaussian profile. In general phase fluctuations prevent the hole formation and help to  keep the density distributions closer to a Gaussian profile.

\subsection{Number Fluctuations}

\begin{figure}[htb]
\vspace{0.5cm}
\includegraphics[width=3.in]{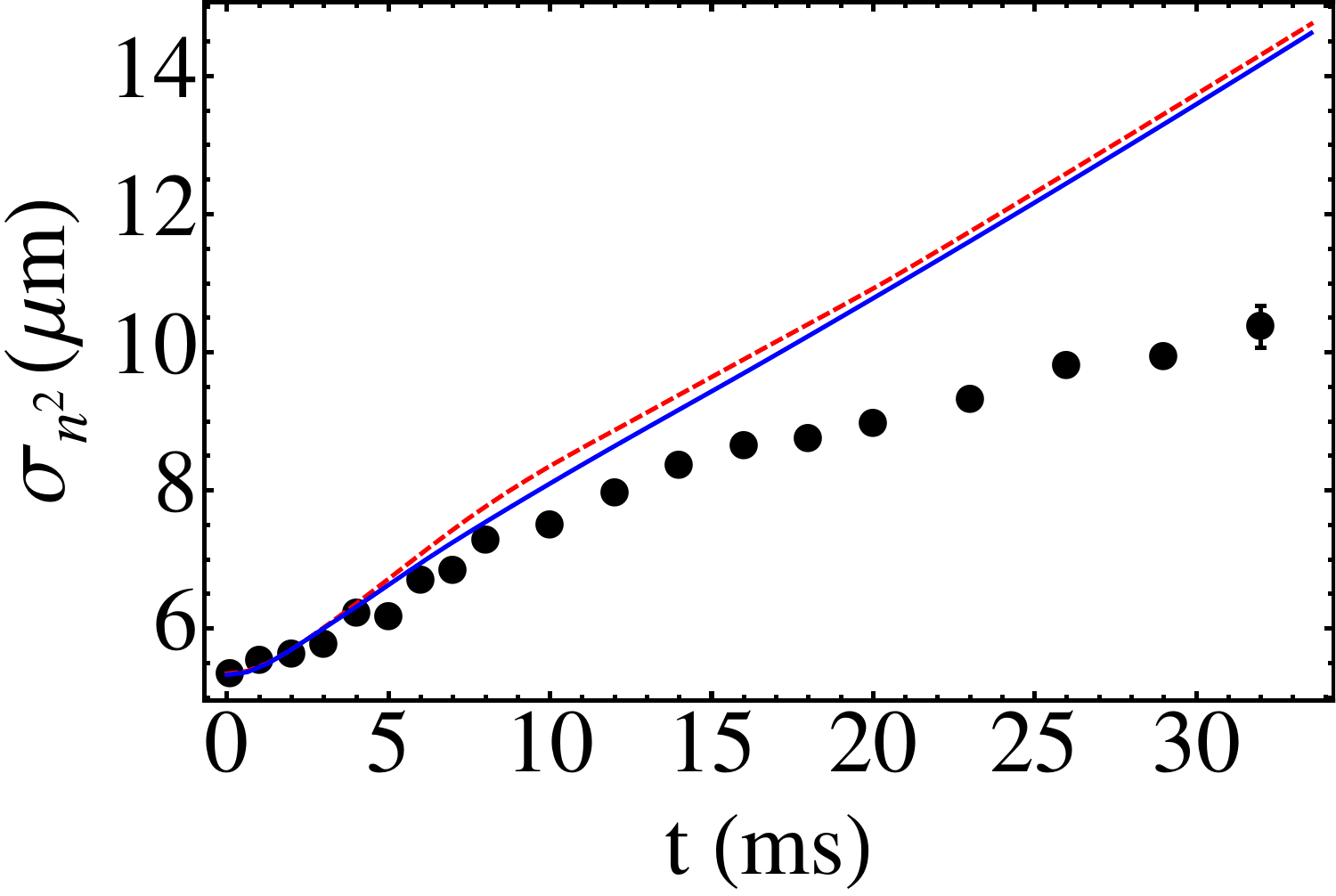}
\caption{(Color online)  Time evolution of  $\sigma_{n^2}$  {\color{black}computed for a $7.25E_R$ lattice and for $\eta=0$.} The red dashed line shows the result when number fluctuations are neglected, and the blue solid  line shows the result when number fluctuations ($\approx 1/\sqrt{N_{mn}}$) are included.  The plot clearly shows that number fluctuations  {\color{black} suppress the expansion, but only by a small amount}. } \label{poisson_only}
\end{figure}

So far we have included {\color{black}only} phase fluctuations. However, we know that in the superfluid regime relative atom number fluctuations  provide  the dominant {\color{black}beyond-mean-field} corrections.  To investigate the role of number fluctuations in the expansion dynamics, we  {\color{black}introduce} number fluctuations in the initial conditions. We implement this by allowing  $\approx 1/\sqrt{N_{nm}}$ fluctuations in the atom number in each tube, with $N_{nm}(0)$ the total number of atoms in a  tube  centered at $\vec{n}=\{n,m\}$.  We focus on the case   $V_0=7.25E_R$. The results are  shown in Fig. ~\ref{poisson_only}, where we can see that number fluctuations  do suppress the expansion rate of the cloud but by {\color{black}only a small amount}.

\subsection{Phase And Number Fluctuations}
\begin{figure}[htb]
\includegraphics[width=3.in]{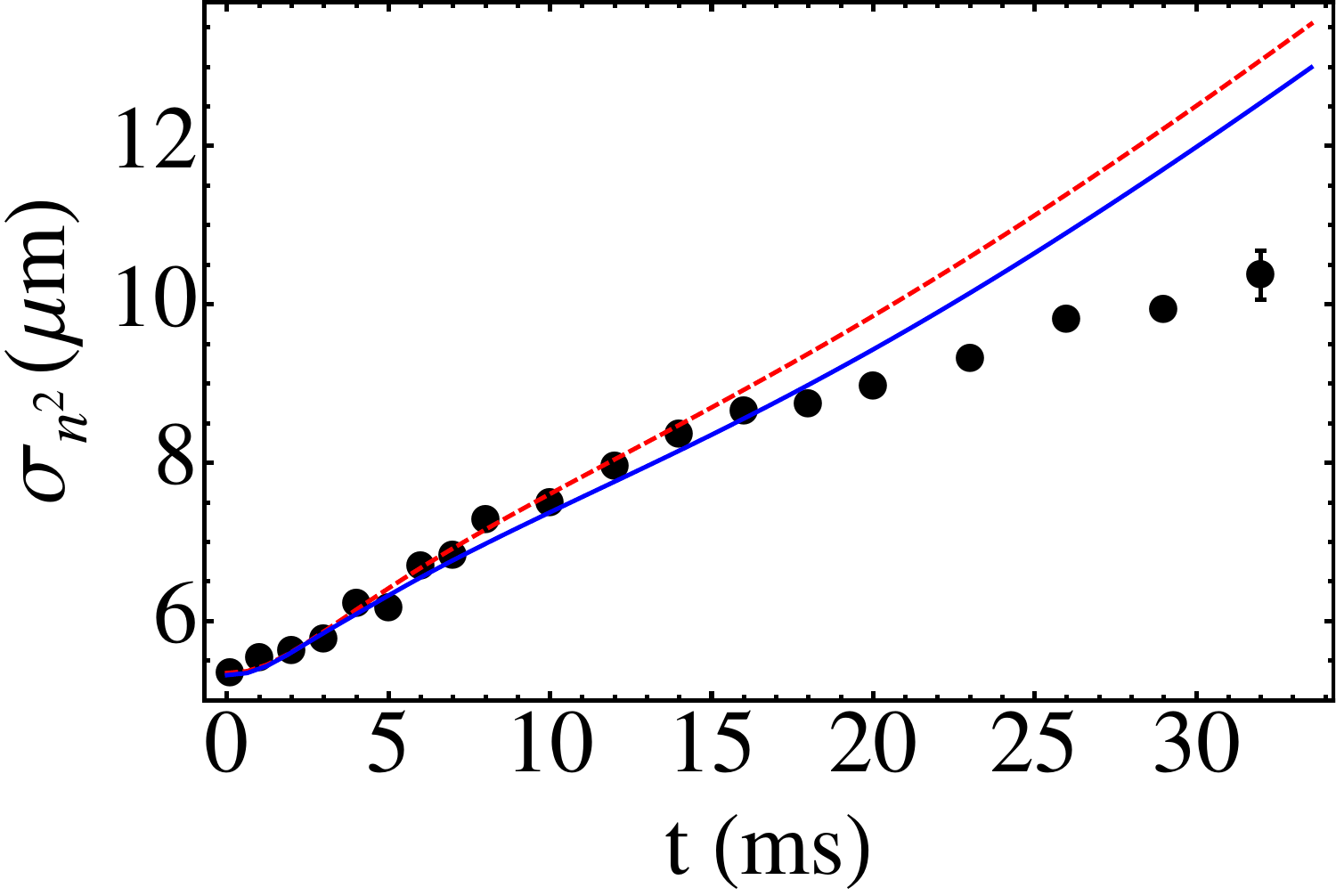}
\caption{(Color online) {\color{black}aTWA results for the} time   evolution of $\sigma_{n^2}$ for  a 7.25$E_R$ lattice and for $\eta=0.35$. The red dashed line shows the result when only phase fluctuations are included, and the blue solid  line shows the result when both number {\color{black} fluctuations ($\approx 1/\sqrt{N_{mn}}$)} and phase fluctuations are included.  The plot  shows that number fluctuations  {\color{black}slightly} suppress the initial expansion rate. } \label{phaseplusnum}
\end{figure}

In the intermediate regime {\color{black} --} between the superfluid and Mott phases {\color{black} --} both quantum and number fluctuations are relevant but   constrained by the corresponding Heisenberg uncertainty relation.  To  get an upper bound of the amount of localization predicted by  the aTWA, we  add both {\color{black}number fluctuations ($\approx 1/\sqrt{N_{nm}}$)} and phase fluctuations in the initial  conditions. The results  are shown in Fig. ~\ref{phaseplusnum}.  Indeed  the addition of  both  phase and  number fluctuations  {\color{black}helps} to suppress  the expansion rate of the cloud for moderate $\eta$  (as the  $7.25E_R$  case) but for large $\eta=1$ the addition of number  {\color{black}fluctuations} is barely noticeable. {\color{black}  In other words, the curves with number and phase fluctuations are quite similar to the curves with phase fluctuations only.}

In summary, so far we have accounted for the quantum correlations developed in {\color{black}the} tubes by  adding an overall {\color{black}random} phase in a phenomenological way. Although this procedure  {\color{black}pushes} the theory closer to the experimental observations, a fundamental problem with the aTWA  is that it misses the correlations within each tube.  Those might be the key ingredient responsible for the observed localization at short times {\color{black} due to the additional energy cost they impose when atoms tunnel from one tube to the next. Those correlations are expected to play a major role at short times when interactions are dominant.}  A significantly better treatment could   be obtained by computing the full Wigner function of the interacting systems. {\color{black}  The latter could, in principle, be achieved by starting with the Wigner function of the non-interacting system (which is known) and then by adiabatically increasing the interactions to the desired value.}  The large number of degrees of freedom in consideration substantially complicates this  implementation.

\section{Beyond The aTWA---Two Coupled Tubes}\label{sec:twotubes}
\begin{figure}
\includegraphics[width=3.3in]{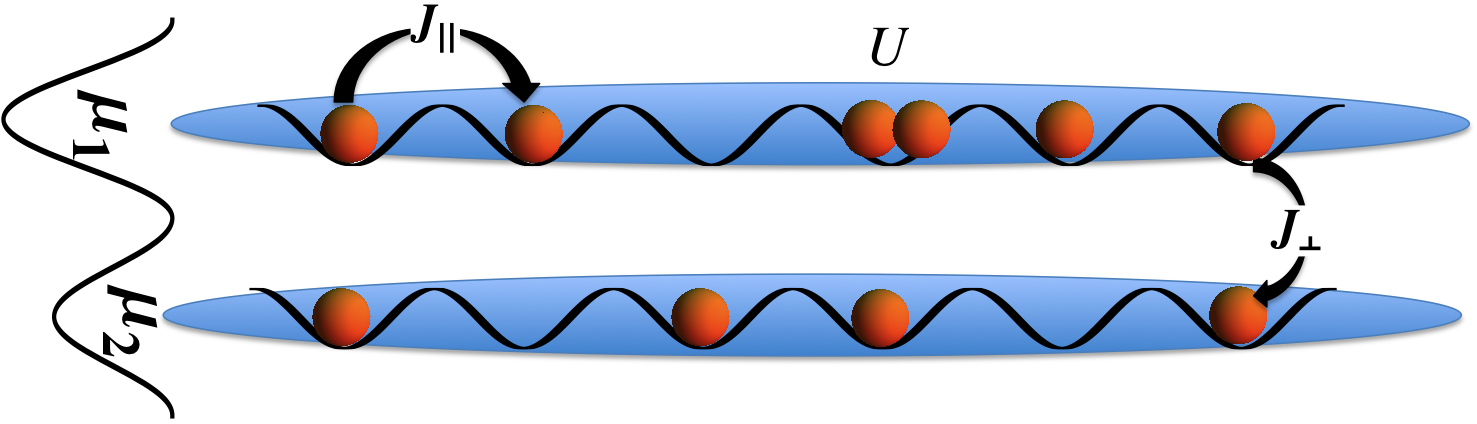}
\caption{(Color online) Schematic configuration of the two-tube system. }\label{2tubes}
\end{figure}

{\color{black}  In order to test the validity of the various approximations used so far, we now turn to a simplified model system which is amenable to a numerically exact treatment using the t-DMRG. 
In the following we will consider a configuration of two coupled tubes (see Fig. \ref{2tubes}) and compare {\color{black}the dynamics of this minimal} system to the one obtained with the pure GPE and the aTWA {\color{black}methods}. 
To be more specific, we replace in our toy two-tube model the axial parabolic trapping potential along the tubes by a box potential which is easier to  treat numerically, and we create the population imbalance at the beginning of the dynamics by choosing different chemical potentials on the two tubes, which we then set to zero instantaneously in order to mimic the time evolution in the experiment. 
Note that when removing the box potential in this setup, the dynamics at the edges of the cloud can dominate the tunneling between the two tubes.
Therefore, we find it more convenient to keep the box turned on throughout the time evolution. 
Moreover, in order to avoid the complications of dealing with a continuum system, we assume that  along each tube  there is a weak lattice (with $L$ lattice sites) and use the tight-binding approximation to describe the dynamics. 
This gives rise to a Bose-Hubbard model,}
\begin{eqnarray}
H=&&-J_{\|}\sum_{l=1,2}\sum_{<i,j>} \left(\hat{a}_{l,i}^{\dagger}\hat{a}_{l,j}+h.c\right)+\sum_{l,i}\frac{U}{2} \hat{n}_{l,i}(\hat{n}_{l,i}-1)\nonumber\\
&&-J_{\perp}\sum_i  \left(\hat{a}_{1,i}^{\dagger}\hat{a}_{2,i}+h.c\right)-\sum_{l,i}\mu_{l} \hat{n}_{l,i}, 
\end{eqnarray}  
where $J_{\|}$ is the tunneling matrix element between nearest neighboring sites within each tube, $J_{\perp}$ is the tunneling matrix element between tubes, $U$ is the on-site interaction strength  and $\mu_{l}$ is the chemical potential on tube $l$ ($l=1,2$). 
{\color{black}The difference in chemical potentials $\mu_1-\mu_2$ equals the bias between the two tubes which is adjusted in order to obtain the desired population on each of the tubes. 
Note that we assume tunneling between the tubes {\color{black}takes} place at nearest neighbouring sites only, so that we obtain a ladder geometry. }

\subsection{The DMRG Method}

DMRG is a numerical method which is capable of obtaining ground-state properties of (quasi)one-dimensional systems with very high efficiency and accuracy for lattices with up to several thousand sites. This is achieved by working in a truncated basis of eigenstates of reduced density matrices obtained for different {\color{black}bipartitions} of the lattice. The so-called discarded weight, which is the sum of the weights of the density-matrix eigenstates {\color{black}that} are neglected and which should be as small as possible \cite{Schollwock2005} quantifies the error of the method. Also, its time-dependent extensions (t-DMRG) can treat the real-time evolution of strongly correlated quantum many-body systems substantially larger than the ones amenable to exactly diagonalizing the Hamiltonian and with an accuracy which can be, at short and intermediate times, similar to {\color{black}what is} achieved in  ground-state calculations. Here, we use this accuracy  to obtain high precision numerical results.  During the evolution, we aim for a discarded weight of $<10^{-9}$ and keep up to $500$ density-matrix eigenstates for systems with $L=40$ sites {\color{black}in each tube}. We apply a time step of {\color{black}$\Delta t=0.01/J_{\|}$} and find a discarded weight of $\lesssim10^{-6}$ at the end of the time evolution displayed in the plots.

1D Bose gases are typically characterized  by the dimensionless parameter $\gamma$ \cite{Dunjko2001}, the ratio {\color{black}of the  mean field interaction energy per particle calculated without correlations to the  kinetic energy calculated with maximal correlations}. 
A  weakly interacting 1D gas ($\gamma\ll1$)  is well described by mean-field theory. {\color{black}In contrast,} in the   strongly interacting regime,  $\gamma \gg 1$, atoms avoid each other  {\color{black}to reduce mean field energy} and the bosons  behave like noninteracting fermions \cite{Girardeau1960}. In the 1D Bose-Hubbard model $\gamma\approx\frac{U}{J}$.

In our ladder configuration  we will use  $\gamma= \frac{U}{J_{\|}}$ to quantify   the interaction strength of the system.   This is expected to be a good characterization when  the intertube  tunneling  is much smaller than {\color{black}the} intratube tunneling ($J_{\perp} \ll J_{\|}$). We set $J_{\|}=1$ in all cases and vary  both $\gamma$  (by changing $U$) and  the density. Table \ref{table1} summarizes the set of parameters  that we use to study the dynamics. $N_{total}$ is the total particle number in the system.

\begin{table}
\begin{tabular}{|c|c|c|c|c|c|c|c|}
\hline
   & \text{UN$_{\text{total}}$} & \text{J}$_\|$ & \text{J}$_{\perp}$& $L$ (\text{ lattice sites})  \\ \hline
 \text{case 1} & 30 & 1 & 0.025  & 40 \\ \hline
 \text{case 2} & 30 & 1 & 0.05  & 40  \\ \hline
 \text{case 3} & 30 & 1 & 0.1 & 40  \\ \hline
  \text{case 4} & 60 & 1 & 0.05 & 40  \\ \hline
& \text{N$_{\text{total}}$} & \text{J}$_\|$ & \text{J}$_{\perp}$& $L$ (\text{ lattice sites}) \\ \hline
 \text{case 5} & 30 & 1 & 0.05  & 40 \\ \hline
\end{tabular}
\caption{Set of parameters and different initial conditions used to study the dynamics. $N_{total}$ is the total particle number in the ladder. $U$ is the on-site interaction strength, {\color{black}$J_\|$} is the nearest neighbor tunneling matrix element within each tube, and $J_{\perp}$ is the tunneling matrix element between tubes. }\label{table1}
\end{table}

\subsection{Short-Time Dynamics }

Even in this simpler two-tube system the many-body dynamics after a  quench can be quite complicated. To compare the differences between {\color{black}results obtained using the} mean-field theory, {\color{black}the} aTWA and {\color{black}the} t-DMRG, and to check the validity of the aTWA, we first focus on the short-time dynamics, {\color{black} for which we can obtain}  analytical expressions and numerical results.

A quantity {\color{black}that} gives relevant information about  the system's dynamics is the total particle number {\color{black}in} each tube,  i.e., $N_{l}= \sum_{i} n_{l,i}$  {\color{black}with} $n_{l,i}=\langle\hat{n}_{l,i}\rangle$. {\color{black} Heisenberg's equations of motion for the atom number in the tubes then read} ($\hbar\equiv 1$),
\begin{eqnarray}
\frac{d}{d t} N_{l}&=&-i  J_{\perp} \sum_{i}\left(\langle \hat{a}_{l',i}^{\dagger}\hat{a}_{l,i}\rangle-\langle\hat{a}_{l,i}^{\dagger}\hat{a}_{l',i}\rangle\right)\label{d1n},
\\
\frac{d^2}{d t^2} N_{l}&=&- \frac{J_{\perp}U}{2} \sum_i\langle  \left\{\hat{n}_{l,i}-\hat{n}_{l',i},\hat{a}_{l',i}^{\dagger}\hat{a}_{l,i}+\hat{a}_{l,i}^{\dagger}\hat{a}_{l',i}\right\}\rangle \nonumber\\
&& -2 J_{\perp}^2 (N_{l}-N_{l'})\label{d2n},
 \end{eqnarray}
where $l' \neq l$ denotes {\color{black}the position in either tube} 1 or 2. 

Generally, for the ground state, $\text{Im}[\langle a_{l,i}^{\dagger}a_{l',j}\rangle]=0$ ({\color{black}there are} no currents  in the system) {\color{black}and} therefore the first derivative of $N_l$ is zero at $t=0$, i.e., $ \frac{d}{d t} N_{l}\big{|}_{t=0}=0$.

The second derivative of $N_l$ is generally non-zero when there is a population imbalance between the two tubes,  {\color{black}due to}  the second term {\color{black}on}  the right hand side of Eq. (\ref{d2n}). The first term accounts for the role of interactions {\color{black}in} the dynamics and   involves four-particle correlation functions{\color{black}, which} can  play a fundamental role in the dynamics. For strongly  {\color{black}repulsive} bosons (hard-core bosons),  {\color{black}on-site occupancies $>1$} are energetically forbidden, {\color{black}so} the four-particle {\color{black}correlations vanish}. {\color{black}Therefore} only the term proportional to the population imbalance contributes to  the dynamics.  {\color{black}Hence, the dynamics of  hard-core bosons are governed by,}
\begin{eqnarray}
\frac{d^2}{d t^2} N_{l}= -2 J_{\perp}^2 (N_{l}-N_{l'}).
\end{eqnarray}
{\color{black}Analytical expressions} for the short time dynamics  can {\color{black}also} be obtained using  the mean-field (GPE)  approximations and the aTWA. 

At the mean-field (GPE) level, the operators $\hat{a}_{l,i}$ {\color{black} are replaced} with c-numbers $\psi_{l,i}$.   At $t=0$,  the ground state corresponds to $\psi_{l,i}=\sqrt{n_{l,i}}e^{i\phi}$, and there is a global phase for all sites.  Without loss of generality,  $\psi_{l,i}(0)$ can be chosen as $\sqrt{n_{l,i}}$.  Under the aTWA,  $\psi_{l,i}$ is chosen as $\sqrt{n_{l,i}}$  times a global phase factor {\color{black}for} each tube,  i.e.,  $\psi_{l,i}(0)\rightarrow \sqrt{n_{l,i}} e^{i \theta_l\cdot\eta}$. {\color{black}As we explained in  section \ref{sec:beyondMFM}},  $\eta$ is a tunable parameter which characterizes the strength of phase fluctuations, and  $\theta_l$  is a uniformly distributed random variable between $0$ and $2\pi$.  The average can be computed analytically{\color{black}, and in the aTWA we obtain that the dynamics is governed by}
\begin{eqnarray}
\frac{d^2}{d t^2} N_{l}\bigg{|}_{t=0}&=&-2 J_{\perp}^2 (N_{l}-N_{l'}) \nonumber\\
 &&-2J_{\perp}U \frac{N_{l}-N_{l'} }{L}\sqrt{N_{l}N_{l'}} \frac{\sin^2(\pi\eta)}{\pi^2\eta^2}.\label{aTWA}
\end{eqnarray} The second term in the above equation equals $-2J_{\perp}U\frac{(N_{l}-N_{l'})}{L} \sqrt{N_{l}N_{l'}}$ when  $\eta=0$, and goes to  $0$ when $\eta=1$, agreeing with the GPE and  hard-core {\color{black}boson limits}.
{\color{black} Since $\frac{d}{d t}  N_{l}\big{|}_{t=0}=0$, to lowest order,  $ N_{l}(t) =\frac{1}{2} \left(\frac{d^2}{d t^2}  N_{l}\big{|}_{t=0}\right) t^2+N_{l}(0) $ at short times.}
Numerically, we can either directly calculate $\frac{d^2}{d t^2}  N_{l}\big{|}_{t=0}$ from the initial wave-function at $t=0$ or extract  it from  $N_{l}(t)$  by fitting it to {\color{black}a quadratic} polynomial.
Here  we use the second (fitting) method  to get $\frac{d^2}{d t^2}  N_{l}\big{|}_{t=0}$.

At $t=0$,  Eq.~(\ref{aTWA}) becomes,
\begin{eqnarray}
\ddot{x}|_{t=0}= f(x)+g(x)  \frac{\sin^2(\pi\eta)}{\pi^2\eta^2}\label{d2x}, 
\end{eqnarray} 
{\color{black}where} $x=\frac{N_{1}}{N_{total}}$, $ f(x)\equiv 2 J_{\perp}^2 (1-2x)$ and $g(x)\equiv2J_{\perp}\frac{UN_{total}}{L}(1-2x)\sqrt{x (1-x)}$. When $x=0$, the population imbalance between the  two tubes is maximum and all particles {\color{black}occupy} the same  tube.  When $x=1/2$, atoms are equally distributed  between the tubes.  

\begin{figure}
\includegraphics[width=3.2in]{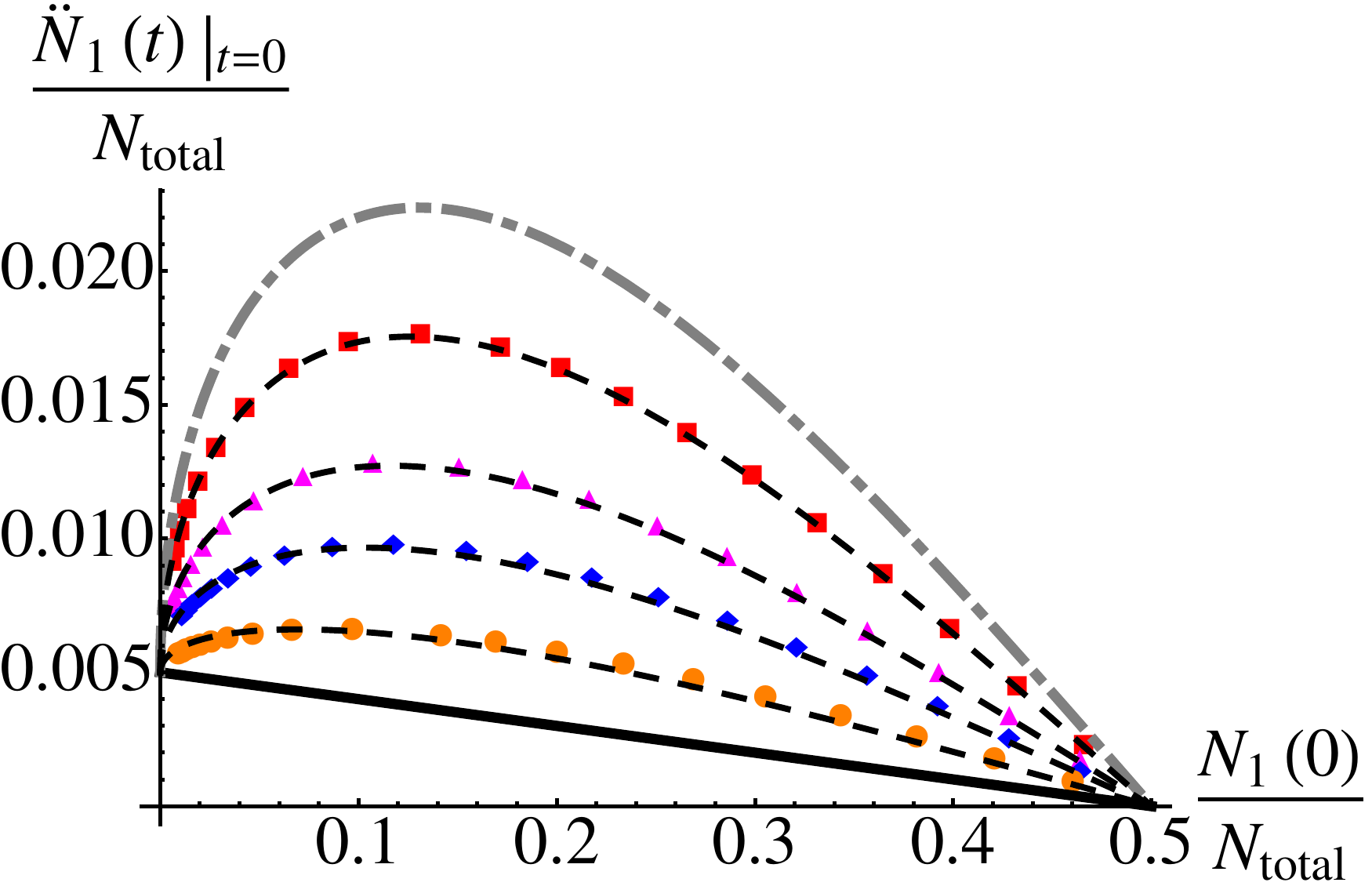}
\caption{(Color online)  $\frac{\ddot{N}_1}{N_{total}}|_{t=0}$ as a function of $\frac{N_1(t=0)}{N_{total}}, J=0.05, UN_{total}=30$. The black solid line shows the  GPE limit  and the gray dash-dotted line the hard-core boson limit [see Eq. (\ref{d2x})].  The red squares,  magenta upper-triangles, blue diamonds and orange disks  show the {\color{black}  initial curvature $\frac{\ddot{N}_1}{N_{total}}|_{t=0}$ extracted from the t-DMRG results} at $U=1$, $U=2$, $U=3$ and $U=5$ respectively. The black dashed lines show the aTWA solutions  using $\eta$ as a fitting parameter.  The values of $\eta$ are listed in TABLE \ref{table2} . {\color{black}This plot is computed using the ``case 2" parameters shown in  Table II.} }\label{alpha_dmrg}
\end{figure}

The fitted values of  $\eta$ {\color{black}that best reproduce the}  t-DMRG {\color{black}calculations obtained for different parameter regimes} are shown in TABLE II.  {\color{black}Fig. \ref{alpha_dmrg} shows comparisons between the aTWA results and the t-DMRG results for the optimal value of $\eta$ found  for one of the parameters displayed in Table II.} In the GPE limit ($\eta=0$), only the product $U N_{total}$ enters in the equations of motion, meaning that,  as long as  $U N_{total}$ is kept constant, the dynamics of the system {\color{black}are independent of} $U$. This feature can be clearly seen  in Eq. (\ref{aTWA}). This is certainly not the behavior predicted by  the t-DMRG solutions which show a slow down of  the dynamics with increasing $U$.
The TWA is expected to fully capture the role of  quantum correlations at short times.   Interestingly, we find that $ \ddot{x}|_{t=0}$ computed  by  t-DMRG  is extremely well described by $f(x)+g(x) \frac{\sin^2(\pi\eta)}{\pi^2\eta^2}$ {\color{black}where} $\eta$ is the only fitting parameter.
The excellent agreement between t-DMRG and aTWA allows us to conclude that  the aTWA  captures well the quantum correlations parameterized by $\eta$ at short times in this simple two-tube model. {\color{black}We also see}, that when $UN_{total}$ {\color{black}is} fixed, {\color{black}in} the parameter regime {\color{black}under} consideration{\color{black},} {\color{black} the fitted values of  $\eta$ that best reproduce the t-DMRG calculations are almost independent of  the initial population imbalance, {\color{black}i.e., 1-2x}, and the tunneling  between tubes, {\color{black}i.e., $J_{\perp}$,} when $J_{\perp}\ll J_{\|}$. }

\begin{table}
\begin{tabular}{|c|c|c|c|c|c|c|}
\hline
& $\eta$ for case 1&$\eta$ for case 2 & $\eta$ for case 3 & $\eta$ for  case 4 &$\eta$ for  case 5 \\ \hline
 \text{U=1} &0.31 &0.30  &0.30 & 0.27 &  0.30\\ \hline
 \text{U=2} &0.46 &0.45  &0.46 & 0.39 &  0.39\\ \hline
 \text{U=3} &0.56 &0.56  &0.57 & 0.48 &  0.45\\ \hline
 \text{U=5} &0.69 &0.70  &0.71 & 0.61 &  0.54\\ \hline
\end{tabular}
\caption{Fitted values of  $\eta$ for different set of parameters used to investigate the dynamics: at short times the t-DMRG dynamics is well captured by the aTWA when $\eta$ is used as a fitting parameter.  }\label{table2}
\end{table}

\subsection{Long Time Dynamics}
\begin{figure}
\subfigure[]{\includegraphics[width=1.65in]{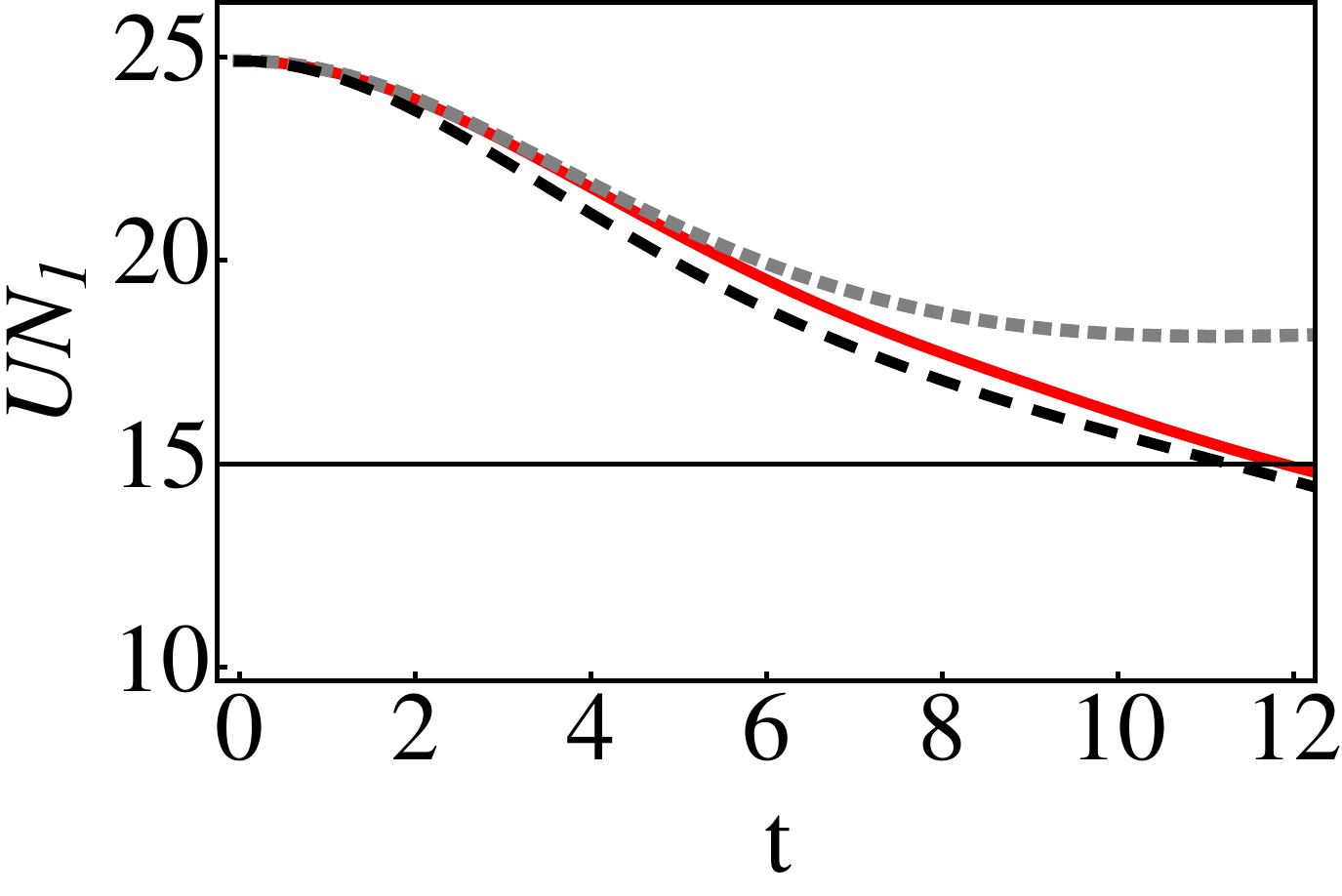}}
\subfigure[]{\includegraphics[width=1.65in]{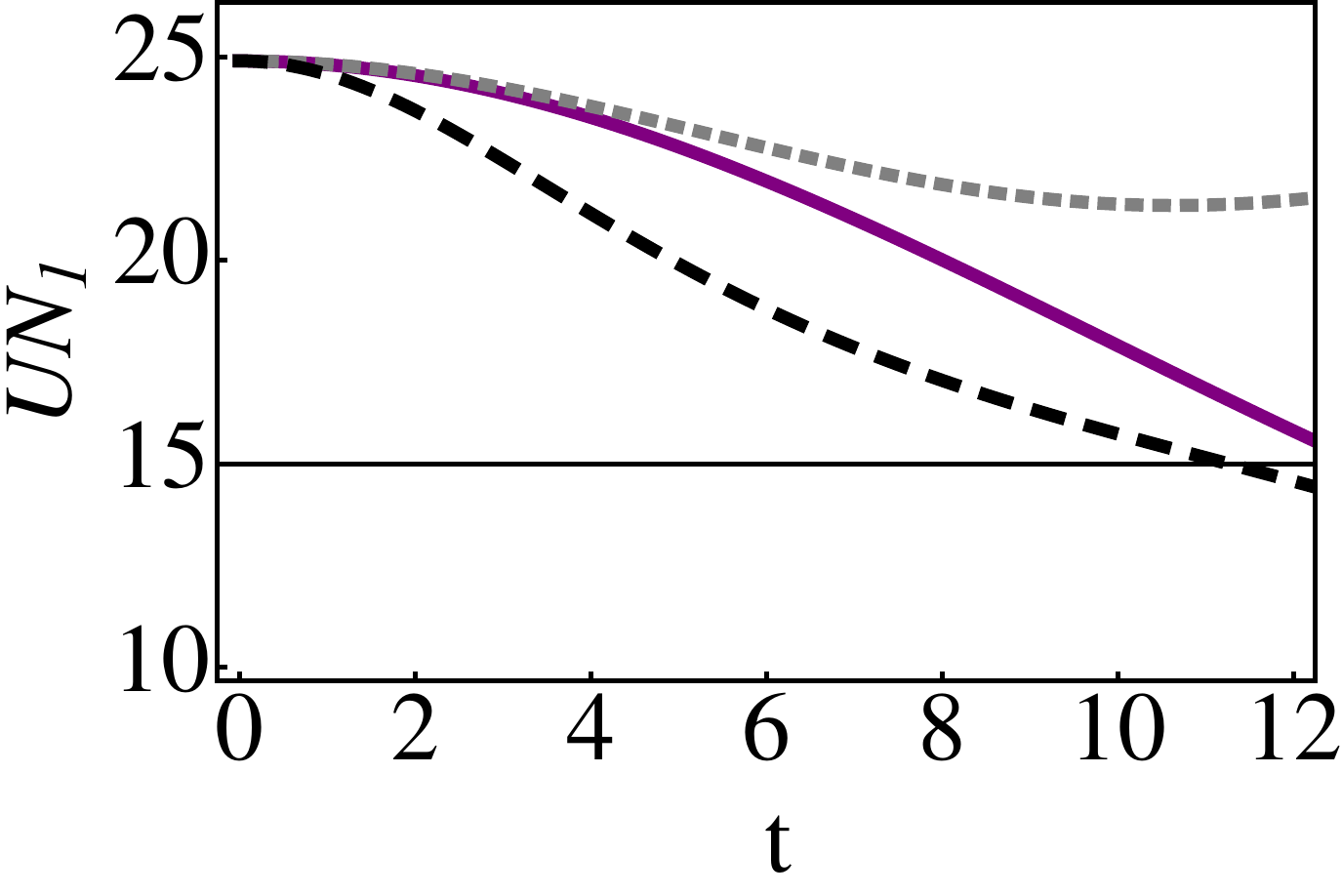}}\\
\subfigure[]{\includegraphics[width=1.65in]{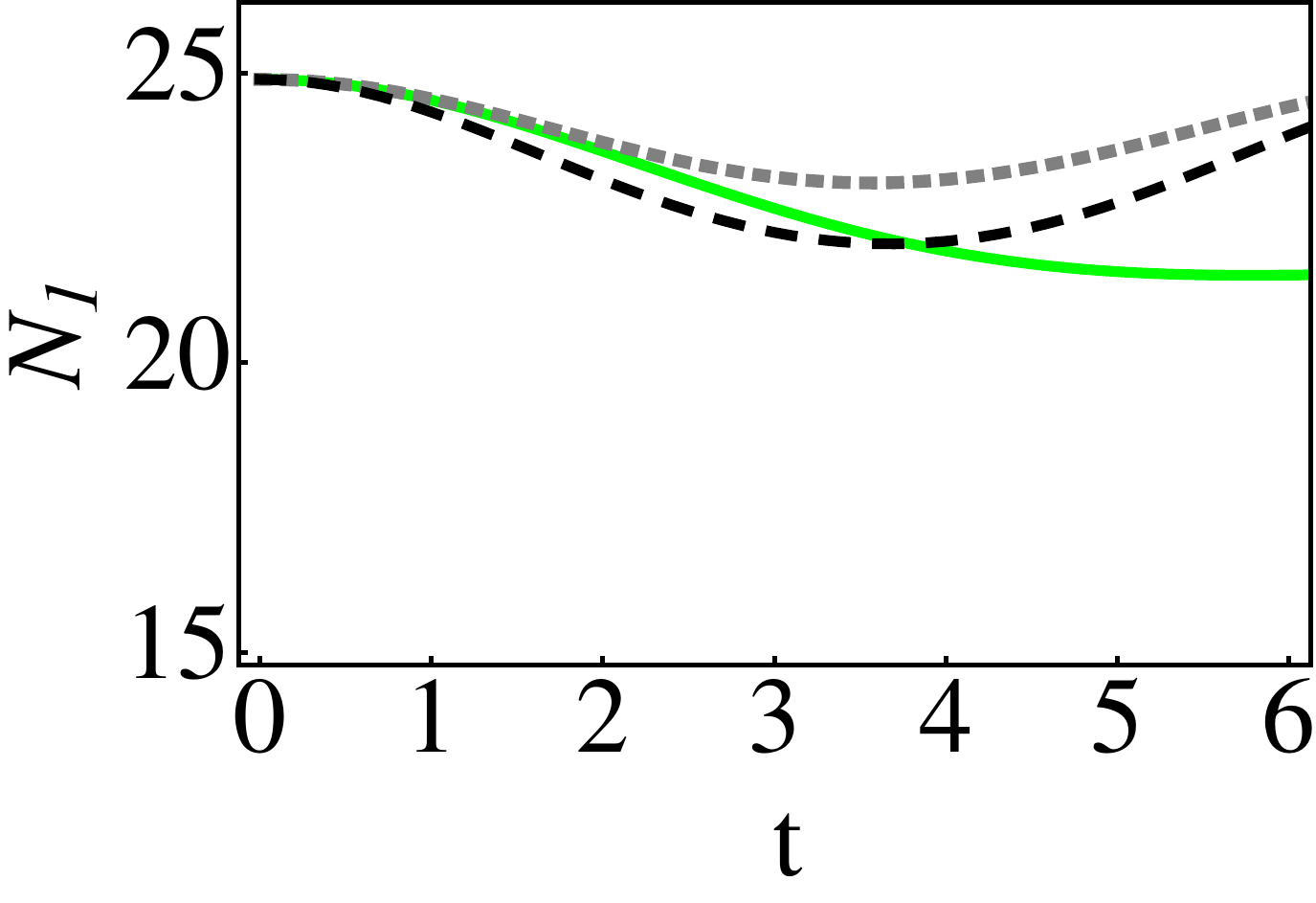}}
\subfigure[]{\includegraphics[width=1.65in]{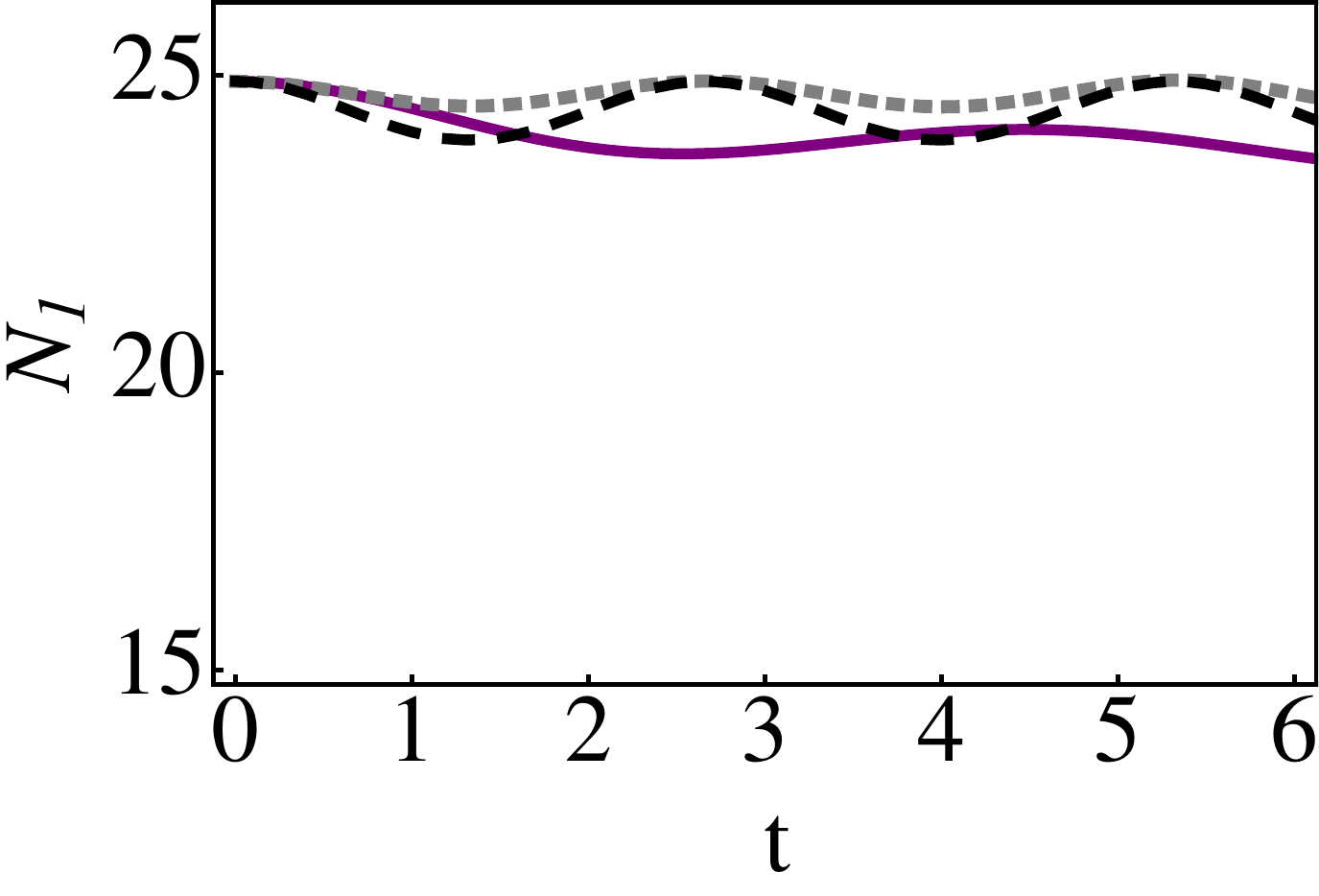}}
\caption{(Color online) Evolutions of $UN_1$ {\color{black}and} $N_1$ as functions of time calculated using the DMRG and GPE methods.  In panel (a) and (b), the values of  $U$ are $1$ and $5$ respectively, while $UN_{total}$ {\color{black}is} fixed to $30$.  In panel (c) and (d), the values of  $U$ are $2$ and $5$ respectively, while $N_{total}$ {\color{black}is} fixed to  $30$.
 In each panel, the solid line is the t-DMRG solution, the black dashed line is the  mean-field calculation with no random phase (GPE). The dotted gray line is for (a)
 $\eta=0.3$, (b) $\eta=0.7$, (c) $\eta=0.4$ and (d) $\eta=0.53$. }\label{N1}
\end{figure}

{\color{black}
In Fig. \ref{N1} we compare the GPE, the aTWA and the t-DMRG dynamics. We see significant deviations between the approximate methods and the t-DMRG at longer times.  
This shows that the dynamics of the Bose-Hubbard model on this ladder geometry is dominated by correlation effects. 
Although at short times the t-DMRG results are well described by the aTWA results with an appropriate value of $\eta$, at longer times the aTWA fails to reproduce the many-body dynamics.  {\color{black}This behavior confirms the relevance of quantum correlations neglected by the aTWA. } Those deviations are consistent with  the deviations we saw when trying to model the experimental data with the aTWA. }

Fig.\ref{N1}  shows that the {\color{black}weakly} interacting regime {\color{black}is the regime where} the GPE and the aTWA (with small $\eta$) {\color{black}predictions are closer to} the t-DMRG results for {\color{black}longer time}. The aTWA gives slightly better agreement {\color{black}at short times}, indicating that for weak interactions it better captures correlation effects.  In the {\color{black}strongly} interacting regime,  tunneling is greatly suppressed{\color{black}, especially} in the presence of a large  initial population imbalance. While in the GPE and the aTWA,  $N_1(t)$  oscillates with more or less constant frequency and amplitude,   in the t-DMRG, $N_1(t)$  oscillates with a {\color{black}lower} frequency and  gradually ``relaxes''  towards an equal distribution of atoms in  the two tubes.  

\begin{figure}
\vspace{0.3cm}
\subfigure[]{\includegraphics[width=1.6in]{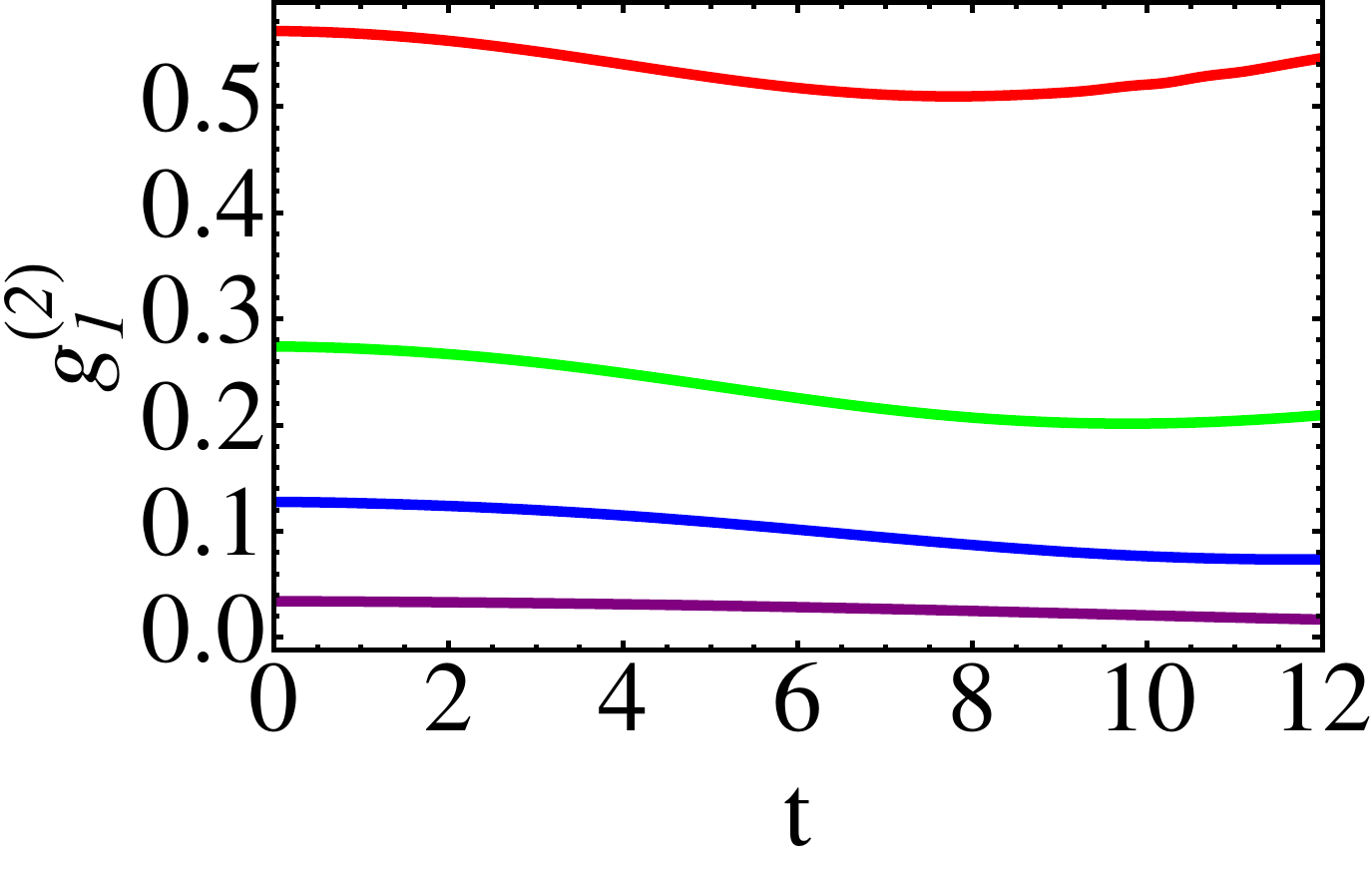}}\quad
\subfigure[]{\includegraphics[width=1.6in]{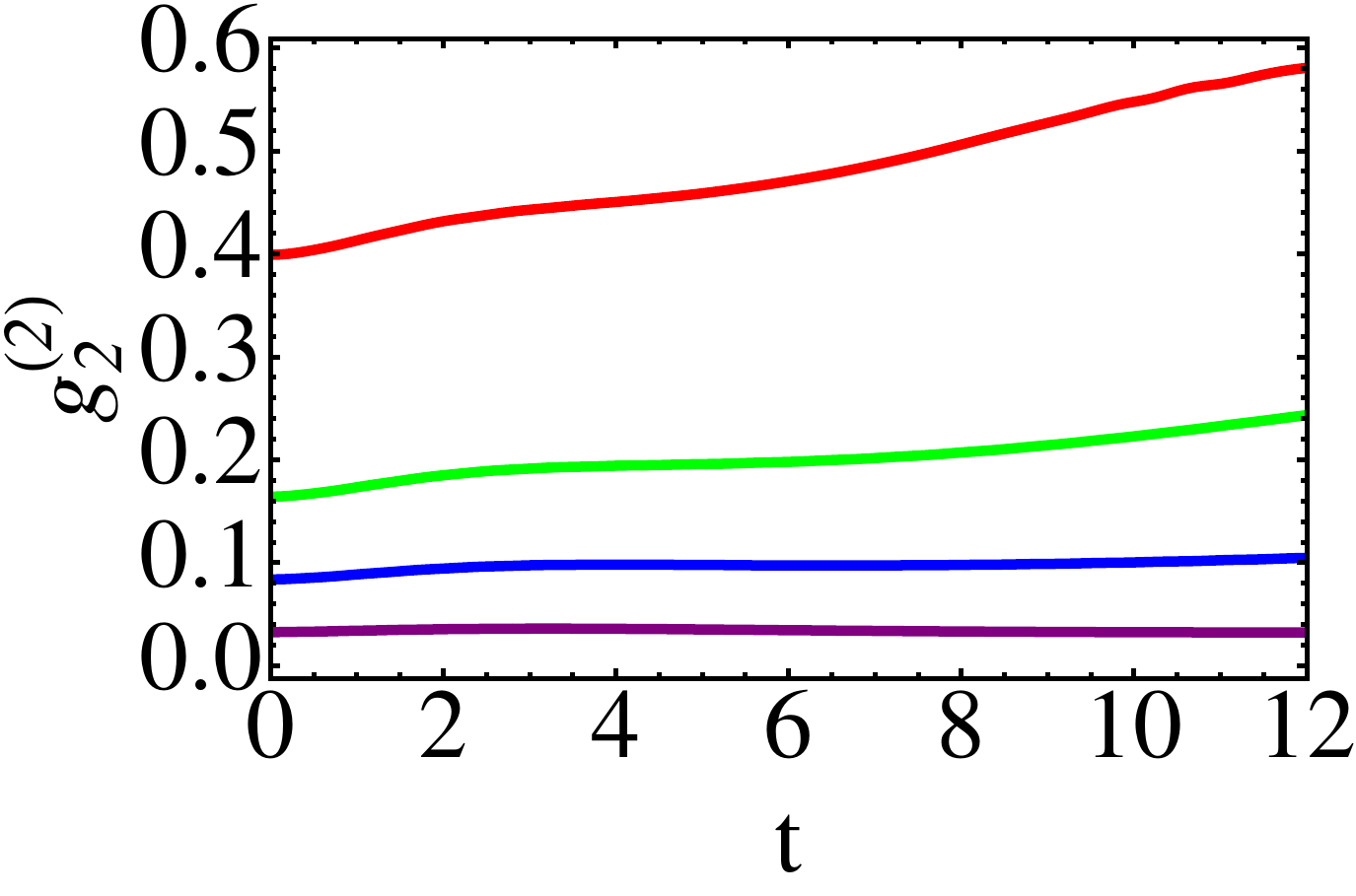}}\\
\caption{(Color online) {\color{black}$g_l^{(2)}$ as function of time in different regimes.} {\color{black}The correlations are  extracted from the t-DMRG {\color{black}results} for the parameters describing case 2.} Panel (a)  {\color{black}shows} $g_1^{(2)}$, $\frac{L}{N_1^2}\sum_{i} \langle\hat{n}_{1i}^2\rangle- n_{1i}$ and  Panel (b) {\color{black}shows}  $g_2^{(2)}$, $\frac{L}{N_2^2}\sum_{i} \langle\hat{n}_{2i}^2\rangle- n_{2i}$.  In each panel, from top to bottom,  {\color{black} the lines correspond to $U=1$, $U=2$, $U=3$ and $U=5$ respectively.}  }\label{g2}
\end{figure}

To further {\color{black}quantify} the quantum correlations (which are not captured in {\color{black}the GPE and are {\color{black}only} partially captured in the aTWA}) in the ladder system, we calculate  the local pair  correlation function, $g^{(2)}$, which is proportional to the probability of observing two particles in the same lattice site \cite{Kinoshita2005,Bloch2008r}.  For a 1D {\color{black}gas}, when $\gamma\ll 1$,  $g^{(2)}\to 1$  and when $\gamma \gg 1$, $g^{(2)}\to 0$.   In our quasi 1D system, we define $g_l^{(2)}$ {\color{black}in a} tube  $l$ as,
\begin{eqnarray}
g_l^{(2)}=\frac{L}{N_l^2}\sum_i\langle \hat{n}_{li}^2\rangle-\langle  \hat{n}_{li}\rangle.
\end{eqnarray} 
Fig. \ref{g2} shows that $g_l^{(2)}$ decreases with increasing  $U$ (larger $\gamma$) {\color{black} and approaches} zero when $\gamma$ goes to infinity.  {\color{black}$g_l^{(2)}$ is a slowly varying function when  compared to the evolution of  the total atomic population in each tube ($N_l$).}

By using t-DMRG as a means to benchmark the  GPE and the aTWA, we clearly see that while {\color{black}the} aTWA is accurate at short times, both approaches break down at longer times. We {\color{black}can} also clearly  {\color{black}observe} the complexity of {\color{black} the non-equilibrium behavior even in this simple coupled-tube model.}  

{\color{black}It is important to keep in mind  that the two-tube  model used in this section  is much simpler than the  multi-tube system of  the experiment, where in addition the  axial external confinement was turned off and  no lattice was present along the tubes.  Consequently,  the conclusions of this analysis cannot be simply compared to the experimentally observed dynamics. However,  this simple two-tube model {\color{black}clearly shows}  that correlations  can play {\color{black}a dominant role} in suppressing the tunneling dynamics,  without the need of  density gradients.}

\subsection{Discussion}
Although we see decent agreement between the aTWA and the DMRG results at short times in the two-tube model, in the real experiment, we find that the aTWA deviates from the measurements almost as soon as the expansion starts.  We will now discuss the various possible explanations for the experiment-theory discrepancy. 

First, although the aTWA adds phase fluctuations that arise from quantum correlations, they are added in an ad hoc way, and not as the result of the system lowering its energy.  In a complete theory (and presumably in the experiment) correlations build up in order to decrease the mean interaction energy. Since the wave function of a tunneling atom will not in general be appropriately correlated with the atoms in an adjacent tube, it would have to pay much of the interaction energy cost that the correlations avoid. If that energy cost is larger than the tunneling energy, then hopping between tubes will be suppressed. Although $\gamma$ is not initially large near the central tubes, the fast density decrease due to the axial expansion could dynamically increase the quantum correlations, so that this mechanism could play an important role in suppressing transverse expansion.

Second, it is possible that the initial phase variations along the tubes, which are left out of the aTWA, could be important. Note, however, that phase variations along tubes are generated dynamically in the aTWA when atoms tunnel between tubes. Still, the resulting phase fluctuations may not be sufficient to capture the quantum dynamics.

Finally, we have seen recent experimental evidence that the initial atom distribution among the central tubes is flatter than a Thomas-Fermi profile. Such a distribution is probably a consequence of the known non-adiabaticity of the 2D lattice turn-on, which leads to much higher initial densities than adiabatic turn-on predicts, in a way that is quite insensitive to turn-on time. We expect that a more flat-top initial distribution would lead to less initial expansion, but it is not clear that it would be decreased enough to explain the experimental results.  The limited spatial resolution of the experiment prevents us from directly checking the distribution of atoms among tubes and performing a calculation based on the result.

In summary, there remain several logical possibilities for the discrepancy between  experiment and theory here. Further investigation is needed.
 
\section{conclusion}\label{sec:conclusion}
We have presented various {\color{black}mean-field and beyond-mean-field} models to describe  {\color{black}our recent} experimental observations of  interaction-induced localization effects during the expansion of  an array of coupled 1D tubes. In contrast to  {\color{black}previous} 1D-lattice  experiments,  where a pure mean field model was able to
capture the observed MQST behavior, in our case we find important corrections induced by quantum fluctuations. {\color{black}  Thermal fluctuations {\color{black}and non-adibatic loading conditions,}  which are not accounted for in our {\color{black}analysis}, may give rise to similar effects, but we have not explored {\color{black}those} in this work.}

The addition of phase {\color{black}fluctuations} in the initial conditions (aTWA) coarsely captured the main self-trapping features seen in the experiment{\color{black},}  but underestimated the observed  localization.  {\color{black} The fact that the aTWA does not {\color{black}properly} account for the {\color{black}full} quantum correlations {\color{black}could be} responsible for the  failure of the theory to reproduce the experiment. {\color{black}Comparisons} with t-DMRG calculations performed in a simple two-tube model  find very good agreement between the aTWA and the t-DMRG at short times but a break down of the aTWA at longer times.  The two-tube calculations support the fact that quantum correlations, not captured by the mean field model,  can play an important role in suppressing tunneling.  Further  comparisons among the various methods used for modeling the dynamics,  and experiments are needed to help shed light on  the complex many-body dynamics of coupled correlated systems.}

{\it Acknowledgements}  R. H.  was supported by the
AFOSR YI. D. S. W. acknowledges support from the NSF
(PHY 11-02737), the ARO, and DARPA. A. M. R and S. Li
acknowledge support from NSF, the AFOSR and the ARO
(DARPA OLE).

\appendix

\section{Variational Method -- For 1D Optical Lattice}\label{var}

In a deep 1D optical lattice, the external potential $V_{ext}$ is composed of a 1D lattice potential $V_{lat}=V_0 \sin^2(\pi x/d)$ and an external  magnetic or dipole confinement
$V_{dip}=\frac{1}{2}M(\omega_x^2 x^2+\omega_y^2 y^2+\omega_z^2 z^2)$ where $V_0$ is the lattice depth and $d$ is the lattice spacing.  In the tight-binding approximation, the macroscopic wave-function can be written as,
\begin{equation}
\Psi=\sum_j\psi_j(t)  \phi_j[\vec{x},N_j(t)]. \label{tightbinding}
\end{equation}
$\phi_j[\vec{x},N_j(t)]$, which {\color{black} is centered} at the minimum of the $j$th well, is a time-dependent function which depends on the particle number on site $j$ at a given instant, and also on the  trapping frequencies  along the  two transverse directions. $\phi_j[\vec{x},N_j(t)]$ is normalized to 1, i.e., $\int d\vec{x} \left|\phi_j[\vec{x},N_j(t)]\right|^2$=1.

By using the tight-binding approximation, Eqn. (\ref{tightbinding}), and after integrating out the spatial degrees of freedom in the GPE, one gets the so called discrete nonlinear Schr$\text{\" o}$dinger equation (DNLS)\cite{Smerzi2003},
\begin{equation}
i\hbar\frac{\partial {\psi_j}}{\partial t}=\epsilon_j\psi_j-J(\psi_{j+1}+\psi_{j-1})+\mu_j \psi_j. 
\end{equation}
The on-site energies $\epsilon_j$ arise from the external potential (along the lattice direction) superimposed on the optical lattice,
\begin{eqnarray}
\epsilon_j&=&\int d\vec{x}\  |\phi_j(\vec{x})|^2  \frac{1}{2}M\omega_x^2 x^2= \frac{1}{2}M\omega_x^2 d^2 j^2.
\end{eqnarray}
 $J$ is the tunneling matrix between adjacent sites,
\begin{equation}
J=\int d\vec{x}\ \phi_j(\vec{x}) \left[\frac{\hbar^2}{2M}\frac{\partial^{2}}{\partial x^{2}}-V_{lat}\right]\phi_{j\pm1}(\vec{x}), 
\end{equation}
and {\color{black}$\mu_j^{loc}$ is} the ``local" chemical potential,
\begin{eqnarray}
\mu_j^{loc}&=&\mu_j^{kin}+\mu_j^{int}+\mu_j^{pot}\nonumber\\
&=&\int d\vec{x}\ \frac{\hbar^2}{2M} \left(\vec{\nabla}  \phi_j\right)^2 +  g \left|\psi_j(t)\right|^2  \int d\vec{x}\ \left|\phi_j\right|^4 \nonumber\\
&+&\int d\vec{x} \  |\phi_j|^2 \frac{1}{2}M\left(\omega_y^2y^2+\omega_z^2z^2\right).
\end{eqnarray}

Depending on the relative {\color{black}values of} the dipole trapping frequencies  and the on-site chemical potential,  the condensates in each well of the lattice can be regarded as $ 0$D, $1$D and $2$D.
The effective dimensionality of the condensate gives a different scaling of the local potential with the number of atoms \cite{Smerzi2003},
$$\mu_j=U_{\alpha} |\psi_j|^{\alpha},$$
where $\alpha=\frac{4}{2+D}, D=0,1,2$ is the dimensionality of the condensate, and $U_{\alpha}$ is a constant which does not depend on the particle number or site  index.

The effective Hamiltonian of the system is \cite{Smerzi2003},
\begin{eqnarray}
H_{eff}&=&-J (\psi_j^*\psi_{j+1}+c.c)+\frac{2}{2+\alpha} U_{\alpha} \left|\psi_j\right|^{\alpha+2}.
\end{eqnarray}

{\color{black} To proceed further,  a variational method can be used.   For this method, we assume  that  the amplitudes $\psi_j$ can be, to a good approximation, parameterized by a Gaussian function, $\psi_n(t)=\sqrt{N}\sqrt[4]{\frac{2}{\pi r(t)^2}}\exp\left[-\frac{n^2}{r(t)^2}+i\frac{\delta(t)}{2}n^2\right]$. From this assumption, we can determine  the equations of motion of $q_i(t)= \{r,\delta\}$.} Here $N$ is the particle number, $r(t)$ is the width of the atom distribution and $\delta(t)$ is its conjugate momenta (in the lattice units). For the case  $\epsilon_j=0$, the effective Hamiltonian under this assumption becomes
\begin{equation}
H_{eff}=2JN\left[- e^{-\sigma}+\left(\frac{2}{2+\alpha}\right)^{\frac{3}{2}}\left( \frac{2}{\pi} \right)^{\frac{\alpha }{4}}  \frac{\Lambda}{r^{\alpha/2}}\right],
\end{equation}
 where $\sigma=\frac{1}{2r^2}+\frac{r^2\delta^2}{8}$, $\Lambda=\frac{N^{\alpha/2}U_\alpha}{2J}$.
 The quasi-momentum dependence of the effective mass,  ${m^*}^{-1}\equiv \frac{\partial^2H}{\partial p^2}$, and group velocity, $\nu_g\equiv\frac{\partial{H}}{\partial{p}}=1/m^*$,  allows a rich variety of dynamical regimes. A diverging effective mass $m^*\rightarrow \infty$ is the signature of wave-packet  self-trapping.

The variational method has been shown to be a great success in characterizing various regimes of the dynamics. It also gives a critical value  of
MQST-to-{\color{black}diffusive} transition (when $ r_0\gg 1$, where $r_0$ is the initial value of $r$),
\begin{equation}
\Lambda_c=r_0^{\alpha/2} \left(\frac{2+\alpha}{2}\right)^{3/2}\left(\frac{\pi}{2}\right)^{\alpha/4}.
\end{equation}

 Based on the variational  method,  when $\Lambda>\Lambda_c$ (MQST regime),  $m^*(t\rightarrow \infty)\rightarrow \infty$.  The ratio between  $r_0$ and the asymptotic width $r_{\text{max}}(t\rightarrow \infty$) is given by
 the relation $\frac{r_0}{r_{max}}=\left(1-\frac{\Lambda_c}{\Lambda}\right)^{\frac{2}{\alpha}}$.   When  $\Lambda\leq\Lambda_c$ (the {\color{black}diffusive} regime), $m^*(t\rightarrow \infty)\rightarrow \frac{\Lambda_c}{\Lambda_c-\Lambda}$,   $r(t\rightarrow \infty)\rightarrow \infty$. The same variational method has also been extended into 2D and 3D optical lattices with 0D  condensates  \cite{Xue2008}.

\bibliography{./ref}

\begin{thebibliography}{30}
\expandafter\ifx\csname natexlab\endcsname\relax\def\natexlab#1{#1}\fi
\expandafter\ifx\csname bibnamefont\endcsname\relax
  \def\bibnamefont#1{#1}\fi
\expandafter\ifx\csname bibfnamefont\endcsname\relax
  \def\bibfnamefont#1{#1}\fi
\expandafter\ifx\csname citenamefont\endcsname\relax
  \def\citenamefont#1{#1}\fi
\expandafter\ifx\csname url\endcsname\relax
  \def\url#1{\texttt{#1}}\fi
\expandafter\ifx\csname urlprefix\endcsname\relax\def\urlprefix{URL }\fi
\providecommand{\bibinfo}[2]{#2}
\providecommand{\eprint}[2][]{\url{#2}}

\bibitem[{\citenamefont{Landau}(1933)}]{Landau}
\bibinfo{author}{\bibfnamefont{L.~D.} \bibnamefont{Landau}},
  \bibinfo{journal}{Z. Phys. (N.Y.)} \textbf{\bibinfo{volume}{13}},
  \bibinfo{pages}{905} (\bibinfo{year}{1933}).

\bibitem[{\citenamefont{Shluger and Stoneham}(1993)}]{Shluger1993}
\bibinfo{author}{\bibfnamefont{A.~L.} \bibnamefont{Shluger}} \bibnamefont{and}
  \bibinfo{author}{\bibfnamefont{A.~M.} \bibnamefont{Stoneham}},
  \bibinfo{journal}{J. Phys.: Condens. Matter} \textbf{\bibinfo{volume}{5}},
  \bibinfo{pages}{3049} (\bibinfo{year}{1993}).

\bibitem[{\citenamefont{Salje et~al.}(1995)\citenamefont{Salje, Alexandrov, and
  Liang}}]{Salje1995}
\bibinfo{author}{\bibfnamefont{E.~K.~H.} \bibnamefont{Salje}},
  \bibinfo{author}{\bibfnamefont{A.~S.} \bibnamefont{Alexandrov}},
  \bibnamefont{and} \bibinfo{author}{\bibfnamefont{W.~Y.} \bibnamefont{Liang}},
  \emph{\bibinfo{title}{Polarons and Bipolarons in High Temperature
  Superconductors and Related Materials}} (\bibinfo{publisher}{Cambridge
  University Press}, \bibinfo{address}{Cambridge, England},
  \bibinfo{year}{1995}).

\bibitem[{\citenamefont{Millis et~al.}(1995)\citenamefont{Millis, Littlewood,
  and Shraiman}}]{Millis1995}
\bibinfo{author}{\bibfnamefont{A.~J.} \bibnamefont{Millis}},
  \bibinfo{author}{\bibfnamefont{P.~B.} \bibnamefont{Littlewood}},
  \bibnamefont{and} \bibinfo{author}{\bibfnamefont{B.~I.}
  \bibnamefont{Shraiman}}, \bibinfo{journal}{Physical Review Letters}
  \textbf{\bibinfo{volume}{74}}, \bibinfo{pages}{5144} (\bibinfo{year}{1995}).

\bibitem[{\citenamefont{Wellein and Fehske}(1998)}]{Wellein1998}
\bibinfo{author}{\bibfnamefont{G.}~\bibnamefont{Wellein}} \bibnamefont{and}
  \bibinfo{author}{\bibfnamefont{H.}~\bibnamefont{Fehske}},
  \bibinfo{journal}{Phys. Rev. B} \textbf{\bibinfo{volume}{58}},
  \bibinfo{pages}{6208} (\bibinfo{year}{1998}).

\bibitem[{\citenamefont{Milburn et~al.}(1997)\citenamefont{Milburn, Corney,
  Wright, and Walls}}]{Milburn1997}
\bibinfo{author}{\bibfnamefont{G.~J.} \bibnamefont{Milburn}},
  \bibinfo{author}{\bibfnamefont{J.}~\bibnamefont{Corney}},
  \bibinfo{author}{\bibfnamefont{E.~M.} \bibnamefont{Wright}},
  \bibnamefont{and} \bibinfo{author}{\bibfnamefont{D.~F.} \bibnamefont{Walls}},
  \bibinfo{journal}{Phys. Rev. A} \textbf{\bibinfo{volume}{55}},
  \bibinfo{pages}{4318} (\bibinfo{year}{1997}).

\bibitem[{\citenamefont{Smerzi et~al.}(1997)\citenamefont{Smerzi, Fantoni,
  Giovanazzi, and Shenoy}}]{Smerzi1997}
\bibinfo{author}{\bibfnamefont{A.}~\bibnamefont{Smerzi}},
  \bibinfo{author}{\bibfnamefont{S.}~\bibnamefont{Fantoni}},
  \bibinfo{author}{\bibfnamefont{S.}~\bibnamefont{Giovanazzi}},
  \bibnamefont{and} \bibinfo{author}{\bibfnamefont{S.~R.}
  \bibnamefont{Shenoy}}, \bibinfo{journal}{Phys. Rev. Lett.}
  \textbf{\bibinfo{volume}{79}}, \bibinfo{pages}{4950} (\bibinfo{year}{1997}).

\bibitem[{\citenamefont{Raghavan et~al.}(1999)\citenamefont{Raghavan, Smerzi,
  Fantoni, and Shenoy}}]{Raghavan1999}
\bibinfo{author}{\bibfnamefont{S.}~\bibnamefont{Raghavan}},
  \bibinfo{author}{\bibfnamefont{A.}~\bibnamefont{Smerzi}},
  \bibinfo{author}{\bibfnamefont{S.}~\bibnamefont{Fantoni}}, \bibnamefont{and}
  \bibinfo{author}{\bibfnamefont{S.~R.} \bibnamefont{Shenoy}},
  \bibinfo{journal}{Phys. Rev. A} \textbf{\bibinfo{volume}{59}},
  \bibinfo{pages}{620} (\bibinfo{year}{1999}).

\bibitem[{\citenamefont{Trombettoni and Smerzi}(2001)}]{Trombettoni2001}
\bibinfo{author}{\bibfnamefont{A.}~\bibnamefont{Trombettoni}} \bibnamefont{and}
  \bibinfo{author}{\bibfnamefont{A.}~\bibnamefont{Smerzi}},
  \bibinfo{journal}{Phys. Rev. Lett.} \textbf{\bibinfo{volume}{86}},
  \bibinfo{pages}{2353} (\bibinfo{year}{2001}).

\bibitem[{\citenamefont{Morsch et~al.}(2002)\citenamefont{Morsch, Cristiani,
  Muller, Ciampini, and Arimondo}}]{Morsch2002}
\bibinfo{author}{\bibfnamefont{O.}~\bibnamefont{Morsch}},
  \bibinfo{author}{\bibfnamefont{M.}~\bibnamefont{Cristiani}},
  \bibinfo{author}{\bibfnamefont{J.~H.} \bibnamefont{Muller}},
  \bibinfo{author}{\bibfnamefont{D.}~\bibnamefont{Ciampini}}, \bibnamefont{and}
  \bibinfo{author}{\bibfnamefont{E.}~\bibnamefont{Arimondo}},
  \bibinfo{journal}{Phys. Rev. A} \textbf{\bibinfo{volume}{66}},
  \bibinfo{pages}{021601(R)} (\bibinfo{year}{2002}).

\bibitem[{\citenamefont{Smerzi and Trombettoni}(2003)}]{Smerzi2003}
\bibinfo{author}{\bibfnamefont{A.}~\bibnamefont{Smerzi}} \bibnamefont{and}
  \bibinfo{author}{\bibfnamefont{A.}~\bibnamefont{Trombettoni}},
  \bibinfo{journal}{Phys. Rev. A} \textbf{\bibinfo{volume}{68}},
  \bibinfo{pages}{023613} (\bibinfo{year}{2003}).

\bibitem[{\citenamefont{Albiez et~al.}(2005)}]{Albiez2005}
\bibinfo{author}{\bibfnamefont{M.}~\bibnamefont{Albiez}} \bibnamefont{et~al.},
  \bibinfo{journal}{Phys. Rev. Lett.} \textbf{\bibinfo{volume}{95}},
  \bibinfo{pages}{010402} (\bibinfo{year}{2005}).

\bibitem[{\citenamefont{Anker et~al.}(2005)}]{Anker2005}
\bibinfo{author}{\bibfnamefont{T.}~\bibnamefont{Anker}} \bibnamefont{et~al.},
  \bibinfo{journal}{Phys. Rev. Lett.} \textbf{\bibinfo{volume}{94}},
  \bibinfo{pages}{020403} (\bibinfo{year}{2005}).

\bibitem[{\citenamefont{Alexander et~al.}(2006)\citenamefont{Alexander,
  Ostrovskaya, and Kivshar}}]{Alexander2006}
\bibinfo{author}{\bibfnamefont{T.~J.} \bibnamefont{Alexander}},
  \bibinfo{author}{\bibfnamefont{E.~A.} \bibnamefont{Ostrovskaya}},
  \bibnamefont{and} \bibinfo{author}{\bibfnamefont{Y.~S.}
  \bibnamefont{Kivshar}}, \bibinfo{journal}{Phys. Rev. Lett}
  \textbf{\bibinfo{volume}{96}}, \bibinfo{pages}{040401}
  (\bibinfo{year}{2006}).

\bibitem[{\citenamefont{Xue et~al.}(2008)\citenamefont{Xue, Zhang, and
  Liu}}]{Xue2008}
\bibinfo{author}{\bibfnamefont{J.-K.} \bibnamefont{Xue}},
  \bibinfo{author}{\bibfnamefont{A.-X.} \bibnamefont{Zhang}}, \bibnamefont{and}
  \bibinfo{author}{\bibfnamefont{J.}~\bibnamefont{Liu}},
  \bibinfo{journal}{Phys. Rev. A} \textbf{\bibinfo{volume}{77}},
  \bibinfo{pages}{013602} (\bibinfo{year}{2008}).

\bibitem[{\citenamefont{W$\text{\" u}$ster
  et~al.}(2012)\citenamefont{W$\text{\" u}$ster, Dabrowska-W$\text{\" u}$ster,
  and Davis}}]{wuster2012}
\bibinfo{author}{\bibfnamefont{S.}~\bibnamefont{W$\text{\" u}$ster}},
  \bibinfo{author}{\bibfnamefont{B.~J.} \bibnamefont{Dabrowska-W$\text{\"
  u}$ster}}, \bibnamefont{and} \bibinfo{author}{\bibfnamefont{M.~J.}
  \bibnamefont{Davis}}, \bibinfo{journal}{Phys. Rev. Lett}
  \textbf{\bibinfo{volume}{109}}, \bibinfo{pages}{080401}
  (\bibinfo{year}{2012}).

\bibitem[{\citenamefont{Reinhard et~al.}(2013)\citenamefont{Reinhard, Riou,
  Zundel, Weiss, Li, Rey, and Hipolito}}]{Aaron}
\bibinfo{author}{\bibfnamefont{A.}~\bibnamefont{Reinhard}},
  \bibinfo{author}{\bibfnamefont{J.-F.} \bibnamefont{Riou}},
  \bibinfo{author}{\bibfnamefont{L.~A.} \bibnamefont{Zundel}},
  \bibinfo{author}{\bibfnamefont{D.~S.} \bibnamefont{Weiss}},
  \bibinfo{author}{\bibfnamefont{S.}~\bibnamefont{Li}},
  \bibinfo{author}{\bibfnamefont{A.~M.} \bibnamefont{Rey}}, \bibnamefont{and}
  \bibinfo{author}{\bibfnamefont{R.}~\bibnamefont{Hipolito}},
  \bibinfo{journal}{Phys. Rev. Lett} \textbf{\bibinfo{volume}{110}},
  \bibinfo{pages}{033001} (\bibinfo{year}{2013}).

\bibitem[{\citenamefont{Girardeau}(1960)}]{Girardeau1960}
\bibinfo{author}{\bibfnamefont{M.}~\bibnamefont{Girardeau}},
  \bibinfo{journal}{Journal of Mathematical Physics}
  \textbf{\bibinfo{volume}{1}}, \bibinfo{pages}{516} (\bibinfo{year}{1960}).

\bibitem[{\citenamefont{Kinoshita et~al.}(2004)\citenamefont{Kinoshita, Wenger,
  and Weiss}}]{Kinoshita2004}
\bibinfo{author}{\bibfnamefont{T.}~\bibnamefont{Kinoshita}},
  \bibinfo{author}{\bibfnamefont{T.}~\bibnamefont{Wenger}}, \bibnamefont{and}
  \bibinfo{author}{\bibfnamefont{D.~S.} \bibnamefont{Weiss}},
  \bibinfo{journal}{Science} \textbf{\bibinfo{volume}{305}},
  \bibinfo{pages}{1125} (\bibinfo{year}{2004}).

\bibitem[{\citenamefont{Paredes et~al.}(2004)\citenamefont{Paredes, Widera,
  Murg, Mandel, F$\text{\" o}$lling, Cirac, Shlyapnikov, H$\text{\" a}$nsch,
  and Bloch}}]{Paredes2004tg}
\bibinfo{author}{\bibfnamefont{B.}~\bibnamefont{Paredes}},
  \bibinfo{author}{\bibfnamefont{A.}~\bibnamefont{Widera}},
  \bibinfo{author}{\bibfnamefont{V.}~\bibnamefont{Murg}},
  \bibinfo{author}{\bibfnamefont{O.}~\bibnamefont{Mandel}},
  \bibinfo{author}{\bibfnamefont{S.}~\bibnamefont{F$\text{\" o}$lling}},
  \bibinfo{author}{\bibfnamefont{I.}~\bibnamefont{Cirac}},
  \bibinfo{author}{\bibfnamefont{G.~V.} \bibnamefont{Shlyapnikov}},
  \bibinfo{author}{\bibfnamefont{T.~W.} \bibnamefont{H$\text{\" a}$nsch}},
  \bibnamefont{and} \bibinfo{author}{\bibfnamefont{I.}~\bibnamefont{Bloch}},
  \bibinfo{journal}{Nature} \textbf{\bibinfo{volume}{429}},
  \bibinfo{pages}{277} (\bibinfo{year}{2004}).

\bibitem[{\citenamefont{Polkovnikov}(2010)}]{Polkovnikov2010}
\bibinfo{author}{\bibfnamefont{A.}~\bibnamefont{Polkovnikov}},
  \bibinfo{journal}{Annals of Physics} \textbf{\bibinfo{volume}{325}},
  \bibinfo{pages}{1790} (\bibinfo{year}{2010}).

\bibitem[{\citenamefont{Schollw$\text{\" o}$ck}(2005)}]{Schollwock2005}
\bibinfo{author}{\bibfnamefont{U.}~\bibnamefont{Schollw$\text{\" o}$ck}},
  \bibinfo{journal}{Rev. Mod. Phys} \textbf{\bibinfo{volume}{77}},
  \bibinfo{pages}{259} (\bibinfo{year}{2005}).

\bibitem[{\citenamefont{Daley et~al.}(2004)\citenamefont{Daley, Kollath,
  Schollw$\text{\" o}$ck, and Vidal}}]{Daley2004}
\bibinfo{author}{\bibfnamefont{A.~J.} \bibnamefont{Daley}},
  \bibinfo{author}{\bibfnamefont{C.}~\bibnamefont{Kollath}},
  \bibinfo{author}{\bibfnamefont{U.}~\bibnamefont{Schollw$\text{\" o}$ck}},
  \bibnamefont{and} \bibinfo{author}{\bibfnamefont{G.}~\bibnamefont{Vidal}},
  \bibinfo{journal}{J. Stat. Mech} p. \bibinfo{pages}{04005}
  (\bibinfo{year}{2004}).

\bibitem[{\citenamefont{Pethick and Smith}(2002)}]{pethick2002}
\bibinfo{author}{\bibfnamefont{C.~J.} \bibnamefont{Pethick}} \bibnamefont{and}
  \bibinfo{author}{\bibfnamefont{H.}~\bibnamefont{Smith}},
  \emph{\bibinfo{title}{Bose-Einstein Condensation in Dilute Gases}}
  (\bibinfo{publisher}{Cambridge University Press},
  \bibinfo{address}{Cambridge}, \bibinfo{year}{2002}).

\bibitem[{\citenamefont{Blakie et~al.}(2008)\citenamefont{Blakie, Bradley,
  Davis, Ballagh, and Gardiner}}]{Blakie2008}
\bibinfo{author}{\bibfnamefont{P.~B.} \bibnamefont{Blakie}},
  \bibinfo{author}{\bibfnamefont{A.~S.} \bibnamefont{Bradley}},
  \bibinfo{author}{\bibfnamefont{M.~J.} \bibnamefont{Davis}},
  \bibinfo{author}{\bibfnamefont{R.~J.} \bibnamefont{Ballagh}},
  \bibnamefont{and} \bibinfo{author}{\bibfnamefont{C.~W.}
  \bibnamefont{Gardiner}}, \bibinfo{journal}{Advances in Physcs}
  \textbf{\bibinfo{volume}{57:5}}, \bibinfo{pages}{363} (\bibinfo{year}{2008}).

\bibitem[{\citenamefont{A.Sinatra et~al.}(2002)\citenamefont{A.Sinatra, Lobo,
  and Y.Castin}}]{Sinatra2002}
\bibinfo{author}{\bibnamefont{A.Sinatra}},
  \bibinfo{author}{\bibfnamefont{C.}~\bibnamefont{Lobo}}, \bibnamefont{and}
  \bibinfo{author}{\bibnamefont{Y.Castin}}, \bibinfo{journal}{J. Phys. B}
  \textbf{\bibinfo{volume}{35}}, \bibinfo{pages}{3599} (\bibinfo{year}{2002}).

\bibitem[{\citenamefont{Hipolito and Polkovnikov}(2010)}]{Rafael2010}
\bibinfo{author}{\bibfnamefont{R.}~\bibnamefont{Hipolito}} \bibnamefont{and}
  \bibinfo{author}{\bibfnamefont{A.}~\bibnamefont{Polkovnikov}},
  \bibinfo{journal}{Phys. Rev. A} \textbf{\bibinfo{volume}{81}},
  \bibinfo{pages}{013621} (\bibinfo{year}{2010}).

\bibitem[{\citenamefont{Dunjko et~al.}(2001)\citenamefont{Dunjko, Lorent, and
  Olshanii}}]{Dunjko2001}
\bibinfo{author}{\bibfnamefont{V.}~\bibnamefont{Dunjko}},
  \bibinfo{author}{\bibfnamefont{V.}~\bibnamefont{Lorent}}, \bibnamefont{and}
  \bibinfo{author}{\bibfnamefont{M.}~\bibnamefont{Olshanii}},
  \bibinfo{journal}{Phys. Rev. Lett} \textbf{\bibinfo{volume}{86}},
  \bibinfo{pages}{5413} (\bibinfo{year}{2001}).

\bibitem[{\citenamefont{Kinoshita et~al.}(2005)\citenamefont{Kinoshita, Wenger,
  and Weiss}}]{Kinoshita2005}
\bibinfo{author}{\bibfnamefont{T.}~\bibnamefont{Kinoshita}},
  \bibinfo{author}{\bibfnamefont{T.}~\bibnamefont{Wenger}}, \bibnamefont{and}
  \bibinfo{author}{\bibfnamefont{D.~S.} \bibnamefont{Weiss}},
  \bibinfo{journal}{Phys. Rev. Lett.} \textbf{\bibinfo{volume}{95}},
  \bibinfo{pages}{190406} (\bibinfo{year}{2005}).

\bibitem[{\citenamefont{Bloch et~al.}(2008)\citenamefont{Bloch, Dalibard, and
  Zwerger}}]{Bloch2008r}
\bibinfo{author}{\bibfnamefont{I.}~\bibnamefont{Bloch}},
  \bibinfo{author}{\bibfnamefont{J.}~\bibnamefont{Dalibard}}, \bibnamefont{and}
  \bibinfo{author}{\bibfnamefont{W.}~\bibnamefont{Zwerger}},
  \bibinfo{journal}{Reviews of Modern Physics} \textbf{\bibinfo{volume}{80}},
  \bibinfo{pages}{885} (\bibinfo{year}{2008}).

\end{thebibliography}
\end{document}